\documentclass[aip,prb,unsortedaddress,superscriptaddress,reprint]{revtex4-1}
\usepackage{hyperref}
\usepackage{graphicx}
\usepackage{amsmath}
\usepackage{amssymb}
\usepackage{times,mathptmx}
\usepackage{dcolumn}

\newif\ifcom
\newif\ifdel
\comtrue               %
\delfalse

\begin{document}

\title{Pure spin currents in magnetically ordered insulator/normal metal heterostructures}
\author{Matthias Althammer}
\email{Matthias.Althammer@wmi.badw-munchen.de}
\affiliation{Walther-Meissner-Institute, Bayerische Akademie der Wissenschaften, 85748 Garching, Germany}
\date{\today}
\begin{abstract}
Pure spin currents, i.e.~the transport of angular momentum without an accompanying charge current, represent a new, promising avenue in modern spintronics from both a fundamental and an application point of view. Such pure spin currents can not only flow in electrical conductors via mobile charge carriers, but also in magnetically ordered electrical insulators as a flow of spin excitation quanta. Over the course of the last years remarkable results have been obtained in heterostructures consisting of magnetically ordered insulators interfaced with a normal metal, where a pure spin current flows across the interface.

This topical review article deals with the fundamental principles, experimental findings and recent developments in the field of pure spin currents in magnetically ordered insulators. We here put our focus onto four different manifestations of pure spin currents in such heterostructures: The spin pumping effect, the longitudinal spin Seebeck effect, the spin Hall magnetoresistance and the all-electrical detection of magnon transport in non-local device concepts. In this article, we utilize a common theoretical framework to explain all four effects and explain important material systems (especially rare-earth iron garnets) used in the experiments. For each effect we introduce basic measurement techniques and detection schemes and discuss their application in the experiment. We account for the remarkable progress achieved in each field by reporting the recent progress in each field and by discussing research highlights obtained in our group. Finally, we conclude the review article with an outlook on future challenges and obstacles in the field of pure spin currents in magnetically ordered insulator / normal metal heterostructures.

This topical review aims to be a useful resource for introducing readers from the outside or just starting in this field, but also for providing perspective to those that already have an established understanding of the underlying physics.
\end{abstract}
\maketitle

\section{Introduction}
\label{Intro}
Modern day electronics, that hinges on charge current transport, is nowadays approaching the fundamental limit and it currently seems that the famous Moore's law is coming to a stop~\cite{waldrop_chips_2016}. The search is now on for novel approaches to store and process information that go beyond simple charge currents~\cite{bourianoff_nanoelectronics_2010}. In the realm of spintronics the so-called pure spin currents, i.e.~the net flow of (spin) angular momentum without an accompanying charge current, represent a new paradigm and are a promising candidate for such novel approaches~\cite{sander_2017_2017}. Over the last decade a lot of fundamental research work has been dedicated to find means to generate and detect such pure spin currents~\cite{hoffmann_pure_2007}.

A very intriguing property of pure spin currents is that they can not only flow in electrical conductors, where the angular momentum is carried by mobile charge carriers, but also in magnetically ordered electrical insulators via magnons (spin excitation quanta). This allows for new interesting device concepts for magnon based information processing~\cite{Chumak2015,Chumak2017,nakata_spin_2017}.

While in the first years of pure spin current centered research the role of magnetically ordered insulators (MOIs) was to rule out spurious contributions in the experiment, they now represent a cornerstone for modern spintronic device concepts. This was only possible due to the progress made both in theory and experiment focused on pure spin currents and requires to interface the MOIs with normal metals~\cite{tserkovnyak_nonlocal_2005,hoffmann_pure_2007,Bauer2012}. A normal metal(NM) is an electrical conductor without magnetic ordering. In such MOI/NM heterostructures most commonly the spin Hall and inverse spin Hall effect caused by spin-orbit interaction in the normal metal are used to electrically detect and even generate pure spin currents~\cite{Dyakonov1971,SHE:Hirsch:PRL:1999,Hoffmann2013,Sinova2015}. In addition, the interface between the MOI and the NM and its transparency for pure spin currents play a crucial role in such type of experiments~\cite{brataas_finite-element_2000,Tserkovnyak2002,tserkovnyak_nonlocal_2005,Bauer2012,Adachi2013,Bender2015}. In the end, this allows to detect the pure spin current flow across the interface using electrical measurement schemes.

A prominent example for the usage of MOI/NM heterostructures are spin pumping experiments, where the excited magnetic order parameter in the MOI pumps a pure spin current across the interface into the NM~\cite{Tserkovnyak2002,brataas_spin-pumping_2004,tserkovnyak_nonlocal_2005,Costache:vanWees:spin-pumping:experiment:PRL2006,spin-pumping:saitoh:APL:2006,woltersdorf_magnetization_2007,Mosendz2010,Ando2010,ando_inverse_2011,Czeschka2011,ando_observation_2012,kapelrud_spin_2013}. An important result from these experiments was that the interface of MOI/NM heterostructures is as transparent as all metallic ferromagnet/NM structures, showing that pure spin current transport across the interface can be very efficient in MOI/NM systems~\cite{Mosendz2010}. In the early stages, spin pumping experiments where mostly dealing with the time-invariant part of the injected spin current in the NM, while quite recently the time-varying part of the pure spin current has been put into focus as it may pave the way to high processing speeds up to the THz regime~\cite{Hahn2013,Kampfrath2013,Cheng2014,Wei2014,Weiler2014,huisman_femtosecond_2016,khymyn_transformation_2016,li_direct_2016,Seifert2016,bocklage_coherent_2017,huisman_spin-photo-currents_2017,Johansen2017,kapelrud_spin_2017,semenov_spin_2017,seifert_ultrabroadband_2017}. Moreover, the reciprocal effect of spin pumping enables us to drive magnetization dynamics and spin waves by applying a DC charge current bias to MOI/NM systems ~\cite{Collet2016,chen_spin-torque_2016,jungfleisch_insulating_2017,evelt_spin_2018}.

The longitudinal spin Seebeck effect showed that also a thermal gradient can be used to generate magnetic excitations in the MOI and drive a pure spin current across the interface into the NM, where it can be detected as an electrical voltage (current)~\cite{Uchida:2010,Uchida2010,Xiao2010,Adachi2013,Uchida2014,Rezende2014,Cahaya2015}. Here, MOIs played a crucial role to rule out other possible sources for the experimentally observed signals~\cite{Uchida:2010,Uchida2010}. The spin Seebeck effect allows to use MOI/NM heterostructures for waste energy recycling by either generating electrical currents by the inverse spin Hall effect, or directly using the spin current for information processing tasks~\cite{Uchida2014}. Another important contribution from spin Seebeck effect research led to a deeper understanding of magnon excitations in MOIs that were previously only experimentally attainable by neutron and inelastic light scattering experiments~\cite{Geprgs2016}.

A next crucial step was the discovery of a novel magnetoresistance effect in MOI/NM heterostructures, where the resistance of the NM depends on the orientation of the magnetic order parameter~\cite{althammer_quantitative_2013,chen_theory_2013,Nakayama2013,Hahn2013SMR,Vlietstra2013,Chen2016SMRReview}. This effect is called the spin Hall magnetoresistance and crucially hinges on the charge based spin current generation and detection via the spin Hall and inverse spin Hall effect and the tunability of the spin current flow across the interface via the orientation of the magnetic order parameter~\cite{chen_theory_2013}. Initially, this effect allowed to infer important spin transport parameters from simple electrical transport experiments~\cite{althammer_quantitative_2013,Vlietstra2013}, but it has also proven its usefulness in detecting complex magnetic phases (e.g.~helical, spin-canting and spin-flop ordering) in MOIs~\cite{Aqeel2015,aqeel_electrical_2016,Ganzhorn2016,Han2014,hoogeboom_negative_2017,ji_spin_2017,hou_tunable_2017,manchon_spin_2017,fischer_spin_2018}.

Last but not least, the experimental observation of the long distance magnon transport using all-electrical techniques two years ago has laid the foundation for interfacing charge based information processing with magnon logic~\cite{Zhang2012,Zhang2012_PRB,Cornelissen2015,Goennenwein2015}. This effect can be thought of as the non-local analogue of the spin Hall magnetoresistance: Two NM layers are separated by a MOI from each other, while a charge current is flowing through one of the normal metal layers, a non-local voltage can be detected in the other NM layer. The magnitude of the non-local voltage depends crucially on the orientation of the magnetic order parameter in the MOI and can be explained by thermally activated inelastic scattering processes at the NM/MOI interfaces, that allow to transfer a fraction of the pure spin current generated from the charge current by the spin Hall effect in the first NM layer via magnons in the MOI to the second NM layer, where the inverse spin Hall effect transfers the spin current back into a charge current for electrical detection.

In this topical review we will cover these 4 different areas of pure spin current in MOI/NM heterostructures. In Section~\ref{theory} we first discuss the theoretical framework that explain the underlying principles of charge based pure spin current generation and detection and the flow of pure spin currents across the MOI/NM interface. This section is followed with a more detailed description of relevant material systems used in the experiment in Sec.~\ref{materials}. As a next step, we discuss the spin pumping effect and highlight the experimental detection of spin pumping using broadband ferromagnetic resonance and electrical detection techniques in Sec.~\ref{spin_pumping}. This is then followed up by a discussion of the longitudinal spin Seebeck effect and its dependence on the magnon bandstructure experimentally observed in compensated rare-earth iron garnets in Section~\ref{spin_seebeck}. The spin Hall magnetoresistance and the application of this magnetoresistance effect to detect non-collinear magnetic phases is covered in Sec.~\ref{spin_hall_magnetoresistance}. In Section~\ref{magnon_mediated_magnetoresistance}, we then review the recent progress in all-electrical spin Hall effect based magnon transport experiments. We summarize the presented results briefly in Section~\ref{conclusion} and give an outlook into interesting future pathways for pure spin current research in MOI/NM heterostructures in Section~\ref{outlook}.

\section{Pure spin currents across MOI/NM interfaces: basic theoretical framework}
\label{theory}
In this section we explain the most fundamental principles of pure spin current physics and their implications for experiments. As a first step the physical concept of a pure spin current is established. This is the followed by discussing the generation and detection via the spin Hall and inverse spin Hall effect. As a last step we look into the transport of a pure spin current across the MOI/NM interface, where elastic and inelastic spin-flip scattering processes at the interface of the charge carriers in the NM give rise to the transfer of angular momentum between the MOI and the NM.

\subsection{Pure spin currents}
In an electrical conductor the charge carriers not only posses a charge, but also a (spin) angular-momentum degree of freedom. From this perspective, the flow of charge carriers not only allows for the transport of charge, i.e. an electrical current, but also enables the transport of angular momentum, i.e. a spin current~\cite{tserkovnyak_nonlocal_2005,hoffmann_pure_2007,Bauer2012}. In addition, charge carriers transport energy leading to the flow of heat currents. In the following we keep the focus on charge and spin currents and assume that the charge carriers are electrons with negative elementary charge $e$.

\begin{figure}[b]
 \includegraphics[width=85mm]{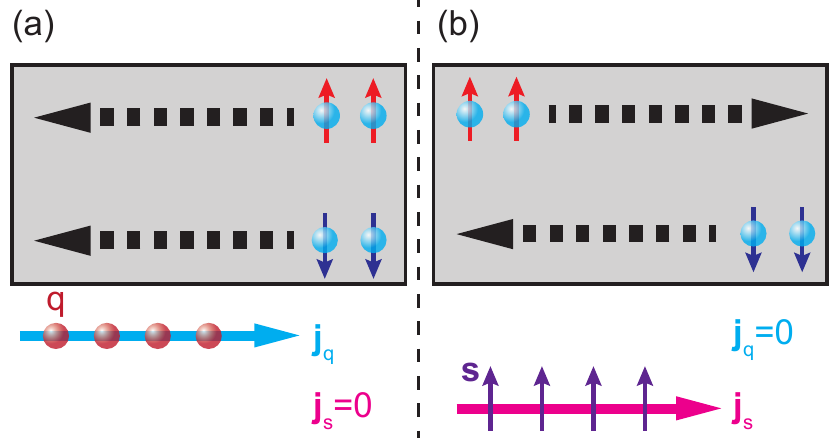}\\
 \caption[Illustration of a pure charge and a pure spin current]{(a) Illustration of a pure charge current, the same number of spin-up and spin-down electrons move in the same direction. This leads to a net charge current density $\mathbf{j}_\mathrm{q}$, while the spin current density $\mathbf{j}_\mathrm{s}$ vanishes. (b) Flow of a pure spin current. The same number of spin-up and spin-down electrons move in opposite directions, such that the net charge current is zero, while a finite amount of angular momentum is transported and a finite pure spin current density flows.}
  \label{figure:PureSpin}
\end{figure}

A very simple picture of of a pure charge and a pure spin current can be drawn using the two spin-channel model established by Juliere in 1975~\cite{julliere_tunneling_1975}. Within this model we consider two independent contributions to the total charge and spin current transport stemming from the spin-up and spin-down electrons. In this way we obtain the charge current densities $\mathbf{j}_\uparrow$ for the spin-up electrons and $\mathbf{j}_\downarrow$ for the spin-down electrons. The total charge current density $\mathbf{j}_\mathrm{q}$ is then the sum of these two quantities: $\mathbf{j}_\mathrm{q}=\mathbf{j}_\uparrow+\mathbf{j}_\downarrow$. In analogous fashion, the pure spin current density $\mathbf{j}_\mathrm{s}$ in units of $\hbar$ per unit area is given by $\mathbf{j}_\mathrm{s}=\hbar/(2e)(\mathbf{j}_\uparrow-\mathbf{j}_\downarrow)$, with $\hbar$ the reduced Planck constant. We first consider a pure charge current in this model as illustrated in Fig.~\ref{figure:PureSpin}(a). Here, the same number of spin-up and spin-down electrons move in the same direction. Thus, the sum of and also $\mathbf{j}_\mathrm{q}$ remain finite while $\mathbf{j}_\mathrm{s}=0$ and no angular momentum is transported in this scenario. On the other hand, Fig.~\ref{figure:PureSpin}(b) depicts the realization of a pure spin current. Now, the same number of spin-up and spin-down electrons move in opposite directions. As a result $\mathbf{j}_\mathrm{q}$ is zero, while we now obtain a finite $\mathbf{j}_\mathrm{s}$ and thus only a flow of angular momentum is realized without an accompanying charge current flow.

In more detail, $\mathbf{j}_\mathrm{q}$ represents the transport of electron charge and can be written as $\mathbf{j}_\mathrm{q}=-en \langle\mathbf{v}\rangle$, where $n$ is the density of the electrons, $\mathbf v$ is the velocity operator, and $\langle...\rangle$ denotes the thermodynamic expectation value for a non-equilibrium state. Similarly, we can write the spin current density as $\mathbf{j}_\mathrm{s}=\hbar/2 n \langle\mathbf v \otimes \boldsymbol{ \sigma}\rangle$, where $\boldsymbol \sigma$ is the vector of Pauli spin matrices~\cite{Bauer2012}. At a first glance these definition seem to represent similar quantities, however, the vector $\mathbf{j}_\mathrm{q}$ describes the transport of a scalar, i.e. the electron charge, and the second rank tensor $\mathbf{j}_\mathrm{s}$ describes the transport of an axial vector, i.e. angular momentum. In this regard, a pure spin current not only has a direction of flow, but also an orientation of (spin) angular momentum. One should also keep in mind that charge is a conserved transport quantity. In contrast, angular momentum is only conserved on the length scale of the spin-flip length $\lambda_{\mathrm{sf}}$, because the angular momentum can be transferred for example to phonons. A very intriguing property of pure spin currents is that they can not only flow in electrical conductors via charge carriers, but also in electrical, magnetically ordered insulators (MOIs) via magnetic excitation quanta (e.g.~magnons). In addition, $\mathbf{j}_\mathrm{q}$ can be driven either by a gradient in the electrochemical potential or a temperature gradient, while for $\mathbf{j}_\mathrm{s}$ a gradient in the spin-dependent electrochemical potential or in temperature act as a driving force. As evident from this discussion, pure charge and pure spin currents are rather different physical entities.

\subsection{Spin Hall effect}

A first major obstacle for the investigation of pure spin currents was to find means to generate and detect pure spin currents. The spin Hall effect (SHE) and inverse spin Hall effect (ISHE) allow to transform a charge current into a spin current and vice versa, enabling all electrical access to spin current physics. While a phenomenological description of the SHE has already been put forward by D'yakonov and Perel'~\cite{Dyakonov1971} in 1971, interest into these effects started to increase 3 decades later, when Hirsch~\cite{SHE:Hirsch:PRL:1999} published his theoretical description of this effect and coined the term spin Hall effect. This was the spark that lead to a series of publications on the SHE in theory and experiment. Two very good review articles that cover all these theoretical and experimental observations have now been published by Hoffmann~\cite{Hoffmann2013} and Sinova \textit{et al.}~\cite{Sinova2015}. It is worth mentioning that the SHE is the more general manifestation of the anomalous Hall effect~\cite{Hall1881,Onoda2008,nagaosa_anomalous_2010}.

The SHE and ISHE originate from spin-dependent transverse velocities that the charge carriers acquire when moving through an electrical conductor with finite spin-orbit interaction. These spin-dependent transverse velocities arise due to extrinsic and intrinsic effects. The term extrinsic effects represents scattering events of the electron with impurities, phonons, etc., which lead to a finite transverse velocity depending on the spin orientation of the electron. Prominent examples are the skew-scattering\cite{Smit1958} and side-jump scattering~\cite{Berger1970}. Intrinsic effects are bandstructure effects that lead to a finite Berry phase and thus in the end also to a spin-dependent transverse velocity~\cite{xiao_berry_2010}.

\begin{figure}[t]
 \includegraphics[width=85mm]{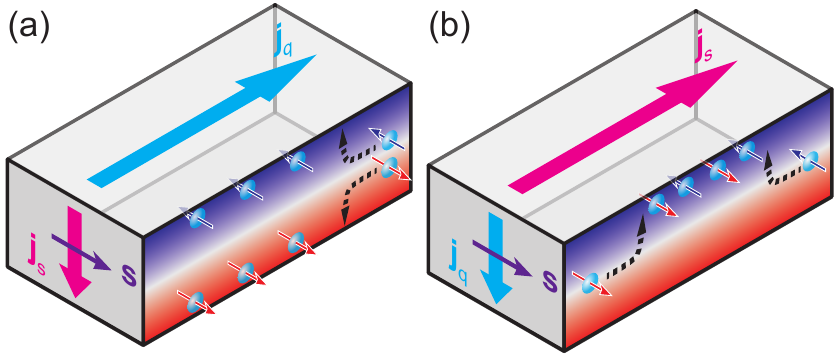}\\
 \caption[Illustration of the spin Hall and inverse Spin Hall effect]{(a) The spin Hall effect transforms a charge current density $\mathbf{j}_\mathrm{q}$ into a pure spin current density $\mathbf{j}_\mathrm{s}$ with spin direction $\mathbf{s}$. In the steady state a spin accumulation is generated at the sample edges, which leads to a gradient in the spin-dependent electrochemical potential and compensates $\mathbf{j}_\mathrm{s}$. (b) A pure spin current density $\mathbf{j}_\mathrm{s}$ with spin orientation $\mathbf{s}$ is transformed into a charge current density $\mathbf{j}_\mathrm{q}$ via the inverse spin Hall effect. The charge current leads to a charge accumulation at the sample surface, which leads to an electric field compensating the charge current flow.}
  \label{figure:SpinHall}
\end{figure}

In more detail, the SHE allows to transform a charge current density $\mathbf{j}_\mathrm{q}$ into a pure spin current density $\mathbf{j}_\mathrm{s}$ with spin orientation $\mathbf{s}$ that is perpendicular to both $\mathbf{j}_\mathrm{q}$ and $\mathbf{j}_\mathrm{s}$. Thus one can write this conversion as
\begin{equation}
\mathbf{j}_\mathrm{s}=\alpha_\mathrm{SH}\left(-\frac{\hbar}{2e}\right)\mathbf{j}_\mathrm{q}\times\mathbf{s}.
\label{equ:SHE}
\end{equation}
The spin Hall angle $\alpha_\mathrm{SH}$ is material dependent parameter that reflects the magnitude of the spin-dependent scattering effects. The SHE thus allows to generate a pure spin current in a material without magnetic order, but finite spin-orbit coupling. Figure \ref{figure:SpinHall}(a) illustrates the relevant process for the SHE. Due to the applied $\mathbf{j}_\mathrm{q}$ the same number of spin-up and spin-down electrons move in the same direction. Due to the spin-dependent transverse velocity acquired while traversing through the material, spin-up and spin-down electrons are deflected into opposite directions and thus a pure spin current flows along this transverse direction with the spin orientation $\mathbf{s}\perp\mathbf{j}_\mathrm{q},\mathbf{j}_\mathrm{s}$. In the steady state and under open circuit boundary conditions, this spin current leads to a spin accumulation at the sample edges and thus a gradient in the spin-dependent electrochemical potential, which counteracts the pure spin current generated via the SHE, such that there is no net transverse spin current flow. The very first optical experiments used these spin accumulations to detect the SHE in semiconducting materials~\cite{spin-currents:Kato:Science:2004,Wunderlich2005}. Non-local spin valve experiments enabled all-electrical measurements of the spin Hall effect~\cite{Valenzuela:2006}.

Due to Onsager reciprocity the inverse process, the ISHE, as illustrated in Fig.~\ref{figure:SpinHall}(b) also has to exist in electrical conductors with finite spin-orbit coupling. We first consider a pure spin current $\mathbf{j}_\mathrm{s}$, with spin-up and spin-down electrons flowing in opposite directions. As both the spin direction and direction of movement are opposite, the spin-up and spin-down electrons are deflected in the same direction, due to the spin-dependent transverse velocity effects, and create a charge current $\mathbf{j}_\mathrm{q}$ given by~\cite{Hoffmann2013,Sinova2015}:
\begin{equation}
\mathbf{j}_\mathrm{q}=\alpha_\mathrm{SH}\left(-\frac{2e}{\hbar}\right)\mathbf{j}_\mathrm{s}\times\mathbf{s}.
\label{equ:ISHE}
\end{equation}
The vector product nature of this transformation thus only leads to a charge current if $\mathbf{j}_\mathrm{s}$ is non-collinear to $\mathbf{s}$. In this way, the ISHE enables the all-electrical detection of a pure spin current in an electrical conductor with finite $\alpha_\mathrm{SH}$~\cite{spin-pumping:saitoh:APL:2006}. Since both SHE and ISHE rely on spin-orbit coupling, large $\alpha_{\mathrm{SH}}$ are expected in heavy elements. Large spin Hall angles have been reported in materials such as platinum (Pt), tantalum (Ta), tungsten (W), gold (Au), or alloys such as CuBi, with $| \alpha_{\mathrm{SH}} | < 0.4$~\cite{Morota:2011,Liu2012Pt,Liu2012,Pai2012,Niimi:2012,Garello2013,Weiler:Solid-state-physics-64:2013}. We here restrict ourselves to the conversion of a charge current driven by a gradient in the electro-chemical potential into a spin current and vice versa. However, it is also possible to generate a pure spin current from a thermal gradient via the spin Nernst effect. This effect has been theoretically postulated to be present in materials with spin-orbit coupling, but only very recently the first experimental observation of this effect was possible using MOI/NM heterostructures~\cite{cheng_spin_2008,liu_spin_2010,meyer_observation_2017}.

While initially it was assumed that only the spin Hall effect can account for a transformation from charge currents to spin currents and back, it is now clear that due to the broken inversion symmetry at the interface (for example at the NM/MOI interface) additional effects like the spin galvanic effect have to be taken into account\cite{edelstein_spin_1990,ganichev_spin-galvanic_2002,sanchez_spin--charge_2013,jungfleisch_interface-driven_2016,seibold_theory_2017}. However, due to the short spin diffusion length in most materials with large spin Hall angle it is very difficult to disentangle the different contributions in the experiment, as they exhibit the very same symmetry. In this way Eqs. (\ref{equ:SHE}) and  (\ref{equ:ISHE}) can still be employed to describe the transformation of a charge current into a pure spin current, but the conversion efficiency given by $\alpha_\mathrm{SH}$ is an effective value of all the different effects contributing to this transformation.

\subsection{Spin currents across MOI/NM interfaces}
In the previous section we discussed the effects of the SHE and ISHE in an electrical conductor with finite spin-orbit coupling and the transformation of a charge current into a spin current and vice versa. As a next step, we look into the pure spin current transport across the MOI/NM interface~\cite{brataas_finite-element_2000,Tserkovnyak2002,tserkovnyak_nonlocal_2005,Bauer2012,Adachi2013,Bender2015}. For this we follow the theoretical framework outlined by Bender and Tserkovnyak~\cite{Bender2015}, which allows to describe spin pumping, spin seebeck effect, spin Hall magnetoresistance as well as magnon mediated magnetoresistance. While Bender and Tserkovnyak in their publication describe the spin and energy transport across the interface, we here mostly focus on the spin transport.

\begin{figure}[t]
 \includegraphics[width=85mm]{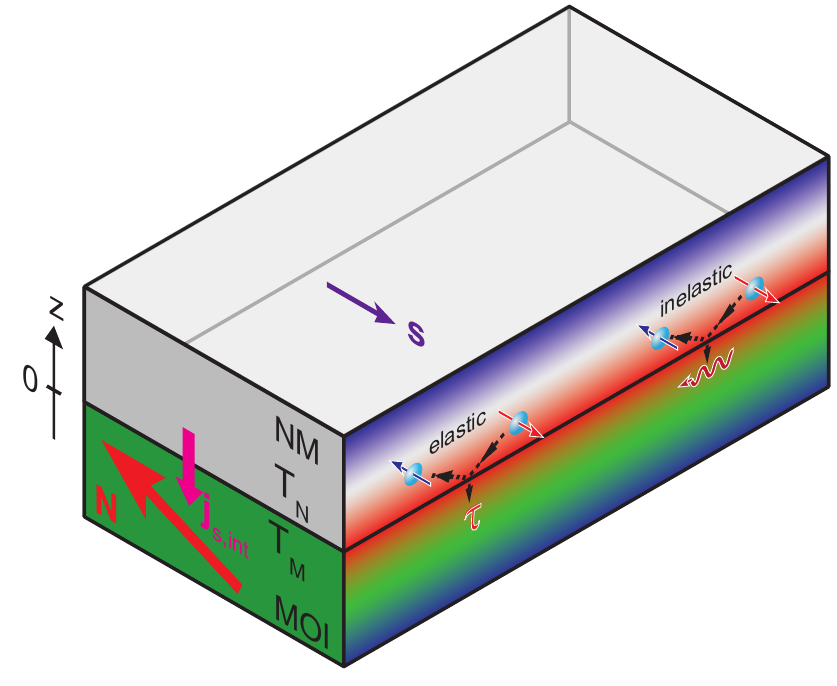}\\
 \caption[Spin currents across interfaces]{Illustration of the interfacial spin current $\mathbf{j}_\mathrm{s,int}$ flowing across a MOI/NM heterostructure. In the NM the orientation of the spin accumulation is given by $\mathbf{s}$. Moreover, the temperature $T_\mathrm{N}$ of the charge carriers in the NM is important for the magnitude of the spin current. In the MOI the orientation of the magnetic order parameter $\mathbf{N}$ and the temperature $T_\mathrm{M}$ influence the spin orientation and magnitude of the interfacial spin current. The microscopic processes relevant for the spin current across the interface are elastic and inelastic spin-flip scattering processes of the charge carriers in the NM at the MOI/NM interface, causing a torque $\boldsymbol{\tau}$ onto $\mathbf{N}$ or transferring angular momentum and energy to the magnetic excitation quanta in the MOI, respectively.}
  \label{figure:SpinCurrentInterface}
\end{figure}
First, we consider a MOI/NM heterostructure as illustrated in Fig.~\ref{figure:SpinCurrentInterface}, where a spin accumulation with spin orientation $\mathbf{s}$ persists in the NM and the magnetic order parameter $\mathbf{N}$ (and a unit vector describing its orientation $\mathbf{n}=\mathbf{N}/N$) is present in the MOI. The spin accumulation in the NM can be parameterized by the spin-dependent chemical potential $\mu_\mathrm{s}(z)$ and the accumulation of magnetic excitation quanta is given by the magnon chemical potential $\mu_\mathrm{mag}$. In addition, we assume different temperatures $T_\mathrm{N}$ and $T_\mathrm{M}$ for the electronic system in the NM and the magnonic system in the MOI, respectively. In this way we can write the pure spin current across the interface flowing from the NM into the MOI as~\cite{Bender2015}
\begin{eqnarray}
 \mathbf{j}_\mathrm{s,int}=&\frac{1}{4\pi}\left(\tilde{g}_i^{\uparrow\downarrow}+\tilde{g}_r^{\uparrow\downarrow}\mathbf{n}\times\right)\left(\mu_\mathrm{s}(0)\mathbf{s}\times\mathbf{n}-\hbar\dot{\mathbf{n}}\right)\nonumber\\
+&\left[g(\mu_\mathrm{mag}+\mu_\mathrm{s}(0)\mathbf{s}\cdot\mathbf{n})+S(T_\mathrm{M}-T_\mathrm{N}) \right]\mathbf{n}\;.
\label{equ:InterfacialSpinCurrent}
\end{eqnarray}
The interface parameters $\tilde{g}_i^{\uparrow\downarrow}$and $\tilde{g}_r^{\uparrow\downarrow}$ are the effective spin mixing conductance parameters, including effects of finite temperature and magnon bandstructure of the MOI. $g$ is the spin conductance and $S$ the spin Seebeck coefficient. All these four parameters can be calculated from the real and imaginary parts of the $T=0$ spin-mixing conductance $g^{\uparrow\downarrow}$~\cite{brataas_finite-element_2000,tserkovnyak_nonlocal_2005,jia_spin_2011,Bauer2012} and taking into account the magnon density of states given by the magnon bandstructure of the MOI (For more details onto this process see Refs.~\cite{Bender2015,Cornelissen2016}). The pure spin current flow across the interface is made up as the sum of two terms. The first term describes the spin current flow due to $\tilde{g}^{\uparrow\downarrow}$ and persists even at vanishing temperatures. The second term in contrast is thermally activated and thus vanishes for $T=0$. Moreover, the first term is responsible for the manifestation of the spin pumping (Sect.~\ref{spin_pumping}) and spin Hall magnetoresistance effect (Sect.~\ref{spin_hall_magnetoresistance}), while the second term causes the spin Seebeck effect (Sect.~\ref{spin_seebeck}) and allows for all-electrical SHE based magnon transport experiments (Sect.~\ref{magnon_mediated_magnetoresistance}). The physical principle behind these two terms are elastic and inelastic spin-flip scattering processes at the interface for the charge carriers in the NM as illustrated in Fig.~\ref{figure:SpinCurrentInterface}. The first term containing $\tilde{g}_i^{\uparrow\downarrow}$and $\tilde{g}_r^{\uparrow\downarrow}$ stands for elastic spin-flip scattering at the interface, here the angular-momentum of the spin-flip is transferred onto the magnetic order parameter $\mathbf{n}$ acting as a torque $\boldsymbol{\tau}$ on it. The second term with $g$ and $S$ represents inelastic electron spin-flip scattering at the interface, the change in energy of the charge carrier is transferred to magnetic excitation quanta in the MOI and thus couples $\mu_\mathrm{s}$ and $\mu_\mathrm{mag}$. The spin orientation of the spin current across the MOI/NM interface caused by the second process is always oriented along $\mathbf{n}$. It is important to note that the interfacial spin current $\mathbf{j}_\mathrm{s,int}$ across the interface is still described as a vector, although the flow direction is fixed by the interface, the spin orientation $\mathbf{s}$ of the spin current still has to be included.

While the detailed calculation/simulation of these 4 interface parameters ($\tilde{g}_i^{\uparrow\downarrow}$, $\tilde{g}_r^{\uparrow\downarrow}$, $g$, and $S$) can be quite complicated, especially if one includes effects from real magnon bandstructures into the model. Eq.~(\ref{equ:InterfacialSpinCurrent}) allows in the experiment to extract these quantities, provided one can measure the other relevant parameters entering the equation independently. It is remarkable that this equation allows to describe all the relevant processes taking place at the interface for the four distinct phenomena covered in this review article.

\section{Materials for MOI/NM heterostructures}
\label{materials}
Before we discuss the effects originating from pure spin current transport across a MOI/NM interface, it is worth to give an short overview of the different materials that have been used in the experiment for the investigation of such pure spin current phenomena. In most of the experiments presented here, the NM acts as the pure spin current detector by making use of the ISHE. For the generation of pure spin currents, non-equilibrium processes either in the MOI using microwave irradiation or temperature gradients or by driving a charge current through the NM and generating a pure spin current via the SHE ar employed. Thus for the NM large values of $\alpha_\mathrm{SH}$ are desirable, which is achieved for example in some transition metal elements. For the MOI, ferro/ferrimagnetic order with a long magnon lifetime, i.e. low damping of the ferromagnetic resonance, are prerequisites to manipulate the orientation of the magnetic order parameter by an external magnetic field and to achieve large spin current values. The material class of the rare-earth iron garnets has with yttrium iron garnet one prototype material of such a MOI. Last but not least, the clean interfaces between the MOI and the NM are required for successful pure spin current transport across the interface.

\subsection{Transition Metals}

For the NM a large spin Hall angle is desirable to increase the efficiency of the transfer from a charge current to a spin current and vice versa. Initial work in this direction was focussed on heavy element metals, as the spin-orbit coupling and thus the spin Hall angle increases with the atomic mass of the element used. Over the course of the recent years especially tungsten, tantalum and platinum have been identified as ideal materials with large spin Hall angles~\cite{Morota:2011,Liu2012Pt,Liu2012,Pai2012,Niimi:2012,Garello2013,Weiler:Solid-state-physics-64:2013,niimi_reciprocal_2015}. While large spin-orbit interaction boosts the spin Hall effect, it is also relevant for the electron spin-flip length $\lambda_{\mathrm{sf}}$ in these materials. Thus all these materials exhibit a small $\lambda_{\mathrm{sf}}$, with values of a couple of nanometers~\cite{Hoffmann2013}. In this way, if one wants to investigate spin currents in these material the dimensions of the samples have to be at least in one direction comparable to this length scale, which in this case requires thin films of these materials.

The growth of transition metals thin films is with the advent of ultra-high vacuum deposition systems nowadays no major challenge. Typically, sputter deposition and thermal evaporation have been employed to fabricate these thin film samples. However, important aspects to keep in mind are excellent thickness control of the deposition process, homogenous thin film growth and avoiding intermixing at the interface. Another point worth mentioning is the role of impurities in the NM. Many ab-initio calculations suggest that the spin Hall angle of the host material can be drastically changed by impurities in the host system. Up to now only few experimental investigations have been conducted to address the role of impurities for the spin Hall effect in these materials. The role of impurities may especially be relevant for other growth methods which are susceptible to leave traces of impurities in the deposited material, like for example atomic layer deposition due to the use of precursors. As already discussed, not only the spin Hall effect can contribute to the transfer of a charge current into a spin current, but also interfacial effects may be relevant. This may provide additional ways of tuning the effective spin Hall angle for example by changing the capping layer, which is needed for easily oxidized materials like W and Ta. However, systematic studies in this direction are very scarce at the moment. Up to now most experiments used polycrystalline or sometimes textured transition metals, such that investigations of the crystalline anisotropy of the spin Hall effect can not be conducted.

Taken all together transition metal elements provide an easily accessible experimental platform that allows to use these materials as the NM acting as an charge-current-based generator and detector of pure spin currents.

\subsection{Rare-earth iron garnets}
\begin{figure}[t]
 \includegraphics[width=85mm]{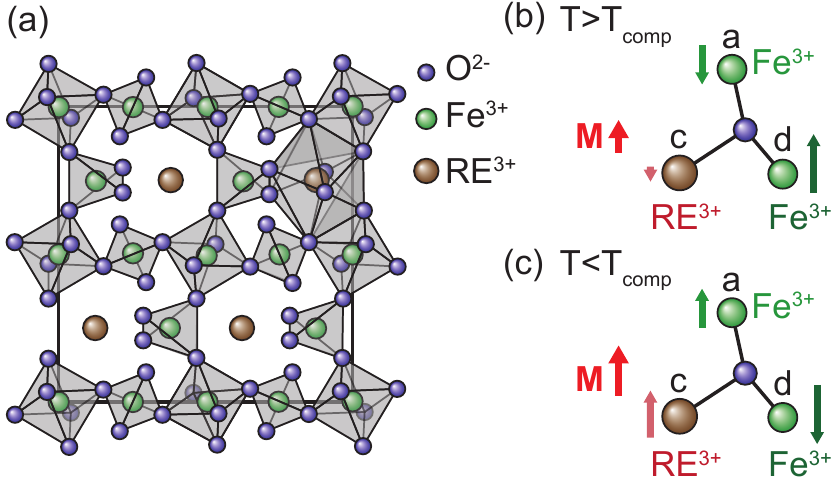}\\
 \caption[Garnet Crystal Structure]{(a) Illustration of the rare-earth iron garnet crystal structure. The 3 rare-earth ions (RE$^{3+}$) per formula unit are dodecahedrally coordinated with oxygen ions, while three Fe$^{3+}$ ions per formula unit are tetrahedrally and two Fe$^{3+}$ ions are octahedrally coordinated with oxygen ions. These three different coordination sites form the three magnetic sublattices of the REIGs. For rare-earth ions with finite magnetic moment it is possible to observe magnetic compensation at the compensation temperature $T_\mathrm{comp}$. The orientation of the magnetizations $\mathbf{M}^\mathrm{Fe,d}$, $\mathbf{M}^\mathrm{Fe,d}$, and $\mathbf{M}^\mathrm{RE}$ are illustrate in (b) for $T>T_\mathrm{comp}$ and in (c) for $T<T_\mathrm{comp}$. Upon crossing the compensation temperature, the sublattice magnetizations reverse their orientation.}
  \label{figure:GarnetCrystal}
\end{figure}

As already mentioned above, rare-earth iron garnets (REIGs) are an ideal MOI material class for pure spin current experiments in MOI/NM heterostructures. Among the rare-earth iron garnets, yttrium iron garnet (Y$_3$Fe$_5$O$_12$, YIG) is the prototype material with unprecedented magnon lifetimes. YIG is an artificial ferrimagnetic insulator with a Curie temperature well above room temperature ($T_\mathrm{C}=560\,\mathrm{K}$~\cite{coey_magnetism_2010}). Since its first fabrication over 50 years ago~\cite{gilleo_magnetic_1958,geller_structure_1957} YIG is widely used in microwave applications, for example as a tunable narrow bandpass filter or resonator~\cite{helszajn_yig_1985}, and in magneto-optical applications, for example as an optical insulator in optical fibre communications~\cite{winkler_magnetic_1981} or even for the ultra fast magneto-optic sampling of current pulses~\cite{elezzabi_ultrafast_1996}. This broad application range is based on the excellent magnetic properties of YIG, such as very low magnetic damping and large Faraday rotation angles when doped with bismuth. In the cubic garnet structure (Ia3d) of YIG (lattice constant $a=1.238\,\mathrm{nm}$) illustrated in Fig.~\ref{figure:GarnetCrystal}(a) three Fe$^{3+}$ ($S=5/2$) ions are tetrahedrally coordinated (24d) by oxygen while the remaining two Fe$^{3+}$ ions are coordinated octahedrally (16a) in one formula unit. This leads to the formation of two oppositely aligned ferroic sublattices with a net magnetization of $5\,\mathrm{\mu_\mathrm{B}/f.u.}$~\cite{geller_crystal_1957,anderson_molecular_1964}. Ga substitution of tetrahedral iron results in a compensation point due to the different temperature dependence of the two sublattices~\cite{hansen_saturation_1974}. Other rare-earth iron garnets exhibit a compensation temperature due to the magnetic moment of the rare-earth element, which is either parallel or antiparallel oriented to the net magnetization of the two iron sublattices (See Fig.~\ref{figure:GarnetCrystal}(b) and (c)). All in all, rare-earth iron garnets are a very versatile magnetic material class and allow to tune their magnetic properties by doping with various elements. This versatility makes the insulating compound YIG an interesting candidate for spin current related experiments. One thing to keep in mind though is that the magnetic unit cell of any REIG is very complex and thus the magnon bandstructure consists of several bands~\cite{Cherepanov1993}, which can make the interpretation of experimental results in this regard quite complex.

In YIG the Y$^{3+}$ ions do not posses a magnetic moment, however, if one replaces Y with a rare-earth element with finite magnetic moment in its ionized state, one has three magnetic sublattices with magnetization $\mathbf{M}^i$~\cite{geller_structure_1957}. We here use $\mathbf{M}^\mathrm{RE}$, $\mathbf{M}^\mathrm{Fe,d}$, and $\mathbf{M}^\mathrm{Fe,a}$ for the magnetizations of the magnetic sublattice for the rare-earth ion, the tetrahedrally coordinated iron ions, and octahedrally coordinated iron ions, respectively. In case of the REIGs the iron magnetic sublattices are coupled strongly antiferromagnetically to each other, such that $\mathbf{M}^\mathrm{Fe,a}$ is always antiparallel aligned to $\mathbf{M}_\mathrm{Fe,d}$~\cite{anderson_molecular_1964,dionne_molecularfield_1976}.  The magnetic sublattice of the rare-earth ion couples antiferromagnetically to the net iron magnetization ($\mathbf{M}^\mathrm{Fe}=\mathbf{M}^\mathrm{Fe,a}+\mathbf{M}^\mathrm{Fe,d}$). As the temperature dependence of $\mathbf{M}^\mathrm{RE}$, $\mathbf{M}^\mathrm{Fe,d}$, and $\mathbf{M}^\mathrm{Fe,a}$ are quite different, it is possible to observe magnetic compensation ($\mathbf{M}^\mathrm{RE}+\mathbf{M}^\mathrm{Fe,d}+\mathbf{M}^\mathrm{Fe,a}=0$) at the so-called magnetic compensation temperature $T_\mathrm{comp}$ for many different rare-earth elements. At this temperature the net magnetization vanishes, but the magnetic sublattices exhibit magnetic order, similar to an antiferromagnet. Most importantly the orientation of each sublattice magnetization is inverted, when going from a temperature above $T_\mathrm{comp}$ to a temperature below $T_\mathrm{comp}$. Thus REIGs with magnetic compensation allow to investigate experimentally the transition from ferrimagnetic ordering to antiferromagnetic by simply tuning the temperature.
\begin{figure}[b]
 \includegraphics[width=85mm]{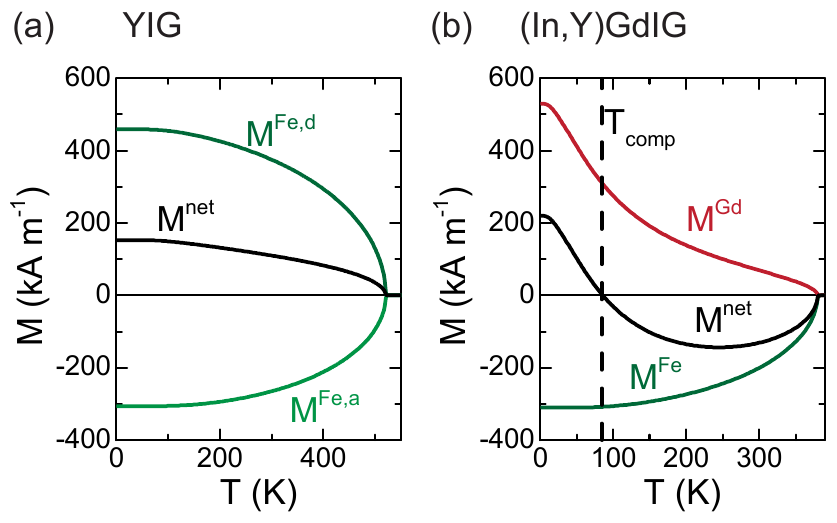}\\
 \caption[Magnetic compensation of rare-earth iron garnets]{Numerical results for the calculation of the sublattice magnetizations for (a) YIG and (b) (In,Y)GdIG. For YIG the net magnetization $\mathbf{M}^\mathrm{net}=\mathbf{M}^\mathrm{Fe,a}+\mathbf{M}^\mathrm{Fe,d}$ increases monotonically with decreasing temperature. For (In,Y)GdIG the drastically different temperature dependence of the rare-earth ion compared to the iron ions leads to a non-monotonic temperature dependence of $\mathbf{M}^\mathrm{net}=\mathbf{M}^\mathrm{Fe,a}+\mathbf{M}^\mathrm{Fe,d}+\mathbf{M}^\mathrm{Gd}$. At $T=T_\mathrm{comp}$ the net magnetization vanishes. Data taken from Ref.~\cite{SchlitzMaster2015}.}
  \label{figure:MagnetizationCompensation}
\end{figure}

A common approach to model the three magnetic sublattices in REIGs is to apply a mean field model with a Brillouin function describing each sublattice magnetization~\cite{anderson_molecular_1964,bernasconi_canted_1971,dionne_molecularfield_1976,eremenko_field-induced_1979,srivastava_exchange_1982}. This approach was first suggested by Bernasconi \textit{et al.}~for gadolinium iron garnet (GdIG)~\cite{bernasconi_canted_1971}. In Fig.~\ref{figure:MagnetizationCompensation} we show the numerical results obtained for such a model for (a) YIG and (b) GdIG doped with indium and yttrium ((In,Y)GdIG) using the parameters from Ref.~\cite{bernasconi_canted_1971}. The doping of GdIG lowers $T_\mathrm{comp}$ from close to $300\,\mathrm{K}$ to below $100\;\mathrm{K}$, which allows in the experiment to more easily gain access to the regimes above and below $T_\mathrm{comp}$. For YIG the temperature dependence of the two iron magnetizations are quite similar and the total magnetization $M^\mathrm{net}$ increases monotonically with decreasing temperature. In contrast, for (In,Y)GdIG the magnetization $M^\mathrm{Gd}$ has a much stronger temperature dependence than the net iron magnetization $M^\mathrm{Fe}$. This leads to a non-monotonic temperature dependence of $M^\mathrm{net}$ and at $T_\mathrm{comp}$ $M^\mathrm{net}=0$. Due to the reversal of the magnetic moment orientations of each magnetic sublattice at $T_\mathrm{comp}$ (upon application of an external magnetic field), compensated REIGs enable us to investigate the role of the sublattice moments for pure spin current physics. It is worth mentioning that in addition to the magnetization compensation point compensated REIGs exhibit an angular momentum compensation point. These two compensation points do not necessarily happen at the very same temperature, which may allow to even more deeply assess the role of sublattice angular momentum and sublattice magnetization in pure spin current experiments.

\begin{figure}[t]
 \includegraphics[width=85mm]{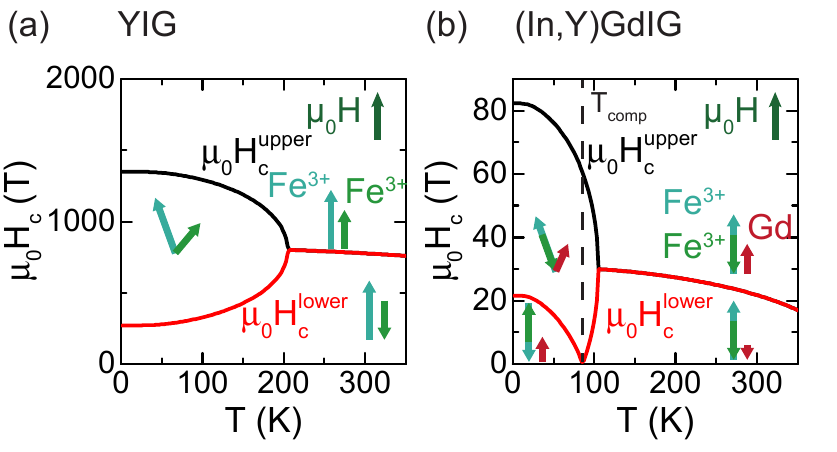}\\
 \caption[spin canting phase in rare-earth iron garnets]{Magnetic phase diagram of the different sublattice configurations for (a) YIG and (b) (In,Y)GdIG determined from mean field calculations. Below a critical Temperature and in between the lower and upper critical fields, a spin canting phase exists, where the sublattice magnetizations are no longer oriented collinear with the external magnetic field. While for YIG the required fields to enter the spin canting phase are not accessible in the lab the magnetically compensated (In,Y)GdIG allows experimental investigation of the spin canting phase for temperatures close to $T_\mathrm{comp}$. Data taken from Ref.~\cite{SchlitzMaster2015}.}
  \label{figure:SpinCanting}
\end{figure}

Another intriguing effect of REIGs is the existence of a spin canting phase upon applying an external magnetic field with sufficient strength, where the sublattice magnetizations are no longer aligned collinear~\cite{clark_neferrimagnets_1968,alben_phase_1970,bernasconi_canted_1971,dionne_molecularfield_1976,eremenko_field-induced_1979}. This effect is similar to the spin-flop transition in antiferromagnets. For YIG the external magnetic field has to be comparable to the exchange field mediating the antiferromagnetic coupling of $\mathbf{M}^\mathrm{Fe,d}$ and $\mathbf{M}^\mathrm{Fe,a}$. Again this phenomena can be modeled using a mean field approach now including an external magnetic field~\cite{bernasconi_canted_1971,eremenko_field-induced_1979}. The net magnetization of the magnetic sublattices has to be oriented parallel to the applied external magnetic field. However, this does not necessarily require the sublattice magnetizations to be oriented parallel or antiparallel to the external magnetic field. From this evaluation one obtains temperature dependent lower ($H^\mathrm{lower}$) and upper critical fields ($H^\mathrm{upper}$) for the existence of the spin canting phase. The numerical result of such calculations for YIG, using molecular field parameters from Ref.~\cite{bernasconi_canted_1971} are shown in Fig.~\ref{figure:SpinCanting}(a). For YIG the spin canting phase is accessible for $T<200\,\mathrm{K}$, but requires external magnetic fields exceeding $250\,\mathrm{T}$, which is well above the magnetic fields that can be realized in the lab. For temperatures above $200\,\mathrm{K}$ the iron sublattice magnetizations are always oriented collinear with the external magnetic field, only $\mathbf{M}^\mathrm{Fe,a}$ reorients from antiparallel to parallel for large enough magnetic fields (again much larger than what is available in the lab). For compensated REIGs like (In,Y)GdIG the necessary magnetic fields to obtain a spin canting phase are significantly reduced as evident from the results of the mean field approach shown in Fig.~\ref{figure:SpinCanting}(b). Especially at $T_\mathrm{comp}$ any applied field directly results in a canting of the sublattice magnetizations. There is also an upper temperature limit (as was the case for YIG) for which spin canting can be observed in the material. It is important to understand that within the spin canting phase the net magnetization, i.e. the vectorial sum of the sublattice magnetizations, is no longer the order parameter of the system. Similar as in the antiferromagnets the N\'{e}el-vector is the order parameter (outside of the spin canting phase the N\'{e}el-vector and the net magnetization vector are oriented collinear to each other). The intriguing aspect of the spin canting phase is that the net magnetization is oriented parallel to the external magnetic field, but the sublattice magnetizations are now in a non-collinear orientation with the external magnetic field, which allows in the experiment to investigate the role of sublattice magnetic moment orientations on pure spin currents in a MOI. It is worth mentioning that from the mean field approach only the canting angle of the sublattice magnetizations with respect to the external magnetic field is determined, which means that in spherical coordinates only one angle is fixed, while the other one is not determined. In this way, there is an infinite number of sublattice magnetization orientations that can realize the required canting angle. Such that it is quite possible that in a real material multiple domains with different magnetic sublattice orientations can form in the spin canting phase~\cite{eremenko_field-induced_1979,seul_domain_1995}. However, the detection of such a multidomain state is experimentally very challenging as it requires techniques that are able to spatially resolve sublattice magnetization orientation. This degeneracy may be lifted by the magnetic anisotropy of the material and can be modeled by a free energy approach as detailed in Ref.~\cite{eremenko_field-induced_1979}.


Single crystals of YIG grown from the melt~\cite{winkler_magnetic_1981} are widely available and substitution with various elements to tailor the magnetic properties of YIG has been extensively studied in the last decades~\cite{hansen_magnetic_1983,gilleo_magnetic_1958,hansen_saturation_1974,helszajn_yig_1985}. Thin film deposition of high quality YIG has been mainly achieved using liquid phase epitaxy~\cite{blank_growth_1972,aichele_garnet_2003}, but also pulsed laser deposition (PLD) allows to grow high quality REIGs thin films~\cite{krockenberger_solid_2008,krockenberger_layer-by-layer_2009,kahl_pulsed_2003,manuilov_submicron_2009,manuilov_pulsed_2010,dorsey_epitaxial_1993,manuilov_pulsed_2009,onbasli_pulsed_2014,hauser_yttrium_2016,zaki_growth_2017}. Even the epitaxial growth of high quality YIG thin films using RF-sputter deposition has been successfully realized~\cite{houchen_chang_nanometer-thick_2014,liu_ferromagnetic_2014,lustikova_spin_2014,du_y_2015,cao_van_effect_2018}. One of the major advantages for the epitaxial growth of YIG is the availability of a well lattice-matched substrate: gadolinium gallium garnet (GGG)~\cite{knorr_lattice_1984}. At room temperature GGG matches nicely the lattice constant of YIG (lattice misfit $0.03\%$) and remains well lattice-matched even at the elevated temperatures needed for epitaxial growth due to similar thermal expansion coefficients. Thus GGG substrates enable the growth of YIG thin films with excellent crystalline and magnetic quality. GGG is also a suitable substrate choice for the deposition of many other rare-earth iron garnet thin films and enables layer-by-layer growth of the REIGs thin films. One drawback when using GGG as the substrate material is the significant paramagnetism of the gadolinium ions, which makes magnetometry measurements of thin ferromagnetic films on top quite challenging due to the large substrate background signal. Another suitable substrate is yttrium aluminium garnet, but the lattice mismatch with YIG is considerably larger, which can significantly reduce the quality of the YIG film on top. In addition, aluminium can diffuse into the YIG film and alter the properties of the YIG layer. The usage of a thin GGG buffer layer may allow for higher quality on other substrates than GGG. Over the last two years the REIG Tm$_3$Fe$_5$O$_12$ (TmIG) has received some attention, as this material allows to realize perpendicular magnetic anisotropy\cite{Avci2016,tang_anomalous_2016,quindeau_tm_2017} in a MOI and magnetization switching in the TmIG layer has been achieved with pure spin currents in TmIG/Pt heterostructures. In the last year even perpendicular magnetic anisotropy has been obtained for YIG thin films~\cite{fu_epitaxial_2017}. Due to these properties REIG thin films can be quite easily produced in decent quality with state-of-the-art thin film equipment. However, due to the large lattice constant of REIGs and their low electrical conductivity, it is not straight forward to grow fully epitaxial MOI/NM heterostructures by using REIGs as the MOI and/or the NM. Such epitaxial thin films may be beneficial for more sophisticated device concepts for pure spin current experiments and applications.

\subsection{Perspectives for materials}

While most of the experimental work dealing with pure spin currents in MOI/NM heterostructures in the last decade has been focused on heavy transition metal elements interfaced with iron garnet materials, some interesting results have been already obtained with other materials, opening up new avenues for future experiments.

For the NM, for example transition metal oxides like IrO$_2$, indium tin oxide, tungsten oxides have been used in the experiment and a sizable spin Hall effect in these materials has been observed~\cite{fujiwara_5d_2013,qiu_experimental_2013,qiu_all-oxide_2015,demasius_enhanced_2016}. Theoretical predictions and experiments suggest large spin-orbit coupling for even more complex transition metal oxides, like SrIrO$_3$, which may also give rise to significant spin Hall effects in combination with the surface states of a topological insulator~\cite{zeb_interplay_2012,chen_topological_2015,nie_interplay_2015,martins_coulomb_2017}. Furthermore, 2-dimensional materials~\cite{feng_intrinsic_2012,qian_quantum_2014,cazalilla_quantum_2014,shao_strong_2016} and topological insulators with surface conduction states have been interfaced with MOIs to investigate pure spin current transport across such interfaces. In this regard, experiments by Jamali \textit{et al.}~yield a large spin Hall angle of 40\%~\cite{Jamali2015} for the topological insulator Bi$_2$Se$_3$. Another intriguing aspect is to employ the charge to spin current transfer due to the anomalous Hall effect in ferromagnetic materials~\cite{Miao2013} or in antiferromagnetic materials~\cite{zhang_spin_2014}. For example, electrically detected spin pumping experiments in SrRuO$_3$ showed an increase in the conversion efficiency around the Curie temperature of the ferromagnet~\cite{Wahler2016}. As already mentioned in Sec.~\ref{theory} not only the "bulk" spin Hall effect of an material allows to generate and detect pure spin currents, but also the spin galvanic effect~\cite{ganichev_spin-galvanic_2002,seibold_theory_2017} and/or Rashba-Edelstein effect~\cite{edelstein_spin_1990,sanchez_spin--charge_2013,sangiao_control_2015,jungfleisch_interface-driven_2016,nakayama_rashba-edelstein_2016,ohshima_strong_2017} due to the broken inversion symmetry can contribute. In this way engineering of the interface during the deposition of the materials might be a possible pathway towards more efficient spin current generators and detectors.

For the MOIs, ferrites and perovskite materials have already been used as an alternative to REIGs. For example, recent experiments showed a significant improvement of nickel ferrite thin films grown by PLD on lattice matched substrates and it would be interesting to see how this affects pure spin current transport~\cite{singh_bulk_2017}. Similarly, La$_{0.7}$Sr$_{0.3}$MnO$_3$ layers grown by pulsed laser deposition have achieved low damping~\cite{qin_ultra-low_2017}. As epitaxial heterostructures are not required for pure spin current experiments in MOI/NM systems, other MOIs with a magnetic order that goes beyond a ferro-/ferrimagnetic ordering have already been investigated in theory and experiment. Here, especially the work on antiferromagnetic MOI has received quite some attention in the recent years. Several groups reported on the spin pumping, spin Seebeck effect and spin Hall magnetoresistance in antiferromagnetic insulators, like NiO and Fe$_2$O$_3$~\cite{hahn_conduction_2014,Han2014,wang_antiferromagnonic_2014,wang_spin_2015,seki_thermal_2015,moriyama_anti-damping_2015,jungwirth_antiferromagnetic_2016,rezende_theory_2016,wu_antiferromagnetic_2016,ji_spin_2017,manchon_spin_2017,holanda_spin_2017,hou_tunable_2017,hoogeboom_negative_2017,fischer_spin_2018}. In a pioneering work by Aqueel \textit{et al.}~pure spin current physics in a topological material as the MOI has been investigated using Cu$_2$OSeO$_2$/Pt heterostructures, which may pave the way towards driving spin dynamics by pure spin currents~\cite{aqeel_electrical_2016}. A next interesting step would be to artificially design of MOIs by growing multilayers of MOIs and other insulators, with the goal to tailor the relevant properties for pure spin current experiments. On the one hand, such engineered multilayers provide a toolset to design the magnonic bandstructure of the system and control exchange interactions. On the other hand, interfacial effects like the interfacial Dzyaloshinskii-Moriya interaction can give rise to topological spin textures in these artificial MOIs, similar to what has already been achieved in metallic multilayers, where nowadays skyrmions at room temperature are readily realized~\cite{Gayles2015,jiang_blowing_2015,Woo2016,fert_magnetic_2017}. A first step into this direction has been carried out quite recently in multilayer structures of perovskites in form of a synthetic antiferromagnet based on La$_{2/3}$Ca$_{1/3}$MnO$_3$/CaRu$_{0.5}$Ti$_{0.5}$O$_3$ stacks~\cite{chen_all-oxidebased_2017} and a 2-dimensional antiferromagnet based on multilayers of SrIrO$_3$ and SrTiO$_3$~\cite{hao_two-dimensional_2017}.

With the multitude of materials available it should be in principle possible to realize fully epitaxial heterostructures of MOI and NM. In such a system effects of crystalline anisotropy on the pure spin currents can be investigated and even be exploited to enhance effects. Moreover, provided 2-dimensional growth modes can be established for the NM as well as for the MOI it is then possible to tune the interface termination and more systematically investigate the influence of termination on the pure spin current transport across the interface. On the other hand, recent experiments conducted on amorphous YIG thin films suggest that even amorphous materials may be suitable for very efficient pure spin current transport mediated by antiferromagnetic exchange interactions.

\section{Spin pumping in MOI/NM heterostructures}
\label{spin_pumping}
\begin{figure}[h]
 \includegraphics[width=85mm]{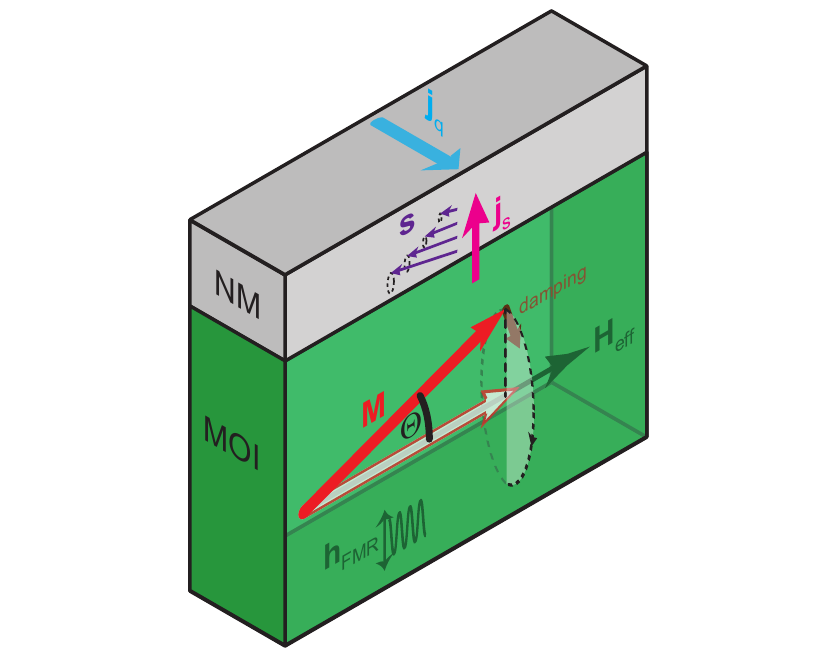}\\
 \caption[Spin pumping]{Illustration of the spin pumping effect. In a ferromagnetic MOI/NM heterostructure the magnetization $\mathbf{M}$ in the MOI is driven into precession with the precession cone angle $\Theta$ around its equilibrium position by applying a microwave drive $\mathbf{h}_\mathrm{FMR}$. The excess angular momentum for $\mathbf{M}$ is relaxed by pumping a pure spin current $\mathbf{j}_\mathrm{s}$ across the interface into the NM with a precessing spin polarization $\mathbf{s}$. In the NM $\mathbf{j}_\mathrm{s}$ is then transformed into a charge current $\mathbf{j}_\mathrm{q}$ via the ISHE.}
  \label{figure:SpinPumpingPrinciple}
\end{figure}

As already discussed in the theory part in Sec.~\ref{theory} one way of generating a spin current in MOI/NM heterostructures is by driving the magnetic order parameter $\mathbf{N}$ out of equilibrium. In terms of Eq.(\ref{equ:InterfacialSpinCurrent}), we then assume that $\dot{\mathbf{n}}\neq0$ and $\mu_\mathrm{mag}=\mu_\mathrm{s}(0)=(T_\mathrm{M}-T_\mathrm{N})=0$, i.e. no magnon accumulation in the MOI, no electron spin accumulation in the NM and no temperature difference between the magnonic system in the MOI and electronic system in the NM. This is then called the spin pumping effect. A time-dependent magnetic order parameter ``pumps'' a pure spin current across the interface~\cite{Tserkovnyak2002,brataas_spin-pumping_2004,tserkovnyak_nonlocal_2005,Mosendz2010,Ando2010,kapelrud_spin_2013}. In a ferro-/ferrimagnetic system this can be achieved by ferromagnetic resonance (FMR). The magnetization is driven into a precessing motion by applying a static and a time-varying external magnetic field, which is a collective excitation of the magnetic moments in the MOI. This leads to the flow of a pure spin current $\mathbf{j}_\mathrm{s}$ across the interface as illustrated in Fig.~\ref{figure:SpinPumpingPrinciple}. To drive the FMR frequencies in the GHz regime are required. Such collective excitations also exist in MOIs with a magnetic order different to ferromagnetism. However, for example in antiferromagnetically ordered MOIs the required frequencies are then in the THz range, which makes an experimental investigation much more demanding~\cite{Cheng2014,Johansen2017}. In the following we will restrict ourselves to ferro-/ferrimagnetically ordered MOIs and discuss the effects of spin pumping only in this regard. Another point worth mentioning is that spin pumping is not limited to MOI/NM heterostructures, but also bilayers of magnetically ordered conductors and NMs can be used for such experiments~\cite{Costache:vanWees:spin-pumping:experiment:PRL2006,spin-pumping:saitoh:APL:2006,woltersdorf_magnetization_2007,Ando2010,ando_inverse_2011,Czeschka2011,ando_observation_2012}. However, a major advantage for the use of MOIs is that one can completely rule out charge current flow in the magnetically ordered material and across the interface, which makes the interpretation of experiments in this regard much easier.

In more detail, the excitation via the microwave field and the relaxation (damping effects) of the magnetization have to balance each other in the steady-state, which leads to a precessional motion of the magnetization $\mathbf{M}$ around its thermal equilibrium state, with a precession cone angle $\Theta$ (See Fig.~\ref{figure:SpinPumpingPrinciple}). For the spin pumping effect we are interested in $\dot{\mathbf{n}}$, which can be modeled using the Landau-Lifshitz-Gilbert (LLG) equation ~\cite{Vonsovskii:FMR:1960,MorrishBuch}
\begin{equation}\label{WMI:eq:LLG}
  \frac{d \mathbf{m}}{dt} = - \gamma \mu_{0} \left( \mathbf{m} \times \mathbf{H}_{\mathrm{eff}} \right) + \alpha \left( \mathbf{m} \times \frac{d \mathbf{m}}{dt} \right)\;,
\end{equation}
with the gyromagnetic ratio $\gamma$, the normalised magnetization direction $\mathbf{m}=\mathbf{M}/|\mathbf{M}|$ (identical to $\mathbf{n}$ for ferromagnetic systems and in good approximation identical for ferrimagnetic systems), and the so-called Gilbert damping parameter $\alpha$ describing viscous magnetization damping. The first term on the right hand side of Eq.(\ref{WMI:eq:LLG}) describes the precession of $\mathbf{M}$ around an effective magnetic field $\mathbf{H}_\mathrm{eff}$, which contains the external magnetic field, as well as contributions from magnetic anisotropy and demagnetizing fields from the sample shape. From the LLG equation the FMR condition can be calculated, i.e. the microwave frequency required to obtain a resonant absorption for a given $\mathbf{H}_\mathrm{eff}$. Following from Eq.(\ref{equ:InterfacialSpinCurrent}), the corresponding pure spin current across the interface is given by~\cite{Tserkovnyak2002,tserkovnyak_nonlocal_2005}
\begin{equation}\label{eq:WMI:Js:spinpumping}
\mathbf{j}_\mathrm{s}^{\mathrm{pump}}=\frac{\hbar}{4 \pi} \left\{ \tilde{g}_r^{\uparrow\downarrow} \left[ \mathbf{m} \times \frac{d \mathbf{m}}{d t}\right] - \tilde{g}_i^{\uparrow\downarrow} \frac{d \mathbf{m}}{d t} \right\} \;.
\end{equation}
We can now add the right hand side of Eq.(\ref{eq:WMI:Js:spinpumping}) to the right hand side of Eq.~(\ref{WMI:eq:LLG}), because a pure spin current is also a change in angular/magnetic momentum ($=d \mathbf{m}/dt$). From this, we find that the first term of Eq.(\ref{eq:WMI:Js:spinpumping}) represents an additional Gilbert damping damping contribution to the magnetization dynamics of the FMR. This can be rationalized by the fact that the flow of $\mathbf{j}_\mathrm{s}$ across the MOI/NM interface represents in this regard a way of removing excess angular momentum from the MOI. Thus, $\alpha$ is changed by the spin pumping contribution. In a FMR experiment a way to quantify the spin pumping effect is then to compare the damping of a bare MOI to a MOI/Pt heterostructure. It should be noted that spin pumping is an interfacial effect, such that to observe sizable changes in the damping parameter nanometer thick layers of the MOI have to be used in the experiment. Moreover, the precession cone angle $\Theta$, which parameterizes $\dot{\mathbf{m}}$, is the relevant parameter that defines the magnitude of the spin current across the MOI/NM interface. By increasing the power of the microwave drive one can increase $\Theta$ and in turn also increase the pure spin current across the interface.

Another pathway to investigate the spin pumping effect is to electrically detect the pure spin current injected into the NM, by exploiting the ISHE in the NM, which transforms the spin current $\mathbf{j}_\mathrm{s}$ into a charge current $\mathbf{j}_\mathrm{q}$ (See Fig.~\ref{figure:SpinPumpingPrinciple}). In most experiments, open electrical circuit boundary conditions are employed, such that the electric field originating from the charge accumulation driven by $\mathbf{j}_\mathrm{s}$ can be detected as a voltage. From Refs.~\cite{Mosendz2010,Mosendz:2010:PRB,Czeschka2011,Weiler:Solid-state-physics-64:2013}, the magnitude of the spin pumping spin Hall voltage $\Delta V$ is given by
\begin{equation}\label{eq:WMI:VSP}
  \frac{\Delta V}{L}=\frac{2e}{\hbar} \alpha_{\mathrm{SH}}
    \frac{ j_{\mathrm{s}}^{\mathrm{pump}} \eta \lambda_{\mathrm{sf}} \tanh\left( \frac{t_{\mathrm{NM}}}{2 \lambda_{\mathrm{sf}}} \right)}{\sigma_{\mathrm{NM}} t_{\mathrm{NM}}}\;.
\end{equation}
Here, $\sigma_{NM}$ and $t_{NM}$ are the conductivities and layer thicknesses of the NM, respectively, $L$ is the distance between the two electrical contacts on the sample, and $\eta$ is the backflow parameter ($0 \leq \eta \leq 1$) accounting for a possible spin current backflow into the FM and is defined as ~\cite{Jiao2013,Chen:SMR:theory:PRB:2013}:
\begin{equation}\label{eq:Backflow}
  \eta=\left[ 1+2\tilde{g}_r^{\uparrow\downarrow} \sigma_\mathrm{NM}\lambda_\mathrm{sf} \frac{e^2}{\hbar}\coth\left( \frac{t_\mathrm{NM}}{\lambda_\mathrm{sf}}\right)\right]^{-1}\;.
\end{equation}
If $t_\mathrm{NM}\gg\lambda_\mathrm{sf}$ then $\eta=1$ and the pumped spin current is completely absorbed in the NM. This backflow parameter has to be introduced if the thickness of the NM is comparable to the spin-diffusion length in the NM, which leads to a finite electron spin accumulation at the interface in the NM and thus $\mu_\mathrm{s}(0)\neq0$. This is in contrast to our initial assumptions for spin pumping in the first paragraph. The electron spin accumulation drives a spin current across the interface, which compensates the spin current driven by the spin pumping effect. The main advantage of this electrical detection approach is that the thickness of the MOI is not relevant for the voltage signal amplitude, in contrast to the damping based detection of spin pumping. However, the thickness of the NM layer has to be as thin as possible to reduce current shunting effects, but thick enough such that $\eta \approx 1$, which requires thicknesses in the range of nanometers for the NM.

\subsection{Broad band ferromagnetic resonance spin pumping}

In the following we exemplarily discuss the results obtained on Gilbert damping parameter based detection of spin pumping for YIG thin films grown by pulsed laser deposition. The results shown here have been first published by Haertinger \textit{et al.}~using YIG thin films grown on GGG (111)-oriented substrates from our group~\cite{haertinger_spin_2015}. In these broadband ferromagnetic resonance (BBFMR) experiments, YIG thin films with various thickness $t_\mathrm{YIG}$ and capped in-situ with $10\,\mathrm{nm}$ thick Pt or without a capping layer have been investigated. In the BBFMR experiments the microwave frequency dependence of the FMR is investigated by applying different fixed microwave frequencies to the sample and recording the FMR spectrum as a function of the applied external magnetic field. In most BBFMR experimental setups this is achieved by placing the sample onto a coplanar waveguide and utilizing the inductive coupling between sample and waveguide for measuring the magnetic susceptibility of the sample. For each microwave frequency $\nu$ one then obtains the FMR resonate field and the FMR linewidth (full width at half maximum) by fitting a Lorentzian line shape to the experimental data. A very detailed explanation of this technique can be found for example in Refs.~\cite{nembach_perpendicular_2011,shaw_determination_2012,shaw_precise_2013}. In the experiments conducted by Haertinger \textit{et al.}~the external magnetic field was applied along the surface normal of the thin film layers, which allows to suppress damping contributions originating from two magnon scattering.

\begin{figure}[h]
 \includegraphics[width=85mm]{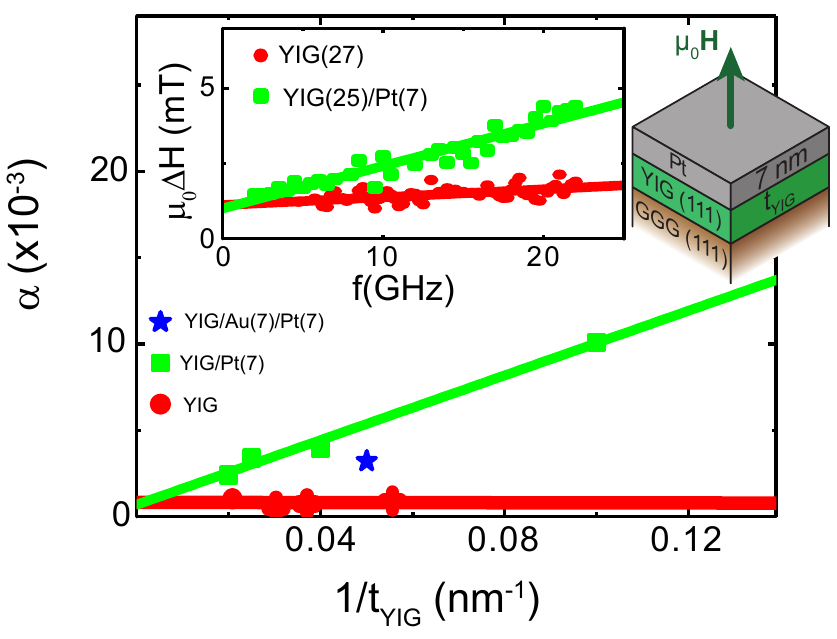}\\
 \caption[Broadband FMR spin pumping]{Experimental results obtained for YIG/Pt bilayers with varying YIG thickness $t_\mathrm{YIG}$ from broadband ferromagnetic resonance experiments. The Gilbert damping parameter shows a linear $t_\mathrm{YIG}^{-1}$ for YIG/Pt bilayers, while it is thickness independent for bare YIG films. The inset shows the frequency dependence of the FMR linewidth for a YIG($26\,\mathrm{nm}$)/Pt($10\,\mathrm{nm}$) bilayer and a bare YIG($25\,\mathrm{nm}$) layer. Adapted with permission from Ref.~\cite{haertinger_spin_2015}. Copyright 2015 by the American Physical Society. }
  \label{figure:SpinPumpingBBFMR}
\end{figure}

For the extraction of $\alpha$ for each sample one then determines the gyromagnetic ratio $\gamma$ from the frequency dependence of the FMR field by a Kittel fit and applying a linear fit to the frequency dependence of the FMR linewidth. The frequency dependence of the FMR linewidth obtained for a YIG($26\,\mathrm{nm}$)/Pt($10\,\mathrm{nm}$) bilayer and a bare YIG($25\,\mathrm{nm}$) layer is shown in the inset of Fig.\ref{figure:SpinPumpingBBFMR}. This process has been repeated for different YIG thicknesses and the extracted $\alpha$ parameters are plotted against $t_\mathrm{YIG}^{-1}$ in Fig.\ref{figure:SpinPumpingBBFMR}. While for the bare YIG layers only a very slight dependence on $t_\mathrm{YIG}$ is observed, for the YIG/Pt bilayers a increases of $\alpha$ with decreasing $t_\mathrm{YIG}$ is observed. For the spin pumping effect one expects that $\alpha$ exhibits a linear $t_\mathrm{YIG}^{-1}$ dependence, which is also the case in the experimental results. A linear function has been used as a fit to both data sets. The difference in the slope for YIG/Pt bilayers and bare YIG is then identical to the damping contribution from the spin pumping effect in YIG/Pt samples. From this data analysis it is then possible to determine $\tilde{g}_r^{\uparrow\downarrow}$, i.e. the transparency of the YIG/Pt interface for pure spin currents. Haertinger \textit{et al.}~find $\tilde{g}_r^{\uparrow\downarrow}=9.7\times10^{18}m^{-2}$ at room temperature, which is comparable to values obtained in metallic ferromagnet/NM heterostructures~\cite{haertinger_spin_2015}.

The observation of $\tilde{g}_r^{\uparrow\downarrow}$ values for MOI/NM bilayers comparable to metallic ferromagnet/NM interfaces is one of the key findings from BBFMR experiments. Large $\tilde{g}_r^{\uparrow\downarrow}$ values are necessary to efficiently transport pure spin currents across the interface and these results thus helped to establish MOI/NM as suitable candidates. First experiments in this direction have been conducted by Heinrich \textit{et al.}~for YIG/Au bilayers~\cite{Heinrich2011}. Ab initio calculations for YIG/Au confirmed the experimentally observed findings~\cite{jia_spin_2011}. Further experiments and optimization of the interface treatment increased the obtained values for YIG/NM interfaces~\cite{Burrowes2012,Sun:2013go,rezende_enhanced_2013}. Similar values have also nowadays been obtained for ferromagnetic perovskites~\cite{Wahler2016}.

\subsection{Electrically detected spin pumping}
\begin{figure}[h]
 \includegraphics[width=85mm]{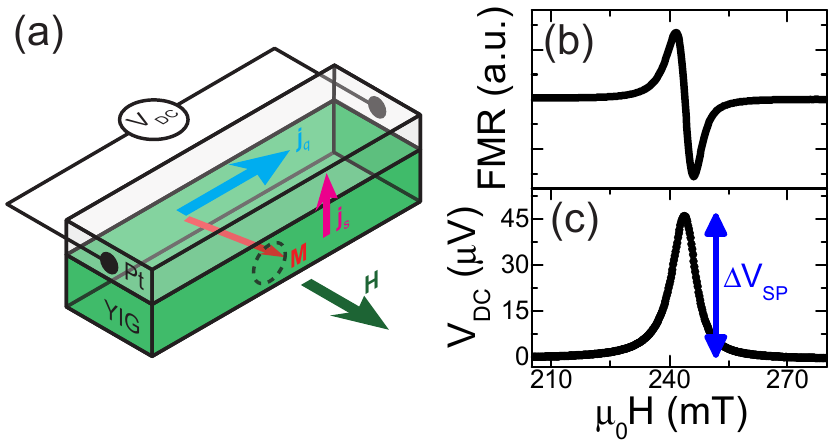}\\
 \caption[Electrically detected spin pumping]{(a) Illustration of the measurement scheme used in electrically detected spin pumping experiments. The static magnetic field $\mathbf{H}$ is applied perpendicular to the long side of the sample and a microwave field drives the magnetization into precession. The ISHE generated charge current is detected as a voltage drop across the long side of the sample with length $L$. (b) Detected FMR response as function of the applied external magnetic field $H$, using a field modulation technique, at the resonance field a peak-dip features is visible. (c) Simultaneously recorded $V_\mathrm{DC}$ signal. A peak appears in $V_\mathrm{DC}$ exactly at the ferromagnetic resonance field. Figure adapted with permission from Ref.~\cite{weiler_experimental_2013}.}
  \label{figure:SpinPumpingElectric}
\end{figure}

In electrically detected spin pumping experiments the goal is to detect the ISHE induced voltage in the NM generated from the pumped pure spin current across the interface as illustrated in Fig.~\ref{figure:SpinPumpingElectric}(a). In most experiments this is realized by attaching electrical connections to the sample and placing it into a microwave resonator within an static external magnetic field. Thus most electrically detected spin pumping experiments are using only one fixed microwave frequency~\cite{Czeschka2011}. The FMR signal of the sample is detected by exploiting the dispersive shift of the resonator at the ferromagnetic resonance. This can be achieved by measuring the reflected power of the resonator as a function of the applied external magnetic field. Most setups enhance the resolution by modulating the external magnetic field and using Lock-In detection techniques. Due to this field modulation the detected FMR signal is the first derivative with respect to the applied external magnetic field of the Lorentzian-shaped FMR absorption peak (See Fig.~\ref{figure:SpinPumpingElectric}(b)). The ISHE voltage from Eq.(\ref{eq:WMI:VSP}) consists of a DC and AC component. In the initial experiments only the DC part of this voltage was measured. For the AC component (oscillating with the FMR frequency/microwave frequency) additional inductive and capacitive crosstalk between microwave field or precessing magnetization makes it very difficult to unambiguously determine the spin pumping contribution to this voltage signal.
In the realm of electrically detected spin pumping experiments, we carried out experiments on YIG/Pt bilayers on GGG substrates grown by pulsed laser deposition and in-situ Pt electron beam evaporation mounted in a $9.3\,\mathrm{GHz}$ microwave resonator as detailed in Ref.~\cite{weiler_experimental_2013}. The results that we obtained for a YIG ($46\,\mathrm{nm}$)/Pt($7\,\mathrm{nm}$) bilayer at $300\,\mathrm{K}$ are shown in Fig.~\ref{figure:SpinPumpingElectric}(b) for the FMR signal and (c) for the DC ISHE voltage $V_\mathrm{DC}$. At the same external field value we observe the FMR of the YIG layer and a maximum in $V_\mathrm{DC}$. From $V_\mathrm{DC}$ we extracted then $\Delta V_\mathrm{SP}$ as illustrated in Fig.~\ref{figure:SpinPumpingElectric}(c). As it is not possible to independently determine $\lambda_\mathrm{sf}$, $\alpha_\mathrm{SH}$, and $\tilde{g}_r^{\uparrow\downarrow}$ from just a single electrically detected spin pumping measurement, we combined the results obtained from spin Hall magnetoresistance and longitudinal spin Seebeck effect experiments to derive a universal set of $\lambda_\mathrm{sf}$, $\alpha_\mathrm{SH}$, and $\tilde{g}_r^{\uparrow\downarrow}$ parameters, which give excellent quantitative agreement with each experiment. For our thin films we found $\lambda_\mathrm{sf}=1.5\,\mathrm{nm}$, $\alpha_\mathrm{SH}=0.11$, and $\tilde{g}_r^{\uparrow\downarrow}=1\times10^{19}\,\mathrm{m^{-2}}$.

First experiments on the electrically detected spin pumping in MOI/NM heterostructures have been conducted by in YIG/Pt bilayers. Initially, it was assumed that a great advantage in using MOIs in these experiments is that any rectification effects leading to additional DC voltage signals can be ruled out, which was a problem in metallic ferromagnetic layers. However, it is now clear that indeed additional rectification effects can be also present in MOI/NM samples, which originate for example from the spin Hall magnetoresistance. Thus, also in the case of MOI/NM samples the line shape of the $V_\mathrm{DC}$ has to be carefully analyzed to extract the relevant contribution from spin pumping. Several groups have investigated electrically detected spin pumping in MOI/NM systems~\cite{jungfleisch_temporal_2011,takahashi_electrical_2012,dallivy_kelly_inverse_2013,du_y_2015,jungfleisch_thickness_2015,takemasa_spatial_2017}. For example, Wang~\textit{et al.}~investigated electrically detected spin pumping in a variety of YIG/NM bilayers and by combining their results with spin Hall magnetoresistance measurements extracted the relevant spin transport parameters~\cite{wang_scaling_2014}. While the FMR is the fundamental collective spin excitation mode, further experiments and theoretical work showed that standing spin waves in the material excited by a microwave drive also lead to $V_\mathrm{DC}$ signals~\cite{Sandweg:2011ig}. One very elusive problem to experimentally tackle was the AC component of the ISHE voltage signal. First experiments tried to exploit parametric pumping of the FMR mode to move the FMR mode to higher frequencies~\cite{Hahn2013}. However, as pointed out by Weiler\textit{et al.}~not only inductive coupling to the microwave field leads to spurious AC contributions, but also the precessing motion of the magnetization itself in the MOI inductively couples AC voltages~\cite{Hahn2013,Wei2014,Weiler2014,li_direct_2016,kapelrud_spin_2017}. Despite these issues, it has been shown that this AC component is considerably larger than the DC part of spin pumping marking its relevance for future applications~\cite{chiba_current-induced_2014,kapelrud_spin_2017}. In recent experiments we showed that the pure spin current allows to establish a dynamical coupling between a ferromagnetic MOI and a ferromagnetic metal, which allow to investigate high order standing spin waves in the MOI~\cite{klingler_spin_2017}. THz emission by spin pumping effects have been studied in metallic ferromagnets/NM bilayers~\cite{Kampfrath2013,Seifert2016,huisman_femtosecond_2016,seifert_ultrabroadband_2017,huisman_spin-photo-currents_2017,bocklage_coherent_2017}, which makes similar experiments in MOI/NM promising~\cite{skarsvag_spin_2014}.

The reciprocal effect, i.e.~driving magnetization dynamics by a charge current in the NM have been already extensively studied as it allows for very interesting device applications~\cite{Liu2011,Jungwirth:2012em,spin-torque:Miron:NatMat:2010,Miron2011,Liu2012,Liu2012Pt}. Initially, experiments focused on the influence of a charge current in the NM on the ferromagnetic resonance properties (changes in the Gilbert damping parameter $\alpha$), for such experiments it is crucial to discern pure spin current effects from spurious contributions like thermal effects and Oersted fields generated by the charge current~\cite{Hamadeh2014}. Taking the concept a step further, spin-transfer torque driven FMR experiments showed that indeed charge currents in the NM can drive magnetization dynamics~\cite{Liu2011,chiba_current-induced_2014,Schreier2015,baumgartner_spatially_2017}. An intriguing application from these studies are spin Hall nano-oscillators, where large localized charge current densities are needed to drive the MOI into auto-oscillations via the SHE generated spin current~\cite{Duan2014,Cheng2016}. Finally, experiments conducted on TmIG with perpendicular magnetic anisotropy showed that pure spin currents can also switch efficiently the magnetization orientation~\cite{Avci2016} similar to results obtained in metallic ferromagnets~\cite{Miron2011,Liu2012Pt,Liu2012,Garello2013}. Similar to the magnetization reversal in metallic heterostructures also in MOI the switching is driven by domain wall nucleation and propagation, such that interfacial Rashba and Dzyaloshinskii-Moriya interactions are also relevant to explain this effect~\cite{Manchon2015}. In addition, pure spin currents can also very efficiently move domain walls~\cite{Miron2011:domainwall,Emori2013,Ryu2013,Ryu2016} and thus might also be used to move topological spin textures like skyrmions~\cite{Muhlbauer2009,sampaio_nucleation_2013,Gayles2015,Bttner2015,jiang_blowing_2015,Woo2016,fert_magnetic_2017}.

\section{Longitudinal spin Seebeck effect}
\label{spin_seebeck}

\begin{figure}[h]
 \includegraphics[width=85mm]{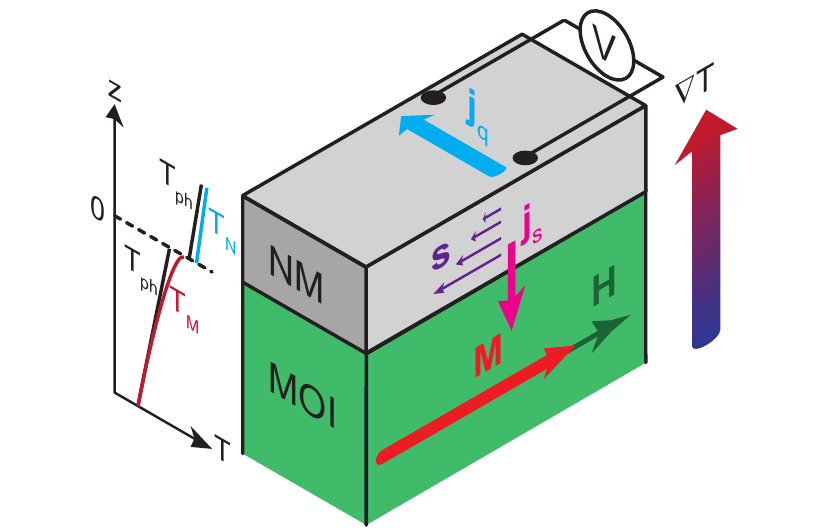}\\
 \caption[Principle Spin Seebeck effect]{Illustration of the spin Seebeck effect. The temperature gradient $\nabla T$ applied along the $\mathbf{z}$-direction gives rise to a temperature difference $T_\mathrm{M}-T_\mathrm{N}$ between the magnons in the MOI and the electrons in the NM at the interface. This drives a pure spin current $\mathbf{j}_\mathrm{s}$ across the interface, where the direction of spin orientation $\mathbf{s}$ depends on the orientation of the magnetic order parameter. In the NM the spin current is transformed into a charge current $\mathbf{j}_\mathrm{q}$ via the ISHE and detected using open circuit boundary conditions as a voltage $V$ between the sample edges.}
  \label{figure:SpinSeebeck}
\end{figure}

Another way of achieving an out-of-equilibrium condition for the magnetic order parameter can be realized by a thermal drive. This can be described using Eq.(\ref{equ:InterfacialSpinCurrent}) and assuming that $(T_\mathrm{M}-T_\mathrm{N})=0\neq0$ and $\mu_\mathrm{mag}=\mu_\mathrm{s}(0)=\dot{\mathbf{n}}=0$, i.e. there is a finite temperature difference between the magnonic system in the MOI and electronic system in the NM (We neglect here any thermally induced magnon accumulation and assume that in the NM the pure spin current is fully absorbed). The thermally driven spin current $\mathbf{j}_\mathrm{s}$ is then transformed into a charge current $\mathbf{j}_\mathrm{q}$ via the ISHE in the NM as illustrated in Fig.~\ref{figure:SpinSeebeck}. Under electrical open circuit conditions the generated $\mathbf{j}_\mathrm{q}$ is compensated by an electrical field, which can then be measured as a voltage drop $V$ across the sample edges over the distance $L$. This then called the longitudinal spin Seebeck effect~\cite{Uchida:2010,Uchida2010,Xiao2010,Rezende2014,Cahaya2015}. For the spin Seebeck effect two nice review articles have been published focused onto experiments by Uchida~\textit{et al.}~\cite{Uchida2014} and theory by Adachi~\textit{et al.}~\cite{Adachi2013}.

We can use a similar approach as for the spin pumping effect, to derive the necessary equations for the longitudinal spin Seebeck effect. First, the spin current flowing across the interface can be calculated straightforwardly from Eq.(\ref{equ:InterfacialSpinCurrent}) by assuming $\mu_\mathrm{mag}=\mu_\mathrm{s}(0)=\dot{\mathbf{n}}=0$:
\begin{equation}\label{eq:WMI:Js:SSE}
\mathbf{j}_\mathrm{s}^{\mathrm{SSE}}=S(T_\mathrm{M}-T_\mathrm{N})\mathbf{n} \;.
\end{equation}
Here, the relevant spin Seebeck parameter $S$ can be expressed as~\cite{Adachi2013,weiler_experimental_2013,Bender2015,Cornelissen2016}
\begin{equation}\label{eq:WMI:S_SSE}
S= \frac{\tilde{g}_r^{\uparrow\downarrow}}{2\pi}\frac{g_J \mu_\mathrm{B}}{M_\mathrm{sat}}\frac{\left(k_\mathrm{B} T\right)^{3/2}}{\left(4\pi D\right)^{3/2}}\zeta\left(\frac{5}{2}\right)k_\mathrm{B}\;,
\end{equation}
when assuming a single parabolic magnon band dispersion $\hbar \omega_k=D\mathbf{k}^2$ in the ferromagnetic insulator, where $\mathbf{k}$ is the magnon wave vector and $\omega_k$ is the corresponding magnon frequency. We neglect the contribution from the magnon band gap, which is a good assumption in case of YIG. Here, $k_\mathrm{B}$ is the Boltzmann constant, $\zeta(x)$ the Riemann zeta function, $D$ is the spin wave stiffness parameter, $g_J$ the Land\'{e} g-factor and $\mu_\mathrm{B}$ the Bohr magneton. $\mathbf{j}_\mathrm{s}^{\mathrm{SSE}}$ is then converted into a charge current in the NM and can then be detected as for an open electrical circuit boundary condition as a voltage $V$ as illustrated in Fig.~\ref{figure:SpinSeebeck}. For this voltage we find in analogy to the case for electrically detected spin pumping:
\begin{equation}\label{eq:WMI:VSSE}
  \frac{V}{L}=\frac{2e}{\hbar} \alpha_{\mathrm{SH}}
    \frac{ j_{\mathrm{s}}^{\mathrm{SSE}} \eta \lambda_{\mathrm{sf}} \tanh\left( \frac{t_{\mathrm{NM}}}{2 \lambda_{\mathrm{sf}}} \right)}{\sigma_{\mathrm{NM}} t_{\mathrm{NM}}}\;.
\end{equation}
We here included the backflow parameter $\eta$ to account for any spin accumulation driven spin current backflow if the thickness $t_\mathrm{NM}$ of the NM is comparable to the spin diffusion length $\lambda_{\mathrm{sf}}$. With these expressions at hand, we can now start to discuss experimental means to measure the longitudinal spin Seebeck effect.

\subsection{Current reversal detection of spin Seebeck effect}

Several means to generate the required out-of-plane temperature gradient have been used in the experiments. A very basic approach is to sandwich the sample between two heater blocks equipped with resistive heaters and thermometers~\cite{Uchida:2010,Uchida2010,kikkawa_longitudinal_2013,meier_longitudinal_2015,sola_longitudinal_2017}. This allows to very accurately determine the temperature gradient across the whole sample and control it by the power applied to the heaters. One important aspect to keep in mind in such experiments is that the substrate used for most multilayers is several orders of magnitude thicker than the thin films responsible for the signal. Thus most of the temperature gradient is picked up by the substrate and not by the thin films itself. A elegant solution is the usage of freely suspended thin film samples and meander shaped metallic strips as a resistive heater~\cite{sultan_thermal_2009,zink_exploring_2010,avery_thermopower_2011}. For example, Avery \textit{et al.}~used such measurement environments to very accurately determine Seebeck coefficients of various metals~\cite{avery_thermopower_2011}. However, the out-of-plane temperature gradient for the longitudinal Seebeck effect is very tough to determine in such freely suspended structures. Another approach also successfully employed by our group is to an intense laser beam to locally heat up the sample~\cite{Weiler2012}. Within the spatial resolution given by the laser spot size, such experiments allow to investigate for example the influence of magnetic domains and domain walls on the spin Seebeck signal. A direct measurement of the temperature gradient is nearly impossible in these optical experiments and only numerical simulations of the heat transport allow to get reasonable estimations of the achieved temperature difference~\cite{schreier_magnon_2013}. Such simulations require for example good knowledge on the interface resistance for heat currents, which up to now is only well know for a very limited set of material combinations.

\begin{figure*}[t]
 \includegraphics[width=170mm]{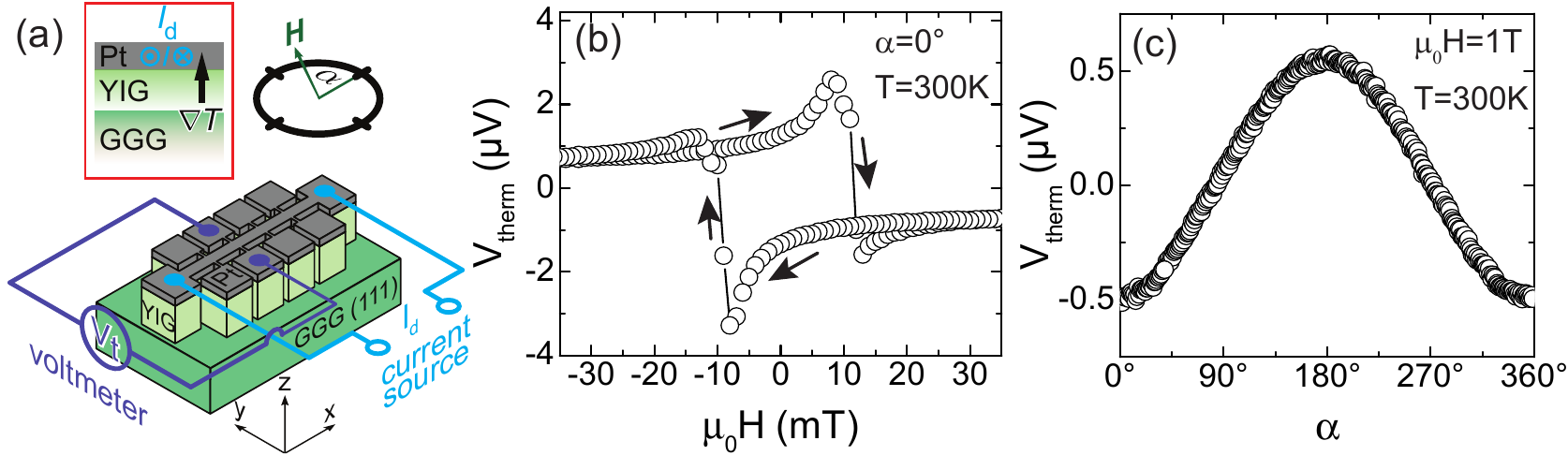}\\
 \caption[Current induced Spin Seebeck effect]{Measurement of spin Seebeck effect using current induced heating of the NM layer. (a) Illustration of the measurement scheme used for the detection of the longitudinal spin Seebeck effect. The transverse voltage $V_\mathrm{t}$ is measured for both polarities of a charge current drive $I_\mathrm{d}$. These measurements can then be conducted as a function of the applied field $\mu_0H$ or field orientation $\alpha$. (b) Magnetic field dependence of $V_\mathrm{therm}$ for a YIG/Pt heterostructure at $T=300\,\mathrm{K}$ and $\alpha=0^\circ$. We observe a hysteretic $V_\mathrm{therm}(H)$ and for large external magnetic field values $V_\mathrm{therm}$ saturates with different polarity for negative and positive fields. (c) Dependence of $V_\mathrm{therm}$ on the magnetic field orientation for the very same sample and temperature as in (b). Figures and data adapted with permission from Ref.~\cite{Schreier2013}. Copyright 2013 by the American Physical Society.}
  \label{figure:SpinSeebeckCurrentHeating}
\end{figure*}

Over the course of our investigations of the longitudinal spin Seebeck effect, we established a measurement scheme, which utilizes a NM patterned into a Hallbar mesa on top of the MOI~\cite{Schreier2013}. The idea is illustrated in Fig.~\ref{figure:SpinSeebeckCurrentHeating}(a), while driving a charge current $I_\mathrm{d}$ across the Hallbar, the transverse voltage $V_\mathrm{t}$ is recorded as a function of the applied field orientation $\mathbf{h}$ (and the in-plane angle $\alpha$ with respect to $\mathbf{j}_\mathrm{q}$) and magnitude $\mu_0 H$. To separate the thermal voltage $V_\mathrm{therm}$ arising from the temperature gradient induced by Joule heating from the resistive voltage response $V_\mathrm{res}$ due to the charge current drive, we utilize the fact that they behave differently upon reversal of charge current polarity. We thus measure $V_\mathrm{t}(I_\mathrm{d}^+)$ for positive current bias polarity and $V_\mathrm{t}(I_\mathrm{d}^-)$ for negative current bias polarity (the absolute value of the applied charge current remains the same). From these two measurements one can then calculate $V_\mathrm{therm}$ as
\begin{equation}\label{eq:SSE_CurrentTherm}
V_\mathrm{therm}=\frac{1}{2}\left(V_\mathrm{t}(I_\mathrm{d}^+)+V_\mathrm{t}(I_\mathrm{d}^-)\right)\;,
\end{equation}
and $V_\mathrm{res}$ as
\begin{equation}\label{eq:SSE_CurrentRes}
V_\mathrm{therm}=\frac{1}{2}\left(V_\mathrm{t}(I_\mathrm{d}^+)-V_\mathrm{t}(I_\mathrm{d}^-)\right)\;.
\end{equation}
The advantage of this measurement technique is that it only requires equipment for electrical magnetotransport measurements and can be employed also quite easily in superconducting magnet cryostats. Of course the direct measurement of the temperature difference at the MOI/NM interface is not possible with this technique making quantitative measurements of $S$ impossible. However, as the resistivity of the NM is simultaneously measured, one can use the NM resistance as a temperature sensor during the measurements. This works for example quite well with Pt as the NM as it exhibits a linear resistance versus temperature curve over a wide range of temperatures. Another important drawback to mention is that due to the temperature dependence of NM resistance the applied constant charge current leads to a variable heating power applied to the sample as a function of temperature, which has to be accounted for in the interpretation of the observed signals.

We exemplarily show in Fig.~\ref{figure:SpinSeebeckCurrentHeating}(b) and (c) the results we obtained for this technique in a YIG($61\,\mathrm{nm}$)/Pt($11\,\mathrm{nm}$) bilayer sample as presented in Ref.~\cite{Schreier2013}. The Hallbar patterned into the blanket film by optical lithography and Ar ion beam milling has a width of $80\,\mathrm{\mu m}$ and a length of $800\,\mathrm{\mu m}$. For the measurements the magnitude of the applied charge current was $10\,\mathrm{mA}$ and the sample was mounted in a superconducting magnet cryostat with a fixed sample temperature of $300\,\mathrm{K}$. We first focus on the evolution of $V_\mathrm{therm}$ as function of $\mu_0 H$ for the external magnetic field applied parallel to the charge current direction $\alpha=0^\circ$ (see Fig.~\ref{figure:SpinSeebeckCurrentHeating}(b)). $V_\mathrm{therm}$ exhibits a hysteresis and a change in sign for large positive and negative external magnetic fields. This sign change for a reversal of is also visible in the measurements of $V_\mathrm{therm}$ as function of magnetic field orientation $\alpha$ for $\mu_0 H=1\,\mathrm{T}$ shown in Fig.~\ref{figure:SpinSeebeckCurrentHeating}(c). Clearly, $V_\mathrm{therm}$ has a $\cos (\alpha)$ dependence. This can be rationalized by ISHE based detection of the thermally driven spin current across the interface. The spin orientation $\mathbf{s}$ of $\mathbf{j}_\mathrm{s}^{\mathrm{SSE}}$ is always aligned collinear to $\mathbf{m}$. The direction of the transformed charge current by the ISHE thus also depends on the magnetization direction, as evident from Eq.(\ref{equ:ISHE}). The voltage we detect due to the longitudinal spin Seebeck effect in the transverse contacts has a maximum, if the ISHE induced charge current is also flowing in the transverse $\mathbf{y}$- direction, which is achieved for $\mathbf{m}$ oriented along $\mathbf{x}$. If $\mathbf{m}$ is oriented along $-\mathbf{x}$ the direction of the ISHE induced charge current flow is also inverted and thus sign of $V_\mathrm{therm}$. These current induced spin Seebeck measurements allow in this regard to qualitatively compare magnetic field dependence and magnetic field orientation dependence from sample to sample and/or as a function of temperature. In the experiment it might be necessary to confirm Joule heating as the sole cause for the extracted $V_\mathrm{therm}$, which can be done by investigating the dependence of $V_\mathrm{therm}$ on the magnitude of $I_\mathrm{d}$ As already mentioned they do not allow to directly infer the spin Seebeck coefficient $S$ from just these measurements.

While this technique allowed us to get a deeper understanding of the underlying physics for the longitudinal spin Seebeck effect in compensated iron garnets as detailed in the next subsection, we also want to briefly highlight other recent interesting discoveries in this field. First it is important to mention that spurious thermal voltages, for example originating from a proximity magnetized Pt layer at the MOI interface need to be account for in the longitudinal spin Seebeck effect~\cite{huang_transport_2012,qu_intrinsic_2013}. Several groups put forward symmetry arguments, which allow to disentangle the different thermal voltages~\cite{kikkawa_longitudinal_2013,kikkawa_separation_2013,bougiatioti_quantitative_2017}. By using a modulated laser power in optical heating experiments the time-dependence of the longitudinal spin Seebeck effect has been investigated by Roschewsky \textit{et al.}~in nm-thick YIG samples grown by pulsed laser deposition~\cite{roschewsky_time_2014}. Within the temporal resolution (bandwidth $30\,\mathrm{MHz}$) of the setup no intrinsic frequency limit of the longitudinal spin Seebeck effect was observed and an upper limit of $5\,\mathrm{ns}$ for reaching a steady-state condition was inferred from these results. Similar experiments on $\mathrm{\mu m}$-thick YIG films grown by the LPE method conducted by Agrawal \textit{et al.}~found an intrinsic limit of $343\,\mathrm{ns}$ for reaching a steady-state in the longitudinal spin Seebeck effect~\cite{agrawal_role_2014}. The MOI thickness dependence of the longitudinal spin Seebeck effect was further explored by Schreier \textit{et al.}~up to microwave frequency, where a thickness dependence of the intrinsic time-constant of the longitudinal spin Seebeck effect was experimentally verified~\cite{schreier_spin_2016}. Quite recently, Seifert \textit{et al.}~conducted THz spectroscopy on YIG/Pt bilayers and found response times in the fs regime~\cite{seifert_launching_2017}. Similarly, optical probing of the thermally  induced spin transfer showed that this transfer happens at ps timescales~\cite{kimling_picosecond_2017}. Furthermore, thickness dependence and temperature dependent investigations of the longitudinal spin Seebeck effect in YIG/Pt heterostructures have been conducted~\cite{kehlberger_length_2015,jin_effect_2015}. All these experiments suggest that magnons with different wavelength and band dispersion contribute differently to the longitudinal spin Seebeck effect in YIG/NM structures. Theoretical models put forward by Ritzmann \textit{et al.}~allowed to calculate the thermal excitation spectrum in longitudinal spin Seebeck effect experiments ~\cite{ritzmann_propagation_2014,ritzmann_thermally_2017} and helped to further understand the observed suppression of the longitudinal spin Seebeck effect for large applied external magnetic fields~\cite{kikkawa_critical_2015,jin_effect_2015,ritzmann_magnetic_2015}. First experiments on the longitudinal spin Seebeck effect in REIGs/NM bilayers have been conducted in Ref.~\cite{uchida_longitudinal_2013}. Experimental observation of the longitudinal spin Seebeck effect in MOIs that are different material class than the REIGs have also been reported in ferrites~\cite{meier_longitudinal_2015,guo_thermal_2016,anadon_characteristic_2016} and perovskite~\cite{wu_longitudinal_2017} materials. Furthermore, inserting an antiferromagnetic MOI between the ferromagnetic insulator and the NM leads to interesting effects for the temperature dependence of the longitudinal spin Seebeck effect~\cite{lin_enhancement_2016,prakash_spin_2016}. Furthermore, first promising results on the longitudinal spin Seebeck effect in antiferromagnetic MOI/NM bilayers have been reported~\cite{seki_thermal_2015,wu_antiferromagnetic_2016,rezende_theory_2016,holanda_spin_2017}. Last, but not least, the non-local spin Seebeck measurements conducted for example by optical heating and current induced heating also confirm that the magnon accumulation due to the magnon chemical potential is also contributing to the thermal voltage detected in the NM~\cite{Giles2015,Cornelissen2015, Cornelissen2016}.

\subsection{Spin Seebeck effect in compensated garnets}

As evident from Eq.(\ref{eq:WMI:Js:SSE}) the spin current across the MOI/NM interface is influenced by the orientation of the order parameter $\mathbf{n}$ and the spin Seebeck coefficient $S$. The orientation of the spin-polarization $\mathbf{s}$ is then determined by $\mathbf{n}$ and the sign of $S$, while the magnitude $j_\mathrm{s}^{\mathrm{SSE}}$ solely depends on $S$, and of course the temperature difference $T_\mathrm{M}-T_\mathrm{N}$ between the magnons in the MOI and the electrons in the NM. We investigated this theoretical conjecture by studying the current induced longitudinal spin Seebeck effect in GdIG/Pt heterostructures published in Refs.~\cite{Geprgs2016,cramer_magnon_2017}. For a ferrimagnetic material, the relevant order parameter $\mathbf{N}$ is the N\'{e}el vector (In the simplified two sublattice model for the REIGs $\mathbf{N}=\mathbf{M}^\mathrm{RE}-\mathbf{M}^\mathrm{Fe}$) and not the net magnetization. From this we know that $\mathbf{N}$ inverts its orientation at $T_\mathrm{comp}$. This change in orientation of $\mathbf{N}$ in turn should lead in the experiment to a change in the voltage sign of the longitudinal spin Seebeck effect (see Eq.(\ref{eq:WMI:Js:SSE})).

\begin{figure*}[t]
 \includegraphics[width=170mm]{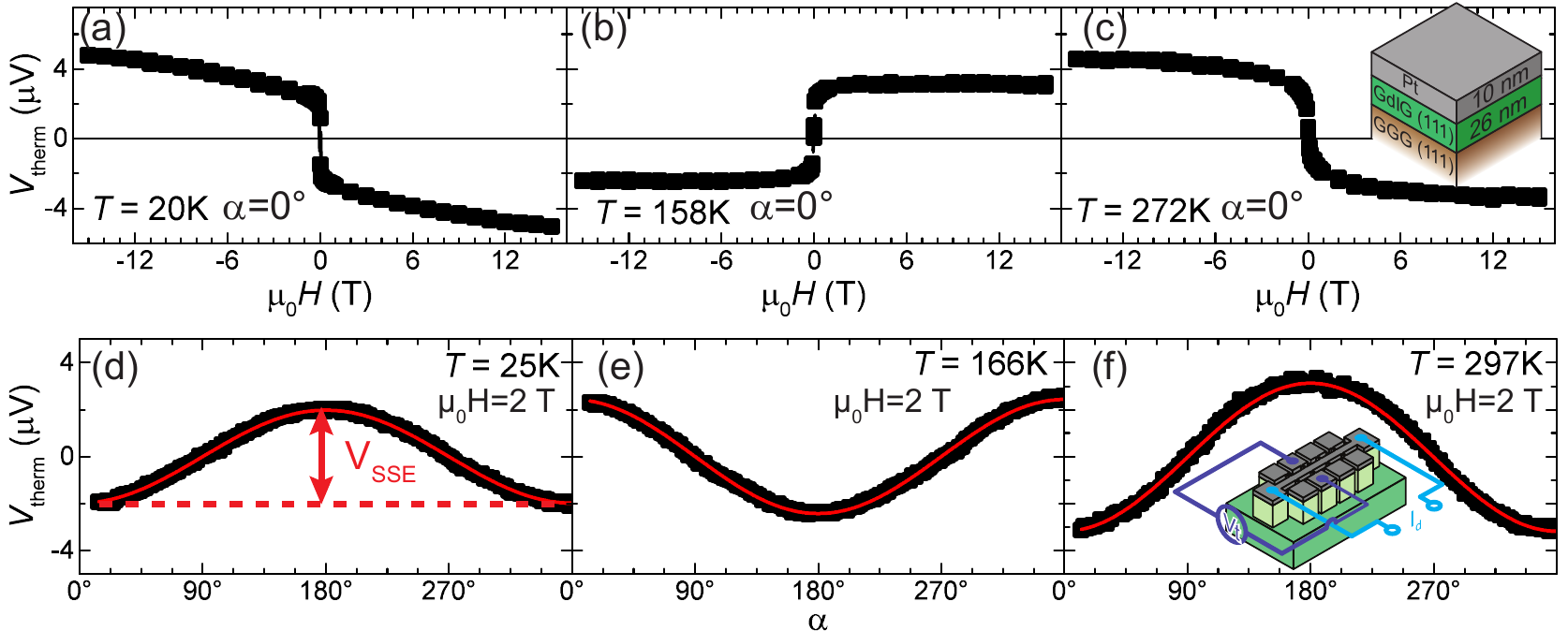}\\
 \caption[Temperature dependence of Spin Seebeck effect in compensated garnets]{Measurement of spin Seebeck effect as a function of temperature in GdIG using current induced heating of the NM layer. (a), (b), and (c) $V_\mathrm{therm}$ as a function of the applied external field $\mu_0 H$ for $\alpha=0^\circ$ at $T=20\,\mathrm{K}$, $T=158\,\mathrm{K}$, and $T=272\,\mathrm{K}$, respectively. The polarity of $V_\mathrm{therm}$ for large positive fields changes its sign for these three different temperatures. (d), (e), and (f) $V_\mathrm{therm}$ as a function of $\alpha$ for $\mu_0 H=2\,\mathrm{T}$ at $T=25\,\mathrm{K}$, $T=166\,\mathrm{K}$, and $T=297\,\mathrm{K}$, respectively. The applied charge current drive is $I_\mathrm{d}=6\,\mathrm{mA}$. Figures and data adapted from Ref.~\cite{Geprgs2016}.}
  \label{figure:SpinSeebeckGdIG_Orient and H}
\end{figure*}

Indeed, the results we obtained for a GdIG/Pt heterostructure grown on a (111)-oriented YAG substrate with $T_\mathrm{comp}=255\,\mathrm{K}$ compiled in Fig.~\ref{figure:SpinSeebeckGdIG_Orient and H} confirm this conjecture, but also added another important finding. From the field dependent current induced spin Seebeck experiments we find for large positive external magnetic fields ($\mu_0 H>1\,\mathrm{T}$) $V_\mathrm{therm}<0$ for $T>T_\mathrm{comp}$ as shown in Fig.~\ref{figure:SpinSeebeckGdIG_Orient and H}(c). For intermediate temperatures below $T_\mathrm{comp}$, we find $V_\mathrm{therm}>0$ at large positive fields (Fig.~\ref{figure:SpinSeebeckGdIG_Orient and H}(b)), but surprisingly a second sign change occurs for even lower temperatures $V_\mathrm{therm}<0$ (Fig.~\ref{figure:SpinSeebeckGdIG_Orient and H}(a)). For angle-dependent current induced spin Seebeck experiments $V_\mathrm{therm}(\alpha)$ at $\mu_0 H=2\,\mathrm{T}$ for comparable temperatures in Fig.~\ref{figure:SpinSeebeckGdIG_Orient and H}(d)-(f), we also observe this double sign change. Here, $V_\mathrm{therm}(\alpha)$, has a $\cos(\alpha)$-dependence as expected for the longitudinal spin Seebeck effect~\cite{Schreier2013} and the amplitude $V_\mathrm{SSE}$ changes its sign. In both measurements, we utilized the Pt resistance measured simultaneously as an on-chip temperature sensor to determine the sample temperatures given here.

\begin{figure}[h]
 \includegraphics[width=85mm]{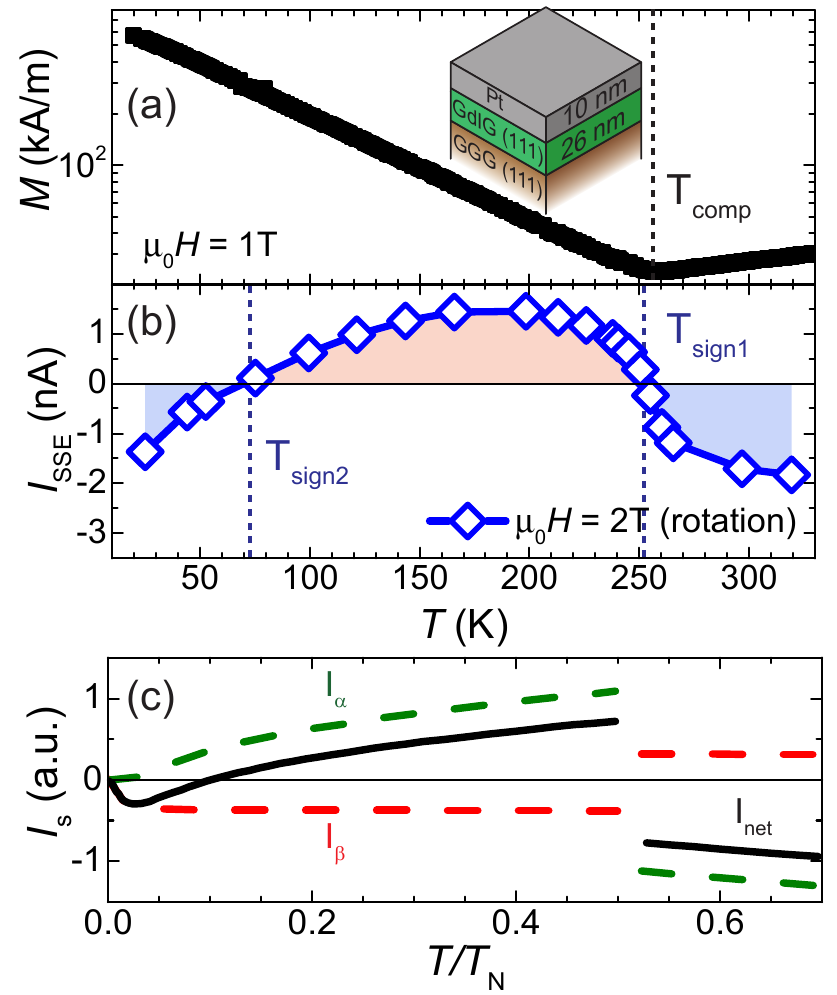}\\
 \caption[Temperature dependence of spin Seebeck effect in GdIG]{(a) Temperature dependence of the magnetization obtained for a GdIG/Pt bilayer sample grown on a YAG substrate for $\mu_0 H=1\,\mathrm{T}$. The observed kink in the temperature dependence yields $T_\mathrm{comp}$. Temperature dependence of $I_\mathrm{SSE}$ extracted from $V_\mathrm{therm}(\alpha)$ measurement at various fixed temperatures. Over the investigated temperature range two sign changes of $I_\mathrm{SSE}$ at $T_\mathrm{sign1}$ and $T_\mathrm{sign2}$ are visible. (c) Temperature dependence of $I_\mathrm{SSE}$ from a theoretical model accounting for contributions from two different magnon modes ($\alpha$ and $\beta$) to the spin Seebeck effect. Figures and data adapted from Ref.~\cite{Geprgs2016}.}
  \label{figure:SpinSeebeck_GdIG_Temperature dependence}
\end{figure}

To further investigate the occurrence of these two sign changes in the spin Seebeck voltage, we conducted $V_\mathrm{therm}(\alpha)$ measurements at various temperatures and extracted $V_\mathrm{SSE}$ for each measured temperature. From these measurements we then extracted $V_\mathrm{SSE}$ (see Fig.\ref{figure:SpinSeebeckGdIG_Orient and H}(d))and calculated the spin Seebeck current $I_\mathrm{SSE}=V_\mathrm{SSE}/R_\mathrm{Pt}$, where $R_\mathrm{Pt}$ is the Pt resistance for each temperature, to correct for the temperature dependent change of $R_\mathrm{Pt}$ in our measurements. The extracted temperature dependence of $I_\mathrm{SSE}$ is shown in Fig.~\ref{figure:SpinSeebeck_GdIG_Temperature dependence}(b) and compared to the temperature dependence of the magnetization for the very same sample in Fig.~\ref{figure:SpinSeebeck_GdIG_Temperature dependence}(a). $I_\mathrm{SSE}$ exhibits two sign changes at $T_\mathrm{sign1}$ and $T_\mathrm{sign2}$. $T_\mathrm{sign1}$ agrees reasonably well with the compensation temperature $T_\mathrm{comp}$ from magnetometry measurements (Fig.~\ref{figure:SpinSeebeck_GdIG_Temperature dependence}(a)). The second sign change $T_\mathrm{sign2}$ at lower temperatures is clearly not related to any changes in the temperature dependence of the magnetization.

Two contributions account for the observation of $T_\mathrm{sign1}$ and $T_\mathrm{sign2}$. As already explained $T_\mathrm{sign1}$ close to $T_\mathrm{comp}$ is explained by the inversion of $\mathbf{N}$ at the compensation temperature and also explains the abrupt temperature dependence of $I_\mathrm{SSE}$ in this temperature regime. The low temperature $T_\mathrm{sign2}$ can be explained by taking into account the contributions of two different magnon bands to $S$~\cite{Xiao2010}: a gapless parabolic magnon band (dominantly spin excitations at the iron sites, $\alpha$-mode), which is thermally occupied at all temperatures, and a gapped optical magnon mode (dominantly spin excitations at the gadolinium sites, $\beta$-mode), which is only populated at higher temperatures (below $T_\mathrm{sign2}$). This simple model takes into account only the lowest energy magnon branches of REIGs and assumes that these two modes contribute differently and with opposite sign to $S$, which could be rationalized by the s-d and s-f coupling being relevant for the spin current flow across the interface for the $\alpha$-mode and $\beta$-mode, respectively. The results of these calculations for $I_\mathrm{SSE}$ are shown in Fig.~\ref{figure:SpinSeebeck_GdIG_Temperature dependence}(c) with an about 1 order of magnitude lower efficiency for the gaped optical mode as detailed in Ref.~\cite{Geprgs2016}. The simulation can nicely explain the observed temperature-dependence of GdIG thin films. Our results highlight the importance of the magnonic bandstructure for the spin Seebeck effect and confirm that in a magnetically ordered system with multiple magnetic sublattices, the individual contribution of each sublattice to the longitudinal spin Seebeck effect can be quite different. One should also emphasize that these experiments provide evidence for the existence of these two magnon bands, which were previously only detectable by neutron scattering and inelastic light scattering experiments.

Taken all together, the investigation of the longitudinal spin Seebeck effect in MOI/NM bilayers has helped to push for more sophisticated theories and experiments and thus to increase the knowledge on the underlying physics. Still many open questions need to be addressed in future especially investigating the longitudinal SSE in MOIs with a more complex spin order, like for example chiral or even topological spin textures, or with engineered magnonic properties.

\section{Spin Hall magnetoresistance}
\label{spin_hall_magnetoresistance}

In the previous two sections we discussed the effects of a non-equilibrium state of the magnetic order parameter, which in turn drives a pure spin current across the MOI/NM interface into the NM. In the following we will look into phenomena arising when one drives a charge current through the NM and generates a pure spin current flowing towards the MOI/NM interface. This gives rise on the one hand to the spin Hall magnetoresistance (SMR), i.e.~the NM resistance depends on the orientation of the magnetic order parameter $\mathbf{N}$ in the MOI. On the other hand, it allows to investigate pure spin current transport via magnetic excitation quanta in all-electrical experiments (see Section~\ref{magnon_mediated_magnetoresistance}).

\begin{figure}[h]
 \includegraphics[width=85mm]{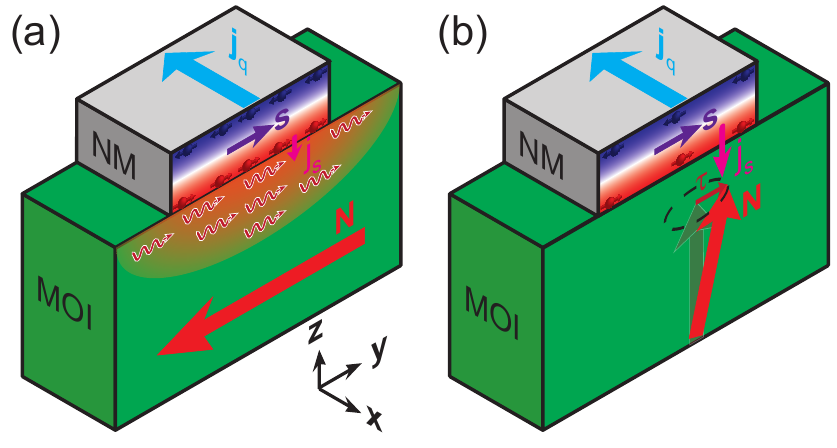}\\
 \caption[SMR Spin current boundary conditions]{Illustration of the different boundary conditions imposed by the orientation of $\mathbf{N}$ in the MOI onto the spin accumulation generated by the SHE due to a charge current $\mathbf{j}_\mathrm{q}$ flowing in the NM. (a) For $\mathbf{N}\parallel\mathbf{s}$, only a small amount of spin current due to the spin convertance $g$ flows across the interface and causes an accumulation of magnons underneath the NM strip. (b) For $\mathbf{N}\perp\mathbf{s}$ a larger spin current flows across the interfaces as in (a) mediated by the spin mixing conductance $\tilde{g}{\uparrow\downarrow}$. The transferred angular momentum acts as a spin transfer torque $\boldsymbol{\tau}$ onto the magnetic order parameter $\mathbf{N}$.}
  \label{figure:SMRBoundaries}
\end{figure}

For both phenomena we need to understand the boundary conditions imposed to the pure spin current flow across the interface by the orientation of $\mathbf{N}$ with respect to the spin polarization $\mathbf{s}$ (The orientation of $\mathbf{s}$ is fixed by the direction of $\mathbf{j}_\mathrm{q}$ and the flow direction of $\mathbf{j}_\mathrm{s}$ towards the interface). If we assume $\mu_\mathrm{s}(0)\neq0$, due to the spin accumulation generated by the SHE (see Fig.~\ref{figure:SpinHall}(a) for the spin accumulation), while putting $\mu_\mathrm{mag}=\dot{\mathbf{n}}=(T_\mathrm{M}-T_\mathrm{N})=0$ we obtain from Eq.(\ref{equ:InterfacialSpinCurrent}) the following expression for the pure spin current $\mathbf{j}_\mathrm{s}^\mathrm{el}$ across the MOI/NM interface:
\begin{equation}\label{eq:WMI:Js:SMR}
\mathbf{j}_\mathrm{s}^{\mathrm{el}}= \frac{1}{4\pi}\left(\tilde{g}_i^{\uparrow\downarrow}+\tilde{g}_r^{\uparrow\downarrow} \mathbf{n}\times\right) \mu_\mathrm{s}(0) \mathbf{s}\times\mathbf{n}+g \mu_\mathrm{s}(0) \left(\mathbf{s}\cdot \mathbf{n}\right)\mathbf{n}\;.
\end{equation}
There are two contributions to $\mathbf{j}_\mathrm{s}^{\mathrm{el}}$. The first term depends on the spin mixing conductance $\tilde{g}^{\uparrow\downarrow}$, which remains finite even for $T=0$. In contrast the second term is governed by the spin convertance $g$, which vanishes for $T=0$. From theoretical calculations one can also estimate that $g=0.06\tilde{g}_r^{\uparrow\downarrow}$ for YIG at room temperature, such that one can safely assume $|\tilde{g}^{\uparrow\downarrow}|\gg g$~\cite{Cornelissen2016} (Although it would be very interesting to find a way to go over to the opposite regime, where the spin convertance dominates). Moreover, we see directly from Eq.(\ref{eq:WMI:Js:SMR}) that the first term vanishes, if $\mathbf{n}\parallel\mathbf{s}$ and the second term is zero for $\mathbf{n}\perp\mathbf{s}$. Thus, by controlling the orientation of $\mathbf{N}$ one can switch between two different conditions for the pure spin current transport as illustrated in Fig.~\ref{figure:SMRBoundaries}. If $\mathbf{n}\parallel\mathbf{s}$ as shown in Fig.~\ref{figure:SMRBoundaries}(a) only the term with $g$ contributes to the spin current across the interface. For $\mathbf{n}\perp\mathbf{s}$ only the term with $\tilde{g}^{\uparrow\downarrow}$ is relevant for $\mathbf{j}_\mathrm{s}^{\mathrm{el}}$ as illustrated in Fig.~\ref{figure:SMRBoundaries}(b). Due to the fact that $|\tilde{g}^{\uparrow\downarrow}|\gg g$, the pure spin current across the interface is much larger for $\mathbf{n}\perp\mathbf{s}$ than for $\mathbf{n}\parallel\mathbf{s}$ (at $T=0$, $\mathbf{j}_\mathrm{s}^{\mathrm{el}}=0$ for $\mathbf{j}_\mathrm{s}^{\mathrm{el}}$). The finite spin current flow across the interface reduces the spin accumulation in the NM at the MOI/NM interface. In the NM the flowing spin current is carried by the angular momentum of the mobile charge carriers, which leads to an effective increase in the path they have to traverse for contributing to the charge current flow along the $\mathbf{x}$-direction and in this way also increases the resistance in the NM. Thus, the larger $\mathbf{j}_\mathrm{s}^{\mathrm{el}}=0$ the larger the resistance increase. From this discussion it follows that for $\mathbf{n}\parallel\mathbf{s}$ the resistance of the NM is smaller than for $\mathbf{n}\perp\mathbf{s}$. This is the spin Hall magnetoresistance.

For the further theoretical discussion of the SMR, we now assume that $g=0$, which is a good approximation for the materials investigated. In addition, we now use the coordinate system defined by the charge current direction $\mathbf{j}$, the surface normal $\mathbf{z}$ and the transverse direction $\mathbf{t}=\mathbf{z}\times\mathbf{j}$ (see Fig.~\ref{figure:SMRADMR}(a)). As the theoretical framework describing the SMR has been nicely described by Chen~\textit{et al.}~in Refs.~\cite{chen_theory_2013,Chen2016SMRReview}, we here focus only on the obtained results from these spin-diffusive transport calculations. For the longitudinal resistivity of the NM $\rho_\mathrm{long}$ one then obtains~\cite{chen_theory_2013,Chen2016SMRReview}
\begin{equation}
\rho_\mathrm{long}=\rho_0+\rho_1\,(1-n_\mathrm{t}^2)\;,
\label{eq:SMR:long}
\end{equation}
where $n_\mathrm{t}=(\mathbf{N}/N)\cdot\mathbf{t}$ is the projection of the magnetic order parameter direction onto $\mathbf{t}$ (see Fig.~\ref{figure:SMRADMR}(a)), $\rho_i$ are the resistivity parameters describing the SMR in the NM. In addition, one can also determine the transverse resistivity $\rho_\mathrm{trans}$ measured along the $\mathbf{t}$-direction as~\cite{chen_theory_2013,Chen2016SMRReview}
\begin{equation}
\rho_\mathrm{trans}=\rho_2 n_\mathrm{z}+\rho_3\,n_\mathrm{j}n_\mathrm{t}\;,
\label{eq:SMR:trans}
\end{equation}
with $n_\mathrm{j}$, $n_\mathrm{z}$ the projections of $\mathbf{n}$ onto $\mathbf{j}$ and $\mathbf{z}$, respectively. The term with $\rho_2$ represents an anomalous Hall~\cite{nagaosa_anomalous_2010} type contribution from the SMR to $\rho_\mathrm{trans}$, while the second term is the transverse resistivity analogue to the $\rho_1$ parameter for $\rho_\mathrm{long}$. Thus, the theoretical description predicts $\rho_1=\rho_3$.

For the the SMR ratio $\rho_1/\rho_0$ one writes for $\tilde{g}_r^{\uparrow\downarrow}\gg \tilde{g}_i^{\uparrow\downarrow}$~\cite{chen_theory_2013,Chen2016SMRReview}:
\begin{equation}
\frac{\rho_1}{\rho_0}=\frac{\alpha_\mathrm{SH}^2\left(2\lambda_\mathrm{sf}^2\rho_\mathrm{NM}\right)(t_\mathrm{NM})^{-1} \tilde{g}_r^{\uparrow\downarrow}
\tanh^2\left(\frac{t_\mathrm{NM}}{2\lambda_\mathrm{sf}}\right)}
{h\,e^{-2}+2\lambda_\mathrm{sf}\rho_\mathrm{NM}\tilde{g}_r^{\uparrow\downarrow}\coth\left(\frac{t_\mathrm{NM}}{\lambda_\mathrm{sf}}\right)}\;.
\label{equ:SMR_Quan_Ratio}\\
\end{equation}
Here, $\rho_\mathrm{NM}$ is the resistivity of the NM. The SMR ratio is proportional to $\alpha_\mathrm{SH}^2$. For $\alpha_\mathrm{SH}\approx0.1$, we expect an SMR ratio of the order of $1\%$. Moreover, Eq.(\ref{equ:SMR_Quan_Ratio}) allows to determine $|\alpha_\mathrm{SH}|$ and $\lambda_\mathrm{sf}$ of the NM, if the SMR is measured as a function of the NM thickness and the spin mixing conductance is known~\cite{Vlietstra2013,althammer_quantitative_2013}. However, the sign of $\alpha_\mathrm{SH}$ can not be determined due to the scaling of the SMR with $\alpha_\mathrm{SH}^2$.

For the anomalous Hall type contribution the theoretical calculations obtain for $\rho_2/\rho_0$ in the limit of $\tilde{g}_r^{\uparrow\downarrow}\gg \tilde{g}_i^{\uparrow\downarrow}$:
\begin{equation}
\frac{\rho_{2}}{\rho_0}=-\frac{\alpha_\mathrm{SH}^2\left(2\lambda_\mathrm{sf}^2\rho_\mathrm{NM}\right)(t_\mathrm{NM})^{-1} \tilde{g}_i^{\uparrow\downarrow}
\tanh^2\left(\frac{t_\mathrm{NM}}{2\lambda_\mathrm{sf}}\right)}
{\left(h\,e^{-2}+2\lambda_\mathrm{sf}\rho_\mathrm{NM}\tilde{g}_r^{\uparrow\downarrow}\coth\left(\frac{t_\mathrm{NM}}{\lambda_\mathrm{sf}}\right)\right)^2}\;.
\label{equ:AHE_SMR_Quan_Ratio}
\end{equation}
This expression is quite similar to the SMR ratio, but scales with $\tilde{g}_i^{\uparrow\downarrow}$ instead of $\tilde{g}_r^{\uparrow\downarrow}$. From theory and experiment, we now know that the imaginary part of the spin mixing conductance is about 2 orders of magnitude smaller than its real part~\cite{althammer_quantitative_2013,Meyer2015,jia_spin_2011}.

From Eqs.(\ref{eq:SMR:long},\ref{eq:SMR:trans}) we find that the SMR leads to a characteristic magnetization orientation dependence for $\rho_\mathrm{long}$ and $\rho_\mathrm{trans}$. This allows to uniquely identify the SMR in a magnetotransport experiment as detailed in the following subsection.

\subsection{Spin Hall magnetoresistance in angle-dependent magnetotransport experiments}

\begin{figure*}[t]
 \includegraphics[width=170mm]{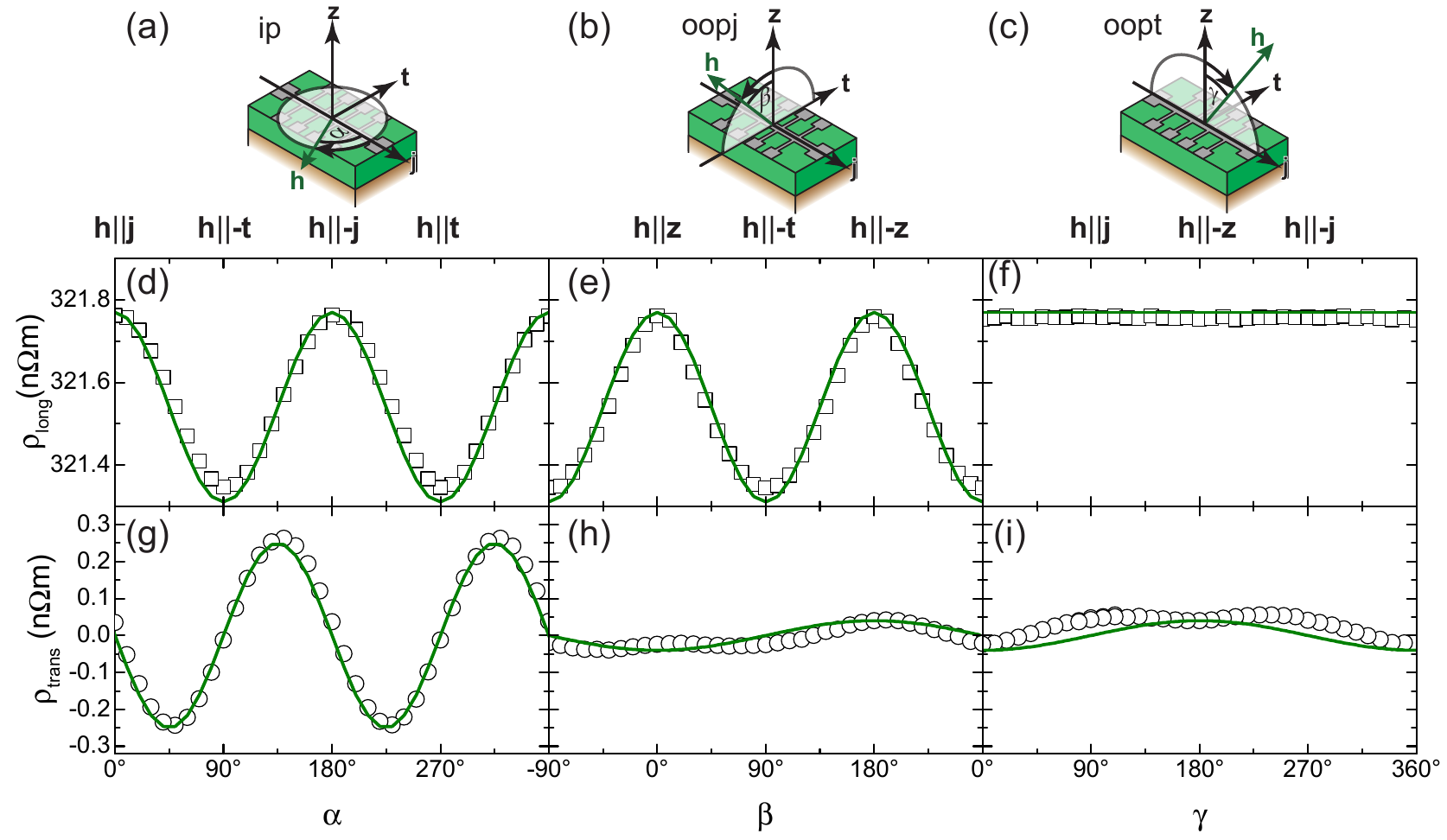}\\
 \caption[Angle-dependent magnetoresistance measurements]{Illustration of the ADMR rotation planes used in our experiments: (a) in-plane, (b) out-of-plane perpendicular to $\mathbf{j}$, and (c) out-of-plane perpendicular to $\mathbf{t}$. Measured ADMR data for $\rho_\mathrm{long}$ ((d)-(f)) and $\rho_\mathrm{trans}$ ((g)-(i)) in the three orthogonal rotation planes for a YIG($45\,\mathrm{nm}$)/Pt($3.5\,\mathrm{nm}$) bilayer at $T=300\,\mathrm{K}$ and $\mu_0H=1\,\mathrm{T}$. There is excellent agreement between experimental data (open symbols) and a simulation using Eqs.(\ref{eq:SMR:long},\ref{eq:SMR:trans}) (olive line), which confirms the SMR as the sole cause for the observed magnetoresistance effect. Adapted with permission from Ref.~\cite{althammer_quantitative_2013}.}
  \label{figure:SMRADMR}
\end{figure*}

The first experimental observation of the SMR has been reported by us for a YIG/Pt heterostructure in the Supplements of Ref.~\cite{Weiler2012}, but it required some more time to find a suitable way to discern the SMR from other spurious effects like an anisotropic magnetoresistance generated by a proximity magnetization induced by the MOI into the NM~\cite{huang_transport_2012}. A very helpful approach in this regard are angle-dependent magnetoresistance (ADMR) experiments~\cite{limmer_angle-dependent_2006}. In these experiments the magnetic field orientation $\mathbf{h}$ is rotated with fixed magnetic field magnitude $\mu_0H$ with respect to the sample and the longitudinal and transverse resistivities are measured simultaneously. For large enough magnetic fields compared to magnetic anisotropy contributions in a ferromagnetic MOI the magnetization direction $\mathbf{m}$ (corresponds to the magnetic order parameter $\mathbf{n}$) is always aligned to the external magnetic field orientation, such that one can safely assume $\mathbf{h}\parallel \mathbf{m}$. In our experiments we now use three orthogonal rotation planes for $\mathbf{h}$ in the $\mathbf{j}$,$\mathbf{t}$,$\mathbf{z}$ coordinate system as illustrated in Fig.~\ref{figure:SMRADMR}(a),(b), and (c), the spin polarization $\mathbf{s}$ is then oriented along $\mathbf{t}$. For the in-plane (ip) rotation, $\mathbf{h}$ is rotated in the $\mathbf{j}$-$\mathbf{t}$-plane (angle $\alpha$, Fig.~\ref{figure:SMRADMR}(a)), for the out-of-plane perpendicular to $\mathbf{j}$ (oopj) rotation, the magnetic field is rotated in the $\mathbf{t}$-$\mathbf{z}$-plane (angle $\beta$, Fig.~\ref{figure:SMRADMR}(b)), and for the out-of-plane perpendicular to $\mathbf{t}$ (oopt) rotation, the magnetic field rotation plane is the $\mathbf{z}$-$\mathbf{j}$-plane (angle $\gamma$, Fig.~\ref{figure:SMRADMR}(c)). Evaluating the measured angle-dependence allows then to analyse the relevant projections of $\mathbf{m}$ contributing to the resistivities. For the SMR this allowed to clearly discern this effect from an anisotropic magnetoresistance effect, as detailed in Refs.~\cite{Nakayama2013,althammer_quantitative_2013}.

We here show in Fig.~\ref{figure:SMRADMR} the ADMR results obtained for a PLD grown YIG layer deposited onto a (111)-oriented GGG substrate interfaced in-situ with a $3.5\,\mathrm{nm}$ thick Pt film deposited by electron beam evaporation at $T=300\,\mathrm{K}$ and $\mu_0H=1\,\mathrm{T}$ taken from data from Ref.~\cite{althammer_quantitative_2013}. We start the discussion with the results obtained for the longitudinal resistivity $\rho_\mathrm{long}$ as shown in the panels (d)-(f) in Fig.~\ref{figure:SMRADMR}. For the ip rotation in Fig.~\ref{figure:SMRADMR}(d), $\rho_\mathrm{long}$ exhibits a $\cos^2(\alpha)$ dependence with maxima for $\mathbf{h}\parallel\mathbf{j}$ ($\alpha=0^\circ$) and $\mathbf{h}\parallel-\mathbf{j}$ ($\alpha=180^\circ$), and minima for $\mathbf{h}\parallel-\mathbf{t}$ ($\alpha=90^\circ$) and $\mathbf{h}\parallel\mathbf{t}$ ($\alpha=270^\circ$). This agrees with our prediction for the spin current flow across the interface. If $\mathbf{h}$ is collinear with the current direction $\mathbf{j}$, $\mathbf{m}$ is perpendicular to $\mathbf{s}$ and a large interfacial spin current flows across the interface due to spin-transfer torque, which then leads to an increase in the NM resistance. The opposite is true for $\mathbf{h}$ collinear to $\mathbf{t}$, now $\mathbf{m}$ is parallel to $\mathbf{s}$ and thus only a very small interfacial spin current flows and thus a lower resistance state is achieved. In the oopj rotation plane (Fig.~\ref{figure:SMRADMR}(e)), $\rho_\mathrm{long}(\beta)$ has a $\cos^2(\beta)$ dependence, with maxima located at $\mathbf{h}\parallel\mathbf{z}$ ($\beta=0^\circ$) and $\mathbf{h}\parallel-\mathbf{z}$ ($\beta=180^\circ$) and minima at $\mathbf{h}\parallel-\mathbf{t}$ ($\beta=90^\circ$) and $\mathbf{h}\parallel\mathbf{t}$ ($\beta=270^\circ$). This is again in agreement with our qualitative model of the SMR: maximum in resistance for $\mathbf{m}\perp\mathbf{s}$ and a minimum in resistance for $\mathbf{m}\parallel\mathbf{s}$. Finally, for the oopt rotation, we do not observe any angular dependence, which is in line with our theoretical expectations of the SMR. In this rotation plane $\mathbf{m}$ is always perpendicular to $\mathbf{s}$, independent of $\gamma$, such that one expects to always obtain the high resistance state, which is clearly observed in the experiment.

For the transverse resistivity $\rho_\mathrm{trans}$ we observe in the ip rotation a $\cos(\alpha)\sin(\alpha)$ dependence with maxima located at $\alpha=135^\circ$, $\alpha=315^\circ$ and minima at $\alpha=45^\circ$, $\alpha=225^\circ$. For the oopj and oopt rotation plane we see only a very weak angle dependence consisting of a $\cos(\beta/\gamma)$ and a $\cos^3(\beta/\gamma)$ contribution, as reported in Ref.~\cite{Meyer2015}. The ordinary Hall effect of very thin Pt layers is rather small such that this $\cos(\beta/\gamma)$-dependence does not dominantly contribute in our measurements.

The fact that our SMR model completely and quantitatively describes the measured data is illustrated by the excellent agreement between the experiment (symbols in the graphs) and an simulation of the data (olive line in the graphs) using Eqs.(\ref{eq:SMR:long},\ref{eq:SMR:trans}) and the very same $\rho_i$ parameter for all rotation planes. This further confirms that the SMR is the sole cause for the magnetoresistance of the YIG/Pt sample. From this fit we extract a SMR amplitude $\rho_1/\rho_0\approx1.4\times10^{-3}$, which is one of the largest values found for the SMR in MOI/NM systems~\cite{althammer_quantitative_2013}. We attribute the large SMR in these samples to the in-situ preparation of the MOI/NM interface, which leads to a large spin mixing conductance, and the large spin Hall angle of Pt used in or experiments~\cite{weiler_experimental_2013}.

Over the course of the last years several groups have reproduced our results on YIG/Pt heterostructures~\cite{Hahn2013SMR,Vlietstra2013,wang_comparative_2017}. The SMR was also observed in other MOI materials interfaced with a NM, confirming the universal nature of this effect~\cite{althammer_quantitative_2013,Han2014,Isasa2014,Aqeel2015,Wu2015,hui_spin_2016,Avci2016,hoogeboom_negative_2017,ji_spin_2017,hou_tunable_2017,manchon_spin_2017} and in different YIG/NM structures~\cite{Vlietstra2013,shang_pure_2016}. Furthermore, it has been used to experimentally investigate the spin Hall and spin transport parameters using magnetotransport experiments~\cite{althammer_quantitative_2013,weiler_experimental_2013,Meyer2014}. Magnetoimpedance and noise spectroscopy experiments suggest that the SMR is also relevant at charge current frequencies of several MHz~\cite{lotze_spin_2014,kamra_spin_2014}. In addition, recent experiments have shown that the SMR is quite sensitive to the magnetic moment orientation at the interface, making it sensitive to magnetic frustration~\cite{Vlez2016}.

While the initial theory of the SMR does not predict dependence on the magnetic field magnitude (neglecting any contributions from magnetic anisotropy), new experiments observe a field dependence, which can be tuned by the interface preparation methods. To explain this effect Velez~\textit{et al.}~have put forward the Hanle magnetoresistance, which accounts for additional spin dephasing of the spin accumulation at the NM interface~\cite{velez_hanle_2016,Vlez2016}. In principle, this effect should also persist in just a thin layer of the NM, but since in thin films spin-flip scattering processes at the interface is a dominant contribution, this could also lead to an additional field dependence.

It is worth mentioning that the SMR also now seems to account for the unusual ADMR dependence in thin film heterostructures of metallic ferromagnets and NMs~\cite{cho_large_2015,kim_spin_2016}, where the symmetry of the SMR was first observed in the experiments by Kobs~\textit{et al.}~\cite{kobs_anisotropic_2011}. In these metallic bilayer systems even larger values of the SMR have been reported reaching values of a few percent~\cite{cho_large_2015,kim_spin_2016}. In these metallic multilayers quite often also oxide interfaces are introduced, which might also play an important role for the magnitude of the SMR effect or additional contributions from this oxide interface, like the spin galvanic effect, may then help to increase the magnetoresistance effect. A very intriguing effect observed in these metallic bilayer systems is the unidirectional SMR, where the resistivity is different for the magnetization being parallel or antiparallel aligned to the charge current flow direction~\cite{avci_unidirectional_2015}. So far it seems that the unidirectional SMR manifests itself only in metallic ferromagnet/NM heterostructures and not in MOI/NM systems.

From the theoretical side, it is proposed already in the initial work by Chen~\textit{et al.}~that the SMR can be further enhanced by interfacing the NM with MOIs on both interfaces, which should also give rise to some spin-valve effects if the two MOIs have different coercive fields~\cite{chen_theory_2013}. Another interesting aspect is the formulation of the SMR in terms of a quantum tunneling phenomena~\cite{Chen2016:tunneling}, where a characteristic thickness dependence for the SMR has been predicted for the case of MOI/NM and an oscillating behaviour of the SMR amplitude for all metallic systems.
\subsection{SMR in compensated rare-earth garnets}

\begin{figure}[h]
 \includegraphics[width=85mm]{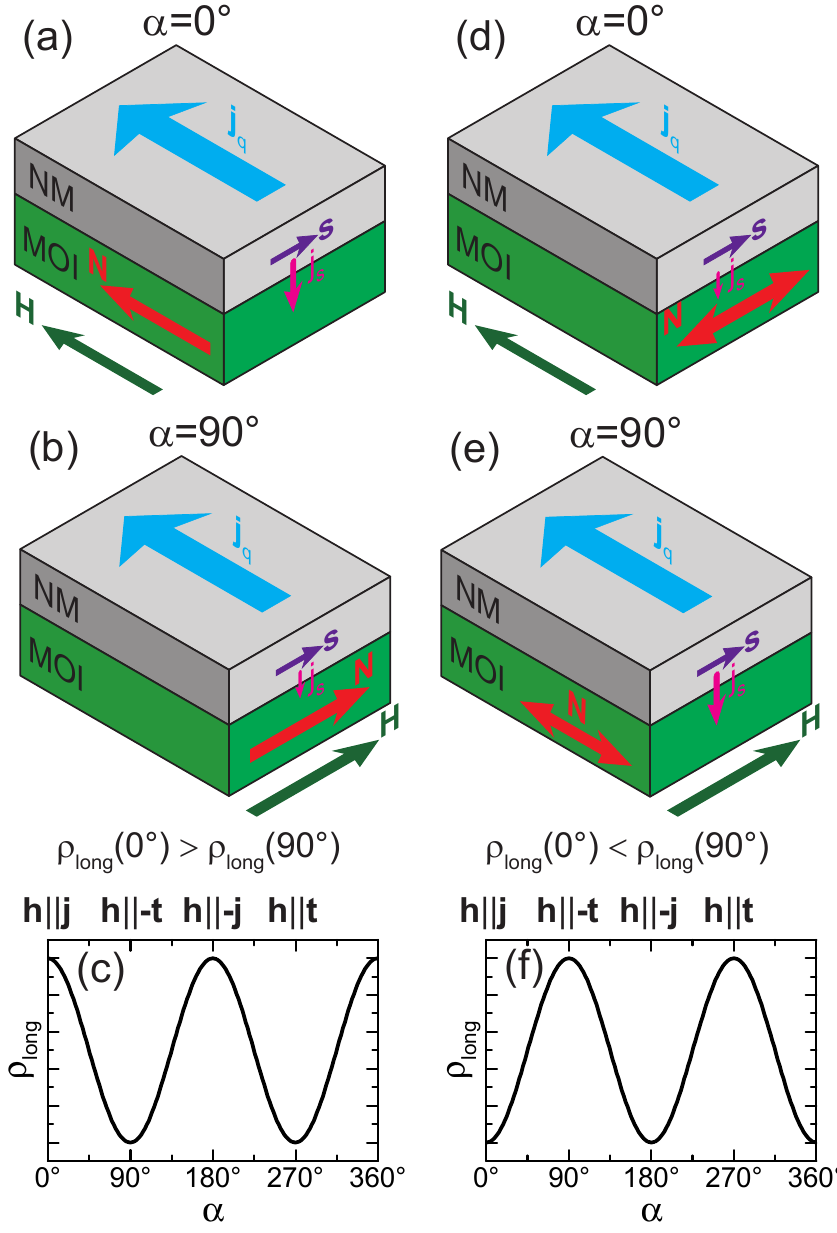}\\
 \caption[Illustration regular vs staggered]{Illustration of the difference in boundary conditions for an ip ADMR rotation for $\mathbf{N}\parallel \mathbf{H}$ and for $\mathbf{N}\perp \mathbf{H}$. For $\mathbf{N}\parallel \mathbf{H}$ and $\alpha=0^\circ$ a large $\mathbf{j}_\mathrm{s}$ flows across the interface (a), while a small $\mathbf{j}_\mathrm{s}$ governed by $g$ is flowing for $\alpha=90^\circ$ (b). (c) The ip ADMR response of $\rho_\mathrm{long}$ has maxima for $\alpha=0^\circ$ and $\alpha=180^\circ$, just like in YIG/Pt bilayers. For $\mathbf{N}\perp \mathbf{H}$ and $\alpha=0^\circ$ a small $\mathbf{j}_\mathrm{s}$ flows across the interface (d), while a large $\mathbf{j}_\mathrm{s}$ governed by $\tilde{g}^{\uparrow\downarrow}$ is flowing for $\alpha=90^\circ$ (e). (f) The ip ADMR response of $\rho_\mathrm{long}$ then has maxima for $\alpha=90^\circ$ and $\alpha=270^\circ$,i.e.~a phase shift by $90^\circ$ as compared to a YIG/Pt bilayer.}
  \label{figure:SMRstaggered}
\end{figure}
In the first experiments on the SMR, YIG was mainly used as the MOI, which has a ferrimagnetic ordering as mentioned in the materials section (Sec.~\ref{materials}). In this regard, the first theoretical model also only included the magnetization orientation as the relevant orientation in the MOI for the SMR. A main open question thus was wether each magnetic moment/sublattice in the MOI at the interface contributes individually to the SMR or if it is enough to just use the net magnetization direction to describe the SMR. We used ADMR experiments with compensated REIGs as the MOI and Pt as the NM to find an answer to this relevant question. As detailed in the publication by Ganzhorn~\textit{et al.}~\cite{Ganzhorn2016}, we used (In,Y)GdIG in our experiments since it allows to get experimental access to the spin-canting phase (See Section~\ref{materials}) in a temperature range well below room temperature. As already discussed the relevant order parameter $\mathbf{N}$ in these REIGs is not the net magnetization, but the N\'{e}el vector. In the ferrimagnetic phase the net magnetic moment direction and the N\'{e}el vector are collinear to each other and to the external magnetic field (see Fig.~\ref{figure:SMRstaggered}(a) and (b)). Thus in the ferrimagnetic phase it is not possible to figure out, which of the two vectors is the relevant contribution to the SMR, since the SMR is $180^\circ$ symmetric (see Fig.~\ref{figure:SMRstaggered}(c)). However, upon entering the spin-canting phase the net magnetic moment will still be aligned parallel to the external magnetic field, while the N\'{e}el vector is aligned to the external magnetic field with a finite angle between $0^\circ$ and $180^\circ$ (see Fig.~\ref{figure:SMRstaggered}(d) and (e) for perpendicular alignment to the external field). This then allows to experimentally verify, which of these two vectors is relevant for the SMR effect (compare Fig.~\ref{figure:SMRstaggered}(c) and (f)). It is important to understand that if only one single magnetic domain would persist in the spin-canting phase, ADMR measurements at different magnetic fields would yield a continuous phase shift of the SMR signal, i.e. the position of maxima and minima in $\rho_\mathrm{long}$ would continuously shift to a different angle position. From our discussion in Sec.~\ref{materials}, we saw that spin-canting leads to the formation of magnetic domains. If one includes multiple domains into the corresponding SMR response, one finds that the phase shift of the SMR vanishes as it is counteracted by the different magnetic domains with mirrored sublattice orientations, but still the amplitude of the SMR should undergo a sign change as in Fig.~\ref{figure:SMRstaggered}(f). The maximum negative amplitude is achieved when the N\'{e}el vector is aligned perpendicular to the external magnetic field. In the ADMR experiment~\cite{Ganzhorn2016} we found a sign inversion of the SMR amplitude in the spin-canting regime as detailed in the following.

\begin{figure}[h]
 \includegraphics[width=85mm]{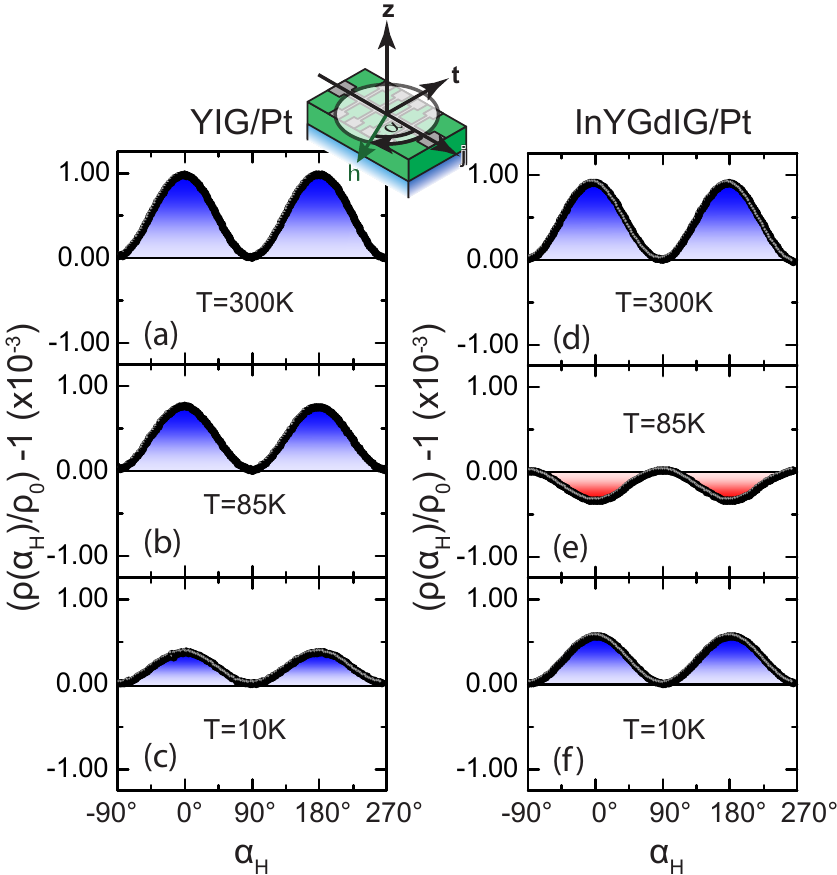}\\
 \caption[Comparison SMR temperature dependence in iron garnets]{ADMR measurements in the ip-configuration for a YIG($40\,\mathrm{nm}$)/Pt($4\,\mathrm{nm}$) (a)-(c) and a (In,Y)GdIG($61.5\,\mathrm{nm}$)/Pt($3.6\,\mathrm{nm}$) (d)-(f) bilayer at $\mu_0H=7\,\mathrm{T}$ for $T=300\,\mathrm{K}$ (a) and (d), for $T=85\,\mathrm{K}$ (b) and (e), and for $T=10\,\mathrm{K}$ (c) and (f). While the SMR in the YIG/Pt bilayer monotonously decreases with decreasing temperature, the SMR in the (In,Y)GdIG/Pt sample has a sign change in amplitude at $T=85\,\mathrm{K}$. As this temperature is also the compensation temperature of the (In,Y)GdIG, this confirms the expected sign change in the spin-canting phase. Reprinted figure with permission from~\cite{Ganzhorn2016}. Copyright 2016 by the American Physical Society.}
  \label{figure:SMRYIGGIG}
\end{figure}

As a first step to confirm this change in the ADMR response, we compared the temperature dependent results obtained for a YIG($40\,\mathrm{nm}$)/Pt($4\,\mathrm{nm}$) bilayer to a (In,Y)GdIG($61.5\,\mathrm{nm}$)/Pt($3.6\,\mathrm{nm}$) sample both grown on (111)-oriented YAG substrates as shown in Fig.~\ref{figure:SMRYIGGIG}. The compensation temperature for the (In,Y)GdIG sample was determined from magnetometry measurements to be at $T_\mathrm{comp}=85\,\mathrm{K}$. At $T=300\,\mathrm{K}$ the ADMR response of the YIG (Fig.~\ref{figure:SMRYIGGIG}(a)) and the (In,Y)GdIG (Fig.~\ref{figure:SMRYIGGIG}(d)) is qualitatively the same with the maxima and minima of $\rho_\mathrm{long}$ at the expected positions. For $T=T_\mathrm{comp}$, the SMR in the YIG sample (Fig.~\ref{figure:SMRYIGGIG}(b)) is opposite to the one obtained for (In,Y)GdIG (Fig.~\ref{figure:SMRYIGGIG}(e)). While for the YIG sample only the amplitude of the SMR has changed, in (In,Y)GdIG we now find a negative SMR amplitude with maxima in $\rho_\mathrm{long}$ at $\alpha=90^\circ$ and $\alpha=270^\circ$, and minima in $\rho_\mathrm{long}$ at $\alpha=0^\circ$ and $\alpha=180^\circ$. Thus the SMR response is shifted by $90^\circ$ with respect to the YIG/Pt sample. For even lower temperatures $T=10\,\mathrm{K}$, the ADMR data is again qualitatively identical for YIG/Pt and (In,Y)GdIG/Pt. This is in agreement with the expected phase diagram of the spin-canting phase in (In,Y)GdIG (compare Fig.~\ref{figure:SpinCanting}(b)) as the applied external magnetic field $\mu_0 H=7\,\mathrm{T}$ is too small to enter the spin-canting phase at low temperatures. From these experiments we see that the SMR changes its sign for temperatures close to the compensation temperature of (In,Y)GdIG.

\begin{figure}[h]
 \includegraphics[width=85mm]{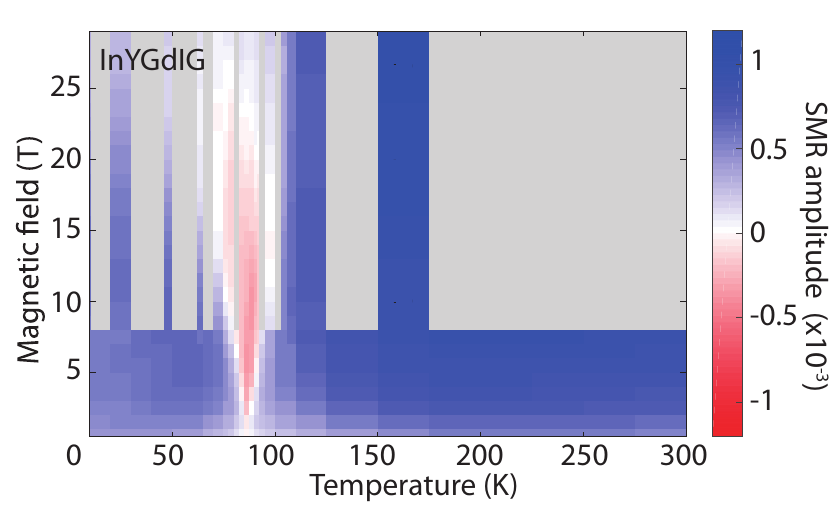}\\
 \caption[Colormap SMR compensated garnets]{Color map of the temperature and field dependence of the SMR amplitude determined from ADMR experiments on the (In,Y)GdIG/Pt sample. Around the compensation temperature the sign of the SMR amplitude is inverted, the color map seems to reasonably well agree with mean field calculations in Fig.~\ref{figure:SpinCanting}(b). Reprinted figure with permission from \cite{Ganzhorn2016}. Copyright 2016 by the American Physical Society.}
  \label{figure:SMRColorMap}
\end{figure}

For a further investigation of temperature vs. magnetic field SMR phase diagram, we conducted further experiments in the high-field magnet laboratory in Grenoble for $\mu_0H\leq29\,\mathrm{T}$. The results are shown as a color map in Fig.~\ref{figure:SMRColorMap}. In this plot we find a cone-shaped area around the compensation temperature ($T=85\,\mathrm{K}$), for which the SMR amplitude has a negative sign. The boundaries of this negative SMR phase (SMR amplitude equal to zero), seem to be in a reasonably well agreement with the phase diagram of the spin canting phase numerically simulated from a mean-field model (Fig.~\ref{figure:SpinCanting}(b)). In Ref.\cite{Ganzhorn2016}, we also conducted additional XMCD measurements on the very same sample to confirm the existence of the spin-canting phase and the phase boundaries from an independent experiment. Thus, we have shown that the SMR changes its sign within the spin-canting phase of a compensated REIG. The sign change may be explained by multiple magnetic domains with mirrored orientations of the N\'{e}el vector, as discussed before. Another possibility is that each magnetic moment contributes individually to the SMR and this leads to the SMR sign change in the spin-canted phase. As the imaging of the magnetic-domains with different sublattice orientations in the spin-canted phase is very challenging, as the net magnetization in each domain is identical, it will be very tough to figure out, which of the two interpretations is the correct one.

We focused here on the results obtained for only the ip rotation. However, Dong~\textit{et al.}~reported on the evolution of the ADMR signal for oopj and oopt rotation planes~\cite{dong_spin_2018}. In these configurations, magnetic anisotropy contributions play a crucial role for the selection of magnetic domains, which leads to a more complex angle and field dependence and makes a quantitative analysis very challenging.

\subsection{SMR in the spin-flop phase of an antiferromagnetic MOI}

Initially, the SMR was only investigated in MOIs with ferro/ferrimagnetic order. In recent years, there were also investigations of the SMR in antiferromagnetic MOIs. Similar to the SMR in compensated garnets in the spin canting phase, one finds an inversion of the regular SMR amplitude in ADMR experiments for external magnetic fields larger than the spin-flop field. This effect is explained by the fact that in the spin-flop phase the magnetic order parameter $\mathbf{N}$ is not collinear to the external magnetic field direction, but encloses a finite angle with it. Due to the degeneracy, the formation of magnetic domains is preferred, especial as now the two sublattices are indiscernible from each other. In the ADMR measurements for the SMR detection we use an external magnetic field to control the orientation of $\mathbf{N}$, thus for antiferromagnetic materials it is necessary to work in the field range close to and above the spin-flop transition, which limits the choice of materials. One possible candidate to investigate this effect is NiO. In the following we discuss the results of our group obtained for NiO films grown by PLD onto (0001)-oriented sapphire substrates with in-situ Pt layers on top of the NiO~\cite{fischer_spin_2018}.

\begin{figure}[h]
 \includegraphics[width=85mm]{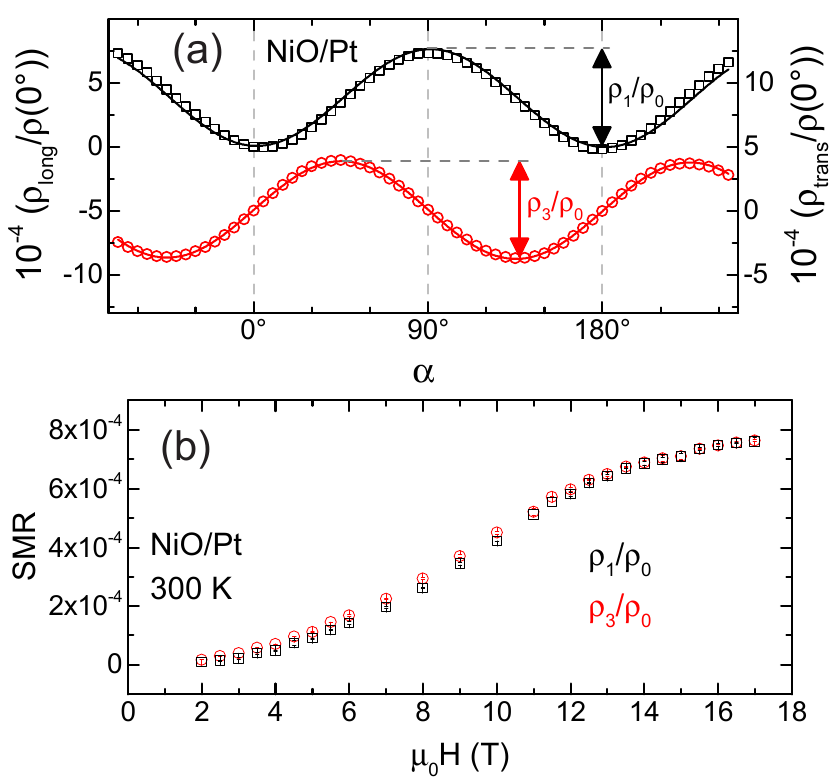}\\
 \caption[SMR in NiO]{SMR results obtained for NiO($120\,\mathrm{nm}$)/Pt($3.5\,\mathrm{nm}$) heterostructure. (a) ADMR results obtained for $\mu_0 H=17\,\mathrm{T}$ in the ip rotation plane for $\rho_\mathrm{long}$ (left axis, black squares) and $\rho_\mathrm{trans}$ (right axis, red squares). For $\alpha=0^\circ$ $\rho_\mathrm{long}$ exhibits a minimum, while for $\alpha=90^\circ$ $\rho_\mathrm{long}$ is maximal. Just opposite to the case for YIG/Pt. (b) Field dependence of the SMR amplitude extracted from the ADMR measurements for $\rho_\mathrm{long}$ (black squares) and $\rho_\mathrm{trans}$ (red squares). For both resistivities we obtain the same field dependence. Adopted figure with permission from~\cite{fischer_spin_2018}. Copyright 2018 by the American Physical Society.}
  \label{figure:SMRNiO}
\end{figure}

In Fig.~\ref{figure:SMRNiO}(a) we show the ADMR response of $\rho_\mathrm{long}$ and $\rho_\mathrm{trans}$ obtained for a NiO($120\,\mathrm{nm}$)/Pt($3.5\,\mathrm{nm}$) sample at $T=300\,\mathrm{K}$ and $\mu_0H=17\,\mathrm{T}$. Similar to the results obtained in the compensated REIGs, we find an inversion of the SMR amplitude as compared to YIG/Pt, with minima in $\rho_\mathrm{long}$ at $\alpha=0^\circ$ and $\alpha=180^\circ$, and maxima in $\rho_\mathrm{long}$ at $\alpha=90^\circ$ and $\alpha=270^\circ$. This finding can be explained by the fact that the magnetic order parameter in the spin-flop phase of the antiferromagnet encloses an angle of $90^\circ$ with applied external magnetic field and thus leads to a shift of the AMDR response by $90^\circ$ (see Fig.~\ref{figure:SMRstaggered}(f)), as the resistivity is recorded with respect to the applied magnetic field orientation.

In a next step we analyzed the magnetic field dependence of the SMR in the NiO/Pt sample by extracting the SMR amplitude from ADMR measurements conducted at various magnetic fields. The results of this procedure are shown in Fig.~\ref{figure:SMRNiO}(b). With increasing magnetic field, the SMR amplitude gradually increases and seems to saturate for very high fields. As detailed in the publication by Fischer~\textit{et al.}~we model the magnetic field dependence by taking into account the multidomain state of the NiO caused by three equivalent magnetocrystalline anisotropy axes present in the plane of the NiO and the nucleation of these domains by magnetoelasitc coupling. Using this model not only allows to quantitatively explain the SMR response in the NiO layer, but also to determine the magnetoelastic coupling strength for this antiferromagnetic compound.

It is worth mentioning that especially in the field of antiferromagnetic MOIs other groups have already achieved similar results or reported them at the same time then our group~\cite{Han2014,hoogeboom_negative_2017,ji_spin_2017,hou_tunable_2017,manchon_spin_2017}. Here, the work by Han\textit{et al.}~was the first that reported on the investigation of the SMR in the antiferromagnetic MOI SrMnO$_3$, where also a sign change of the SMR was observed compared to MOIs with ferro/ferrimagnetic order~\cite{Han2014}. However, Han~\textit{et al.}~could not give a reasonable explanation for the occurrence of this sign change. From our results, the work of Hoogeboom~\textit{et al.}~and Ji~\textit{et al.}~it is now clear that also in the antiferromagnetic MOIs magnetic domains are relevant in the spin-flop phase and have to be accounted for by the model that describes the magnetic field dependence of the SMR amplitude~\cite{hoogeboom_negative_2017,ji_spin_2017,fischer_spin_2018}.

A further step into underlining the importance of the SMR for investigating complex magnetic structures was the work by Aqeel~\textit{et al.}, where the SMR in a MOI with chiral magnetic order was studied~\cite{Aqeel2015,aqeel_electrical_2016}. However, a detailed analysis requires a sophisticated model, which takes into account the formation of magnetic domains and interaction with magnetic anisotropy effects in these chiral systems.

Taken all together over the course of the last few years the SMR has established itself as a very versatile technique to even study complex magnetic ordering phenomena with interface sensitivity. Moreover, it is now clear that magnetic domains are crucial in understanding the SMR response in multidomain magnetic systems, and have to be accounted for in any quantitative modeling. Combined with other methods like the current heating induced longitudinal spin Seebeck effect, this provides means to study magnetic order in MOIs by only using magnetotransport experiments.

\section{All-electrical magnon transport experiments}
\label{magnon_mediated_magnetoresistance}
Last but not least, we now focus onto the all-electrical detection of magnon transport by the SHE and ISHE. Magnon transport was previously mostly investigated by spatially resolved inelastic light scattering and/or inductive magnon generation and detection mechanisms~\cite{Demokritov2001,Demokritov2008,Sebastian2015}. The main advantage of the all-electrical SHE based approach is that it only requires magnetotransport setups for investigating spin excitation transport in the MOI and does not require sophisticated microwave and optical equipment.

\begin{figure}[h]
 \includegraphics[width=85mm]{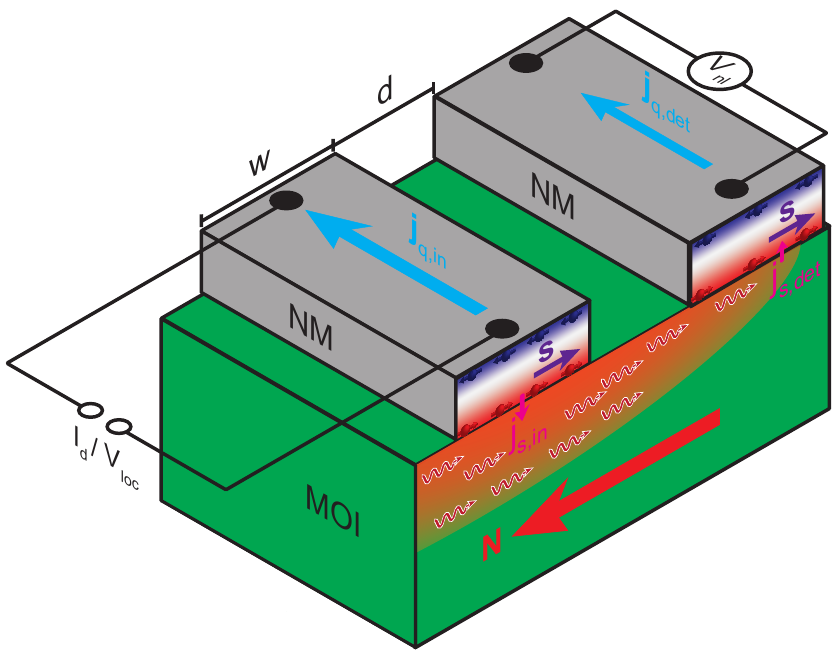}\\
 \caption[Illustration of the principle of magnonmediated Magnetoresistance]{Illustration of the magnon mediated magnetoresistance. Two NM strips are in contact with a magnetically ordered insulator. A charge current $\mathbf{j}_{q,\mathrm{in}}$ flowing in one strip generates an electron spin accumulation due to the spin Hall effect in the NM. If the magnetic order parameter $\mathbf{N}$ and the spin orientation $\mathbf{s}$ of the spin accumulation are collinear, due to inelastic electron-magnon scattering at the interface the electron spin accumulation is also causing a magnon accumulation underneath the NM injector strip. The magnons diffusive away from the injector strip. At the second NM detector strip, the magnons transfer angular momentum into the NM strip by injecting a pure spin current $\mathbf{j}_{s,\mathrm{det}}$ along the surface normal. This pure spin current is then transformed into a charge current $\mathbf{j}_{q,\mathrm{det}}$ via the inverse spin Hall effect. As open electric charge circuit boundary conditions are imposed, an electric field is generated, which counteracts the generated charge current and allows to detect the transport of magnons as a non-local voltage $V_\mathrm{nl}$ in the second strip.}
  \label{figure:MMR}
\end{figure}

As shown in Fig.~\ref{figure:MMR} a minimal device to investigate this effect consists of two NM strips (injector and detector) on top of the MOI. In the injector strip a charge current density $\mathbf{j}_\mathrm{q,in}$ is driven through it by applying a charge current bias $I_\mathrm{d}$ to it. As already discussed the pure spin current generated by the SHE leads to spin accumulation with spin polarization $\mathbf{s}$ at the interface. As $\mathbf{N}$ and $\mathbf{s}$ are collinear to each other, the spin current across the interface is governed by the spin convertance $g$~\cite{Zhang2012,Zhang2012_PRB,Cheng2017}. The inelastic electron-magnon scattering at the interface leads to the generation or absorption of magnons in the MOI. In the end, this also leads to a magnon accumulation (depletion) in the MOI at the NM/MOI interface underneath the injector strip. This magnon accumulation then diffuses on the length scale of the magnon diffusion length $\lambda_\mathrm{mag}$ into the MOI. The diffusing magnons can then be detected in the second, separate NM detector strip. The magnons arriving at the second NM strip inject a pure spin current $\mathbf{j}_\mathrm{s,det}$ into the NM. By means of the inverse spin Hall effect this pure spin current is then transformed into a charge current $\mathbf{j}_\mathrm{q,det}$ and can then be electrically detected as an open circuit voltage $V_\mathrm{nl}$. For the first experimental observation, yttrium iron garnet (YIG) grown via liquid phase epitaxy (LPE) has been used as the MOI as it features a rather long magnon diffusion length. The long diffusion length in this material allows to use lateral device schemes in the experiment as illustrated in Fig.~\ref{figure:MMR}. For shorter $\lambda_\mathrm{mag}$ one can then use vertical transport designs, where the MOI is sandwiched between two NM layers~\cite{Li2016_Riverside}.

For the modeling of the interface transport we now need to account for $\mu_\mathrm{s}(0)=\mu_\mathrm{mag}=(T_\mathrm{M}-T_\mathrm{N})\neq0$, while using $\dot{\mathbf{n}}=0$ in Eq.(\ref{equ:InterfacialSpinCurrent}), model the magnon diffusion in the MOI and account for the heat current transport by magnetic excitations in the MOI~\cite{Cornelissen2016}. Due to the diffusion process of magnons in three dimensions it is not possible to give analytical expressions for the detected $V_\mathrm{nl}$. Even reducing the diffusion to one dimension gives a rather lengthy expression for $V_\mathrm{nl}$, which we omit here and refer the interested reader to Refs.~\cite{Zhang2012,Zhang2012_PRB,Cornelissen2016,Cheng2017} for more details onto these calculations and numerical solutions to the diffusion problem. It is important to understand that $V_\mathrm{nl}$ originates from two contributions in the injector, a first contribution from the SHE induced spin accumulation by the charge current (odd with respect to charge current polarity), and a second contribution originating from the Joule heating of the injector by the charge current (even with respect to charge current polarity).

Due to the fact that the charge current $\mathbf{j}_\mathrm{q,in}$ through the injector strip not only leads to a spin accumulation at the interface, but also to a temperature gradient and heat transport due to Joule heating, one needs to find means in the experiment to single out these two contributions to $V_\mathrm{nl}$. In our experiments we use a current reversal method to separate these two effects. A DC charge current $I_\mathrm{d}$ is applied to the injector and we record the local voltage $V_\mathrm{loc}$ and $V_\mathrm{nl}$ for both current polarities $I_\mathrm{d}^{+}$ and $I_\mathrm{d}^{-}$. Using Eqs.(\ref{eq:SSE_CurrentRes},\ref{eq:SSE_CurrentTherm}), we determine from these measurements the resistive response $V_\mathrm{loc,res}$ and $V_\mathrm{nl,res}$ and the thermal response $V_\mathrm{loc,therm}$ and $V_\mathrm{nl,therm}$~\cite{Goennenwein2015}. Contributions from the spin accumulation are visible in the resistive response and contributions from the Joule heating are in the thermal response. One should note that one can also use a sinusoidal modulated AC $I_\mathrm{d}$ with a fixed frequency and then use Lock-In detection on the first and second harmonic signal of $V_\mathrm{nl}$ to obtain the resistive and thermal response of $V_\mathrm{nl}$, as for example done by Cornelissen~\textit{et al.}~\cite{Cornelissen2015}.

\begin{figure*}[t]
 \includegraphics[width=170mm]{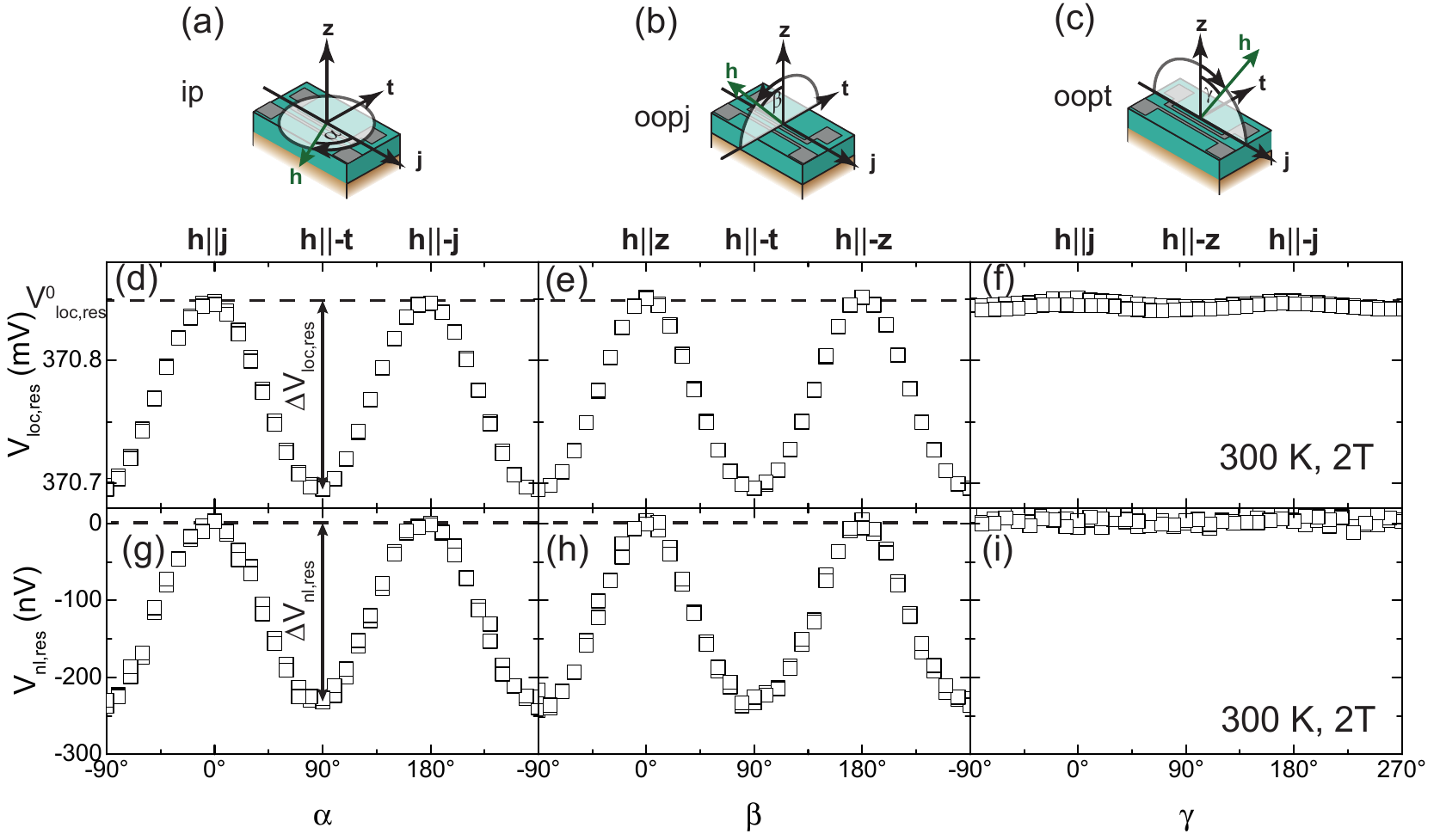}\\
 \caption[Resistive local and non-local voltages]{Extracted angle-dependent data of the local $V_\mathrm{loc,res}$ and non-local $V_\mathrm{nl,res}$ resistive voltages obtained for Pt strips with $d=200\,\mathrm{nm}$, $w=500\,\mathrm{nm}$, and a thickness of $10\,\mathrm{nm}$ deposited onto a $3\,\mathrm{\mu m}$ thick LPE YIG layer. (a)-(c) Illustration of the three orthogonal rotation planes of the external magnetic field: in-plane (a), out-of-plane perpendicular to $\mathbf{j}$ (b) and out-of-plane perpendicular to $\mathbf{t}$ (c). (d)-(f) Measured angular dependence of $V_\mathrm{loc,res}$ at $T=300\,\mathrm{K}$ and $\mu_0H=2\,\mathrm{T}$ for ip(d), oopj(e), and oopt(f). $V_\mathrm{loc,res}$ exhibits the angle-dependence expected for SMR. (g)-(i) Simultaneously measured angular dependence of $V_\mathrm{nl,res}$ for ip(g), oopj(h), and oopt(i). $V_\mathrm{nl,res}$ goes to 0 for $\mathbf{h}\parallel\mathbf{j}$ and $\mathbf{h}\parallel\mathbf{z}$, while a maximum negative voltage is obtained for $\mathbf{h}\parallel\mathbf{t}$. Figures and data adapted with permission from Ref.~\cite{Goennenwein2015}.}
  \label{figure:MMRMeasure}
\end{figure*}

We first focus on the local and non-local resistive response $V_\mathrm{loc,res}$ and $V_\mathrm{nl,res}$ in ADMR measurements. For our first set of experiments we used $10\,\mathrm{nm}$ thick Pt strips with a width $w=500\,\mathrm{nm}$, a length of $100\,\mathrm{\mu m}$ and an edge-to-edge separation of $d=200\,\mathrm{nm}$ deposited onto a $3\,\mathrm{\mu m}$ thick LPE YIG layer at $T=300\,\mathrm{K}$. For these measurements we rotated the external magnetic field with a fixed magnitude $\mu_0H=2\,\mathrm{T}$ in the three orthogonal rotation planes ip, oopj and oopt as illustrated in Fig.~\ref{figure:MMRMeasure}(a)-(c).

For $V_\mathrm{loc,res}$ we observe the ADMR fingerprint of the SMR (see Fig.~\ref{figure:MMRMeasure}(d)-(f)). For $\mathbf{h}\parallel\mathbf{j}$ and $\mathbf{h}\parallel\mathbf{z}$ we find a maximum in $V_\mathrm{loc,res}$ and for $\mathbf{h}\parallel\mathbf{t}$ a minimum in $V_\mathrm{loc,res}$. As explained in the previous section this change in the injector resistance can be explained by the change in the pure spin current across the interface due to the orientation of the magnetization with respect to $\mathbf{s}$. For field rotations in the ip and oopj rotation planes, we observe an angle-dependence of $V_\mathrm{loc,res}$, while $V_\mathrm{loc,res}$ is constant for a field rotation in the oopt plane. This is the local SMR with an relative amplitude of $4\times10^{-4}$ as expected for a YIG/Pt heterostructure. From this we can conclude that the interface is transparent enough for pure spin current transport across it.

In case of the non-local voltage shown in Fig.~\ref{figure:MMRMeasure}(g)-(i) we observe a qualitative similar angle-dependence as for the local voltage. However, a closer look reveals that for $\mathbf{h}\parallel\mathbf{j}$ and $\mathbf{h}\parallel\mathbf{z}$ $V_\mathrm{nl,res}=0$, and for $\mathbf{h}\parallel\mathbf{t}$ $V_\mathrm{nl,res}<0$. Thus we only observe a resistive non-local signal in the detector strip if the external magnetic field is not aligned collinear to the surface normal or the charge current direction. This is in agreement to our model for the spin current flow across the interface, as only for a non vanishing projection of the order parameter $\mathbf{N}$ in the MOI onto the spin polarization direction $\mathbf{s}$ of the electron spin accumulation in the NM inelastic scattering processes lead to magnon accumulation underneath the detector strip and the diffusion of magnons is detected at the same time via the ISHE in the detector strip. The negative sign we observe is consistent with the reciprocity of the SHE and ISHE in the Pt layer, as the charge current $\mathbf{j}_\mathrm{q,det}$ induced in the detector strip due to the injected spin current by the magnons is flowing in the same direction as $\mathbf{j}_\mathrm{q,in}$ (compare Fig.~\ref{figure:MMR}), thus the electrical field counteracting the charge current is oriented in the opposite direction and the detected open circuit voltage is negative. Similar to the SMR, we find an angle-dependence of $V_\mathrm{nl,res}$ for the ip and oopj rotation planes of our ADMR experiments and no angle dependence for the oopt rotation plane. The observed symmetry agrees with our phenomenological expectation of the all-electrical detection of magnon transport via SHE and ISHE.

\begin{figure*}[t]
 \includegraphics[width=170mm]{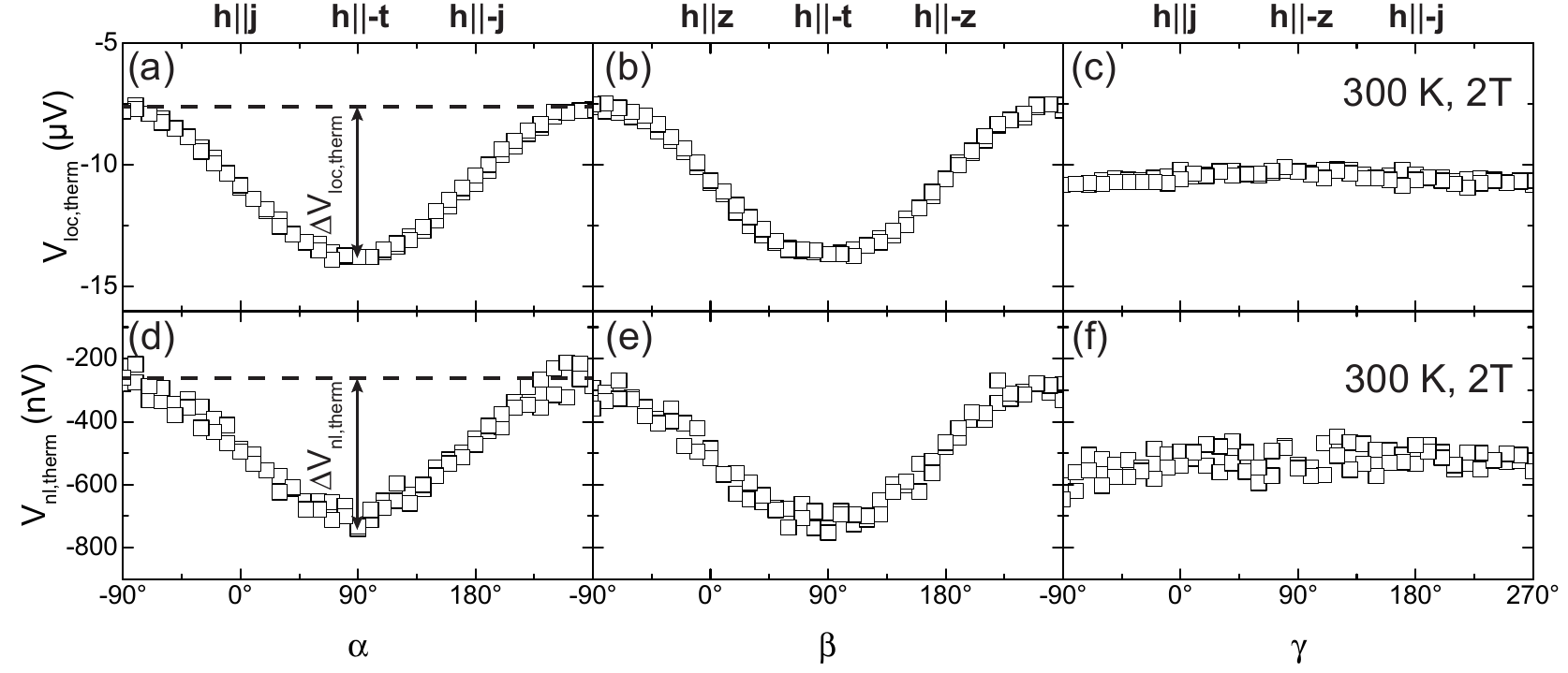}\\
 \caption[Thermal local and non-local voltages]{Extracted angle-dependent data of the local $V_\mathrm{loc,therm}$ and non-local $V_\mathrm{nl,res}$ thermal voltages for the very same sample as in Fig.~\ref{figure:MMRMeasure}. (a)-(c) Measured angular dependence of $V_\mathrm{loc,therm}$ at $T=300\,\mathrm{K}$ and $\mu_0H=2\,\mathrm{T}$ for ip (a), oopj (b), and oopt (c). $V_\mathrm{loc,therm}$ exhibits the angle-dependence expected for current heating induced longitudinal SSE. (d)-(f) Simultaneously measured angular dependence of $V_\mathrm{nl}$ for ip (d), oopj (f), and oopt (g). The non-local signal has the very same angle-dependence as the local thermal response. Figures and data adapted with permission from Ref.~\cite{Goennenwein2015}.}
  \label{figure:MMRSSEMeasure}
\end{figure*}

Similarly, one can look into the local and non-local thermal response, which is simultaneously measured by the current reversal method we use in these measurements~\cite{Schreier2013}. The results obtained for the very same sample in the same measurement runs as in Fig.\ref{figure:MMRMeasure} are depicted in Fig.~\ref{figure:MMRSSEMeasure}.

For the local response we expect to observe for the ip rotation a $\sin(\alpha)$-dependence of $V_\mathrm{loc,therm}$ and for the oopj rotation a $\sin(\beta)$-dependence. For the oopt rotation plane we predict no angle-dependence from our model for the current heating induced spin Seebeck effect (See Section~\ref{spin_seebeck}). Indeed, in Fig.~\ref{figure:MMRSSEMeasure}(a) and (b) we find for $V_\mathrm{loc,therm}$ a sinusoidal angle-dependence for the ip and oopj rotation plane. No angle-dependence is observed for the oopt rotation plane in Fig.~\ref{figure:MMRSSEMeasure}(c). The extracted amplitude $\Delta V_\mathrm{nl,therm}=6.2\,\mathrm{\mu V}$ for the angle-dependence of the ip and oopj rotation. This amplitude is comparable to the results obtained from current heating induced experiments in YIG/Pt heterostructures (see Fig.~\ref{figure:SpinSeebeckCurrentHeating}), if one accounts for the higher charge current densities originating from the smaller width of the Pt strip in the experiments here. In addition, a comparable offset signal is visible in $V_\mathrm{loc,therm}$, which originates from additional thermal voltages, like for example Seebeck contributions from the electrical contacts to the sample, and quite possibly an additional voltage offset of the voltmeter.

The thermal non-local voltage exhibits the very same angle-dependence for the three orthogonal rotation planes investigated in the experiment. We find an amplitude $\Delta V_\mathrm{nl,therm}=450\,\mathrm{nV}$, which is about an order of magnitude smaller than the local amplitude in the injector. For the non-local spin Seebeck voltage two contributions need to be accounted for. On the one hand the thermally generated magnon accumulation, described by the magnon chemical potential $\mu_\mathrm{mag}$ in the MOI. On the other hand, heat transport in the MOI can also cause a finite temperature difference at the interface between NM detector and MOI. Both of these contributions give the very same angular dependence and can thus not be separated from just one ADMR measurement.

\begin{figure}[h]
 \includegraphics[width=85mm]{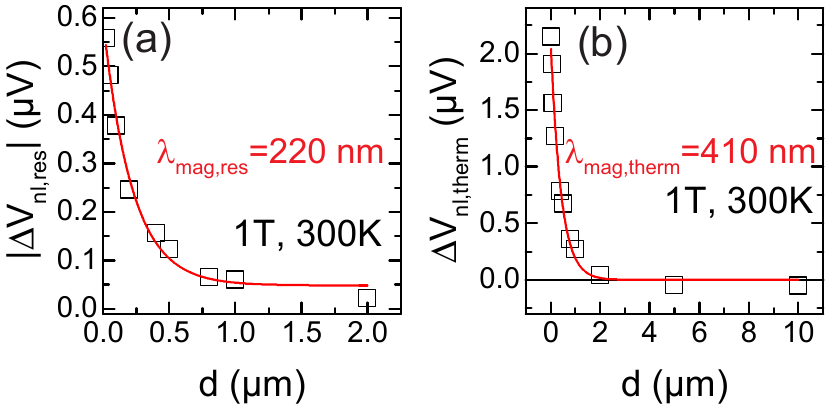}\\
 \caption[distance dependence of the non-local voltages]{(a) $|\Delta V_\mathrm{nl,res}|$ as a function of the edge-to-edge separation $d$ for $T=300\,\mathrm{K}$ and obtained for Pt strips with $w=500\,\mathrm{nm}$, and a thickness of $10\,\mathrm{nm}$ deposited onto a $2\,\mathrm{\mu m}$ thick LPE YIG layer. (b) Separation dependence of $\Delta V_\mathrm{nl,therm}$ for the same sample and measurements as in (a). For large $d$ the extracted $\Delta V_\mathrm{nl,therm}$ changes its sign. Figures and data adapted from Ref.~\cite{WimmerMaster2016}.}
  \label{figure:MMRDistance}
\end{figure}
For both non-local responses $V_\mathrm{nl,res}$ and $V_\mathrm{nl,therm}$ we extracted the voltage amplitudes $\Delta V_\mathrm{nl,res}$ (compare Fig.~\ref{figure:MMRMeasure}(g)) and $\Delta V_\mathrm{nl,therm}$ (compare Fig.~\ref{figure:MMRSSEMeasure}(d)) for the ip ADMR measurements as a function of the edge-to-edge separation $d$. We plotted both quantities as a function of the edge-to-edge separation $d$ for $T=300\,\mathrm{K}$ and $\mu_0 H=1\,\mathrm{T}$ for another YIG/Pt sample, with identical dimensions for the Pt strips, but only a thickness of $2\,\mathrm{\mu m}$ for the LPE YIG layer in Fig.~\ref{figure:MMRDistance}.

For $\Delta V_\mathrm{nl,res}$ we find a monotonic decrease with increasing $d$. Fitting the edge-to-edge separation dependence with an exponential decay function ($\propto \exp(-d/\lambda_\mathrm{mag,res})$), we obtain the effective magnon diffusion length $\lambda_\mathrm{mag,res}=220\,\mathrm{nm}$. It is important to keep in mind that in the all-electrical SHE and ISHE based magnon transport experiments one does not selectively excite magnetic excitations with a defined wavelength and wave vector, but rather excites all thermally available modes. Thus the measured diffusion length is only an average over all thermally available modes. As $\lambda_\mathrm{mag}$ is dependent on the $q$-vector of the magnon, one would expect to find different decay constants in the $d$-dependence of $|\Delta V_\mathrm{nl,res}|$. However, as a multitude of magnons may contribute to the non-local signal and as we are limited in the spatial resolution of our experiment, we do not observe such an effect here. However, at larger $d$ the single exponential decay fitted to the data seems to be not enough to describe the obtained data and at least a second contribution with a longer diffusion constant may be required. Similar findings have been reported by Cornelissen~\textit{et al.}~\cite{Cornelissen2016_temperaturedependence} and Shan~\textit{et al.}~\cite{shan_influence_2016}.

The thermal signal amplitude also decreases with increasing distance, but for $d>3\,\mathrm{\mu m}$, we observe a sign change in the non-local thermal voltage. As nicely explained by Shan~\textit{et al.}~\cite{shan_influence_2016} and Ganzhorn~\textit{et al.}~\cite{Ganzhorn2017} this sign change originates from the finite thickness of the MOI, which influences the distribution of the magnon chemical potential introduced by the thermal gradient at the injector and leads to a sign change either as a function of the MOI thickness for fixed $d$ or to a sign change of $\Delta V_\mathrm{nl,therm}$ as a function of the edge-to-edge separation $d$ for a fixed thickness of the MOI. This observed sign change highlights the importance of the magnon chemical potential for the non-local thermal voltage, as the thickness of the MOI is relevant for the critical $d$-value, where $\Delta V_\mathrm{nl,therm}$ crosses 0. This length scale is only relevant for magnons. Heat transport by phonons is also possible in the substrate and thus a much larger thickness value would play a role there. Similar to the resistive response one can fit the $d$-dependence with an exponential decay function ($\propto\exp(-d/\lambda_\mathrm{mag,therm})$) and obtains a value for the thermal magnon diffusion length $\lambda_\mathrm{mag,therm}=410\,\mathrm{nm}$.

Given the very simplistic approach by fitting with just an exponential decay function for describing the diffusive process in the MOI, we get reasonable agreement between $\lambda_\mathrm{mag,res}$ and $\lambda_\mathrm{mag,therm}$. This underlines the importance of the magnon chemical potential to describe the diffusion process for both resistive and thermal response of the non-local voltage signal. As $\Delta V_\mathrm{nl,therm}$ is nearly an order of magnitude larger as $\Delta V_\mathrm{nl,res}$ for a given $d$ it is also possible to investigate $V_\mathrm{nl,therm}$ for much larger values of $d$, while we are limited to shorter values for $d$ for $V_\mathrm{nl,res}$ as we are approaching the noise limit of our setup there much sooner. This might also explain the different values we obtain for the magnon diffusion length for the resistive and the thermal non-local response.

\begin{figure}[h]
 \includegraphics[width=85mm]{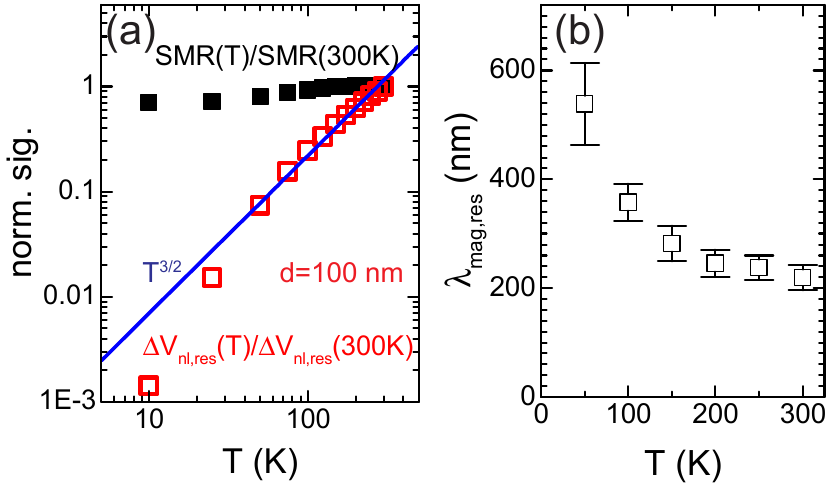}\\
 \caption[distance dependence of the non-local voltages]{(a) Normalized temperature dependence of the local SMR and non-local $\Delta V_\mathrm{nl,res}$ response for $d=100\,\mathrm{nm}$ and obtained for Pt strips with $w=500\,\mathrm{nm}$, and a thickness of $10\,\mathrm{nm}$ deposited onto a $3\,\mathrm{\mu m}$ thick LPE YIG layer. The local SMR response is only reduced by a factor of 4 from room temperature down to liquid Helium temperatures, while the non-local voltage is reduced by more than 2 orders of magnitude over that same temperature range. The blue line is a simulation of a $T^{3/2}$ temperature dependence, which agrees well with the experimental data of the non-local signal. (b) Extracted temperature dependence of the effective magnon diffusion length $\lambda_\mathrm{mag,res}$ extracted from an exponanetial fit to $|\Delta V_\mathrm{nl,res}|(d)$ for different temperatures. With decreasing temperature the magnon diffusion length increases. Figures and data adapted with permission from Ref.~\cite{Goennenwein2015}.}
  \label{figure:MMRTemp}
\end{figure}

Another important aspect is the temperature dependence of the non-local voltage, which confirms the postulated origin of it. We here focus on the results obtained for $V_\mathrm{nl,res}$ published in Ref.~\cite{Goennenwein2015}. In Fig.~\ref{figure:MMRTemp}(a) we compare the local SMR response normalized to the SMR value at $300\,\mathrm{K}$ to the normalized non-local $\Delta V_\mathrm{nl,res}(T)/\Delta V_\mathrm{nl,res}(300\,\mathrm{K})$ values as a function of temperature. For the local SMR signal, the SMR decreases with decreasing temperature and is reduced by $30\%$ from $300\,\mathrm{K}$ down to $10\,\mathrm{K}$. For the non-local voltage measured at $d=100\,\mathrm{nm}$ we also observe a decrease with decreasing temperature, but the signal is decreased by more than 2 orders of magnitude from $300\,\mathrm{K}$ down to $10\,\mathrm{K}$. This drastic difference in the local and non-local temperature dependence is explained by the different processes at interface relevant for the SMR and the non-local resistive voltage. For the SMR the relevant spin current across the interface is determined by the spin-mixing conductance $\tilde{g}_r^{\uparrow\downarrow}$, which has only a weak temperature dependence and remains finite for $T=0$. In this case inelastic electron scattering processes at the interface are most relevant for the occurrence of SMR. In contrast, for the non-local resistive voltage $\Delta V_\mathrm{nl,res}$ the spin current across the MOI/NM interface is given by $g$, which varies as $T^{3/2}$~\cite{Zhang2012,Zhang2012_PRB} and vanishes for $T=0$. For the non-local effect, inelastic electron scattering processes at the MOI/NM interface are relevant, which are suppressed for $T=0$. The blue line in Fig.~\ref{figure:MMRTemp}(a) is a $T^{3/2}$ simulation for the non-local data obtained for $T \leq 50\, \mathrm{K}$, which can very well describe the observed temperature dependence and confirms that our model based on pure spin current transport across the interface is correct.

The temperature dependence of $V_\mathrm{nl,therm}$ is much more complex and presently not fully understood over the whole temperature range investigated. Results from our measurements are published and discussed in Ref.~\cite{Ganzhorn2017}. Here, the temperature dependence of the heat conductivity and the resulting temperature profile for magnons and phonons in the MOI at different temperatures have to be modeled quantitatively, to check if the proposed models can fully describe the observed temperature dependence. Such a task is very difficult to successfully implement, as many parameters need to be known precisely as a function of temperature.

From the measurements of the $d$-dependence and an exponential decay fit for each temperature, we can also determine the temperature dependence of the magnon diffusion length $\lambda_\mathrm{mag,res}$ for our LPE YIG layer as depicted in Fig.~\ref{figure:MMRTemp}(b). With decreasing temperature $\lambda_\mathrm{mag,res}$ increases. As $\lambda_\mathrm{mag,res}$ is influenced by magnon-magnon and magnon-phonon scattering it is reasonable that it increases for lower temperatures. The larger error bars at low temperatures are due to the lower detected non-local signal (see Fig.~\ref{figure:MMRTemp}(a)), which reduces the number of data points that can be used for evaluation of $\lambda_\mathrm{mag,res}$.

In the following we want to shortly compare the results presented in this section to published data from other groups and mention new findings obtained by them. The all-electrical SHE based magnon transport experiments shown in this section are based on the theoretical predictions by Zhang and Zhang~\cite{Zhang2012,Zhang2012_PRB} in 2012. On the experimental side, the first experimental observation of this effect was reported by Cornelissen~\textit{et al.}~\cite{Cornelissen2015} in the group of Bart van Wees in Groningen in 2015. In this experimental work Cornelissen~\textit{et al.}~observed $\lambda_\mathrm{mag,res}=9400\;\mathrm{nm}$ in a $210\,\mathrm{nm}$ thick LPE YIG film, which is quite promising for future applications. Compared to our results discussed here this value is at least an order of magnitude larger. Shan~\textit{et al.}~have shown that the extracted magnon diffusion length depends on the thickness of the LPE YIG and decreases for thicker YIG layers~\cite{shan_influence_2016}. However, the change observed by Shan~\textit{et al.}~was only a reduction by $45\%$ for a $12\,\mathrm{\mu m}$ thick LPE YIG film compared to a $210\,\mathrm{nm}$ thick LPE YIG film. Moreover, the extracted $\lambda_\mathrm{mag,res}$ sensitively depends on the method used for the extraction, for example which range of edge-to-edge separations are used for the exponential decay fit. Another possibility might be that the surface treatment of the LPE YIG layer during fabrication of the NM strips also plays a crucial role for $\lambda_\mathrm{mag,res}$, but this requires furtehr systematic investigations. Similar to our results in Fig.~\ref{figure:MMRTemp}(a), Cornelissen~\textit{et al.}~showed that the temperature dependence of the all-electrical SHE based magnon transport experiments proofs the thermally activated nature of the process responsible for the magnon accumulation, such that the signal vanishes for temperatures $T\leq30\;\mathrm{K}$~\cite{Goennenwein2015,Cornelissen2016_temperaturedependence}. After the first experimental results by the group of Bart J. van Wees, we and a couple of other groups could reproduce the obtained results and helped to build a more fundamental understanding of the underlying physics.~\cite{Goennenwein2015,Ganzhorn2016Logic,Ganzhorn2017,Vlez2016,Li2016_Riverside}. Another important aspect first studied by Cornelissen~\textit{et al.}~was the influence of the external magnetic field magnitude onto $\Delta V_\mathrm{nl,res}$ and $\Delta V_\mathrm{nl,therm}$~\cite{Cornelissen2016_fielddependence}. Both voltages are reduced with increasing external magnetic field, which can be explained by a field induced change of the spin convertance $g$ and/or a field dependence of $\lambda_\mathrm{mag,res}$ and $\lambda_\mathrm{mag,therm}$ by field induced changes in the magnon band structure. To better quantify these two contributions more systematic studies are still required. Utilizing the all-electrical magnon transport effect further studies showed that is possible to establish electronically accessible magnon logic circuits~\cite{Ganzhorn2016Logic} and provided evidence for anisotropic magnon transport in MOIs~\cite{Liu2017}. A confirmation by an independent measurement technique for the presence of a magnon chemical potential driven by the SHE electron spin accumulation has been achieved last year by detecting the stray field of the magnetic excitations by nitrogen-vacancy centers in diamond~\cite{Du2017}. First experiments with higher charge current densities for the local drive also provide evidence for accessing nonlinear magnon effects with this all-electrical technique~\cite{thiery_nonlinear2017}.

From a materials point of view the first experiments used high quality LPE YIG thin films for the MOI, further development of vertical transport structures allowed also to investigate the all-electrical magnon transport effect in sputtered YIG thin film samples~\cite{Li2016_Riverside,Wu2016_SputteredYIGVertical}. In a recent publication Shan~\textit{et al.}~showed that sputtered nickel ferrite layers can be used as the MOI and the all-electrical magnon transport effect can still be observed, which is further evidence for the universal nature of this effect~\cite{Shan2017_NFO}. Means to increase the spin convertance $g$ by surface treatments have already been investigated~\cite{Vlez2016}, but more systematic studies are still required.

Taken all together, the all-electrical SHE based magnon transport experiments are interesting from both a fundamental and an application point of view. The magnon transport experiments enable us in combination with nano-structuring processes to investigate magnetic excitation transport in the diffusive and ballistic regime in a MOI by magnetotransport techniques. Further experiments and theoretical work in this direction will deepen the knowledge on these transport phenomena in MOI. For applications, the all-electrical scheme allows to bridge the gap between magnon logic~\cite{Chumak2015,Chumak2017} and conventional charge based logic elements, which allows to fabricate new device concepts~\cite{Ganzhorn2016Logic}.

\section{Conclusion}
\label{conclusion}

This topical review dealt with the pure spin current transport in MOI/NM heterostructures and means to detect these currents via the SHE and ISHE. The spin current across the MOI/NM interface is either driven by a non-equilibrium condition of the magnetic order parameter $\mathbf{N}$ or by an accumulation of electron spins in the NM and magnetic excitation quanta in the MOI. For the electron spin accumulation one can exploit the pure spin current generated by a charge current in the NM. The interfacial pure spin currents across the interface (Eq.(\ref{equ:InterfacialSpinCurrent})) give rise to the four effects covered in this review article: the spin pumping effect, the longitudinal spin Seebeck effect, the spin Hall magnetoresistance and the all-electrical magnon transport effect.

For the spin pumping effect, we showed that the magnetic order parameter is driven out-of-equilibrium by the application of an RF field, driving it into magnetic resonance. We covered two experimental techniques to quantify the spin pumping effect. On the one hand broadband ferromagnetic resonance allows to detect the change in Gilbert damping from spin pumping and can quantify the transparency of the interface (the spin mixing conductance) for pure spin current, which makes it an interesting tool for material/device characterization. On the other hand, it is possible to detect the pumped spin current as an electrical signal via the ISHE. Such experiments allow to also determine the spin Hall angle and spin diffusion length in the NM.

The longitudinal spin Seebeck effect hinges on the temperature difference at the interface between the electrons in the NM and the magnetic excitation quanta in the MOI. We showed that one can investigate qualitatively the longitudinal effect by magnetotransport measurements and utilizing a current reversal measurement technique. We discussed the temperature dependence of the current heating induced longitudinal spin Seebeck effect in GdIG/Pt heterostructures. In these samples we found two temperatures, where the detected spin Seebeck effect voltage changed its sign and explained this by the influence of the different magnon modes onto the longitudinal spin Seebeck effect. Our obtained results establish the longitudinal spin Seebeck effect as a powerful tool to investigate the magnonic band structure, provided a reasonable model to explain the observed signals is available.

The pure spin currents across the interface are also responsible for the spin Hall magnetoresistance. The resistance of the NM on top of the MOI depends on the orientation of the magnetic order parameter with respect to the charge current induced spin polarization at the MOI/NM interface. We discussed how angle-dependent magnetoresistance effects allow to obtain the fingerprint of the SMR and to separate it from other possible contributions. The SMR allows to study pure spin current transport across the interface with simple magnetotransport techniques and a quantitative extraction of the relevant parameters, i.e.~the spin Hall angle and spin diffusion length in the NM and the spin mixing conductance of the MOI/NM interface, is possible with this magnetoresistance effect. As a next step we showed that SMR even allows to detect the spin canting phase in compensated REIGs, as in this regime the SMR even changes its sign. Furthermore, we looked into the SMR effect in antiferromagnetic MOI/NM heterostructures and also observed a $90^\circ$ phase shift for the resistive response. The results obtained in the antiferromagnet NiO can be quantitatively explained by taking into account the multidomain state present in the spin-flop phase of the MOI. This shows that the SMR does not depend on the orientation of the net magnetic moment, but rather on the orientation of the corresponding magnetic order parameter (e.g.~the N\'{e}el vector) or on the orientation of the individual magnetic moments at the interface. Thus the SMR is a powerful and surface sensitive magnetotransport technique to study complex spin textures in MOIs.

Finally, we covered the important aspects and recent findings of the all-electrical magnon transport experiments. Here, a non-local electrical measurement concept with tow NM strips allows to investigate the transport of spin excitation quanta in the MOI. The effect crucially hinges on the spin convertance $g$ at the interface and the inelastic electron scattering at the interface associated with it. We showed that, similar to the SMR, all-electrical magnon transport experiments have a characteristic fingerprint in angle-dependent magnetotransport experiments. Most prominently, due to the charge current drive used in the experiment thermal effects due to Joule heating and effects driven by the electron spin accumulation coexist and can be separated by the current reversal technique. In addition, we showed that a systematic investigation of the edge-to-edge distance dependence of the non-local signal enables us to extract the magnon diffusion length in the MOI, which reaches several hundred nanometers in YIG. Finally, the experimental investigation of the temperature dependence of the non-local signal confirms the theoretical predicted concept of the spin convertance and even allows to investigate the magnon diffusion length as a function of temperature. This novel technique allows to study by electrical measurement techniques magnetic excitation transport in MOIs and may even pave the way towards ballistic magnon transport. For applications, this effect allows to combine charge based logic circuits with magnon logic circuits.

\section{Outlook}
\label{outlook}
Last but not least, we want to give an insight into future investigations based on pure spin currents in MOI/NM bilayers, which we think might be worthwhile to pursue.

For the spin pumping effect it has been already shown that the spin mixing conductance at the interface can be tailored in more sophisticated multilayer structures, where between a ferromagnetic MOI and a NM an antiferromagetic MOI is inserted. On the one hand this provides additional tools to engineer the spin mixing conductance at the interface. On the other hand, an more importantly this result suggests that pure spin currents can be transported in antiferromagnetic MOIs. The spin pumping effect in antiferromagnetic MOIs has been theoretically investigated~\cite{Cheng2014,khymyn_transformation_2016,Johansen2017,semenov_spin_2017}. From an experimental point of view these experiments are rather challenging as it requires microwave fields of several hundred GHz or even THz to drive antiferromagnetic resonances in these MOIs. Similarly, first experiments showed that excited standing spin waves also drive a pure spin current across the interface. Here, more insight into the relevant parameters has to be gained by theory and experiment. It seems to very interesting to also investigate spin pumping in magnetically ordered systems with more exotic collective excitations like chiral spin structures and topological spin textures. First experiments in this realm have been carried out by Hirobe~\textit{et al.}~in Cu$_2$OSeO$_3$/Pt heterostructures~\cite{hirobe_generation_2015}, but more experiments in different materials are necessary to better understand the relevant processes. Going a step further, investigations of the dynamical coupling between two MOIs via pure spin currents generated by spin pumping could be interesting to mediate coherent coupling between the MOIs. Such experiments may allow to gain a deeper insight into the AC component of the spin pumped pure spin current. Another intriguing avenue in spin pumping would be to experimentally verify proposed spin noise signatures of squeezed magnon states by dipolar interactions~\cite{kamra_super-poissonian_2016}.

The reciprocal effect of driving magnetization dynamics with SHE generated spin currents is another interesting field. In these experiments, the main challenge is to separate pure spin current transferred via the SHE from a charge current from other spurious effects caused by the charge current like Joule heating. For applications, auto-oscillations driven by SHE generated pure spin currents are a promising approach for tunable microwave and spinwave emitters~\cite{Collet2016,chen_spin-torque_2016,jungfleisch_insulating_2017,evelt_spin_2018}, but high microwave output powers can only be achieved by synchronizing multiple auto-oscillators with each other. For this task pure spin currents may also be beneficial. In addition, spin excitations created from auto-oscillations may allow to study magnon transport in an SHE based all-electrical fashion, with the advantage to rely solely on the spin mixing conductance and not the temperature sensitive spin convertance. Using antiferromagnetic MOIs it might be even possible to achieve THz emission in such devices~\cite{Cheng2016}. Thermal gradients may also be used to drive spin dynamics in MOI/NM heterostructures~\cite{safranski_spin_2017}, which opens up a new avenue for engineering these devices.

The longitudinal spin Seebeck effect has still to prove its usefulness in energy harvesting applications. For such applications it is important to obtain a useful figure-of-merit. Recent theoretical work in this direction seems to indicate that the efficiency at room temperature is not competitive enough with already established Seebeck based devices~\cite{Cahaya2015}. However, at low temperatures the longitudinal spin Seebeck effect may allow to fabricate very sensitive thermometers. From a fundamental point of view it is still very important to find means in the experiment to separate contributions originating from the magnon chemical potential and temperature differences at the interface and single out the corresponding relevant length scales. In the same direction it is worthwhile to investigate how one can enhance the longitudinal spin Seebeck effect by tailoring the magnonic bandstructure in artificial systems. In addition, longitudinal spin Seebeck effect experiments in antiferromagnetic and topological spin-textured MOIs can be interesting from a fundamental point of view. A major challenge in such experiments will be to change the orientation of the magnetic order parameter by an external control parameter, to separate the spin Seebeck contributions from other thermally induced voltages. Recent experimental work utilizing the spin-flop transition at large external magnetic allowed to study the longitudinal spin Seebeck effect in the antiferromagnetic MOIs Cr$_2$O$_3$, MnF$_2$ and NiO~\cite{seki_thermal_2015,wu_antiferromagnetic_2016,rezende_theory_2016,holanda_spin_2017}. These first results show that a thermal gradient can also drive a pure spin current across the interface for a antiferromagnetic MOI.

In the realm of the spin Hall magnetoresistance a lot of experiments already showed that this technique allows for surface sensitive magnetic moment detection in MOIs and studies on a variety of different MOIs are to be expected. An increase of this effect might be possible by NM with larger spin Hall angle like 2-dimensional materials and surface conduction states in topological insulators. Furthermore, investigations of the size dependence in MOIs where the magnetic texture size is comparable with feature sizes of lithography processes allow to investigate the suitability of the SMR for electrical detection of magnetic domains or domain walls. A similar aspect is a more profound study of the SMR in MOIs with topological spin texture and exploring the suitability for detection of for example Skyrmions. From a fundamental point of view, it would be great to include contributions of $g$ to the SMR model and figure out means to extract these influences in the experiment. This might be possible in nanostructured, ultrathin MOI/NM bilayers, as one can then quite possibly enhance the magnon chemical potential due to the spin convertance $g$ at the interface and then be able to detect those. Further progress in MOI/NM multilayers may provide means to study the spin-valve like SMR effect in the future and more insight into engineering the MOI/NM interface to enhance for example the imaginary part of the spin mixing conductance. As already mentioned in Section~\ref{spin_hall_magnetoresistance}, quantum tunneling aspects of the SMR provide for interesting aspects, here especially in metallic heterostructures where an oscillatory behavior is expected in theory.


For the all-electrical spin Hall effect based magnon transport experiments, one possible scenario is to simultaneously conduct all-electrical measurements and spatially resolved inelastic-light scattering experiments to independently resolve the magnon transport. Another interesting experiment would be to utilize nanostructuring processes and/or artificial magnonic crystals to study on the one hand the effect of reduction in dimensionality on the magnon transport and on the other hand allow for single frequency magnon transport in these experiments. Similarly, investigating the all-electrical magnon transport in MOIs with antiferromagnetic, or topological, or chiral magnetic order are interesting steps in evolving a deeper understanding of the magnon transport in these MOIs. In addition, it would be interesting to investigate the temporal evolution of the non-local voltage signal, which should allow to extract relevant timescales for the transport process. However, electrical crosstalk between the two separated strips may make such an approach quite challenging. Investigating theoretical predictions for a spin superfluid magnon current utilizing all-electrical magnon transport may be interesting from both a fundamental and applications point of view~\cite{Takei2015_SpinSuperfluid}. Especially in these all-electrical magnon transport experiments it is now possible to study interactions induced by additional drives, like microwaves exciting standing spin waves or thermal excitations and their influence on the magnon transport in MOIs. Accomplishing ballistic spin excitation transport in magnon transport experiments would allow to extract the transported quanta ($\hbar$) and to investigate spin current transport in a regime different to diffusive transport, as already studied in the thermal transport of antiferromagnetic MOIs~\cite{li_ballistic_2005}. To achieve ballistic transport, 1-dimensional transport channels with lengths shorter than the magnon diffusion length are necessary. Last but not least, the ultimate goal would be to achieve coherent spin current transport by these spin Hall based approaches, which requires further work in both theory and experiment.


From this topical review it is evident that pure spin current physics in MOI/NM systems have already extended our knowledge and brought major breakthroughs in the field of pure spin currents and understanding the underlying physics. Still many exciting things are yet to be discovered in this research area, making it an sensational time to be active in this field.

\begin{acknowledgments}
Financial support from the DFG via SPP 1538 ``Spin Caloric Transport'', Project No. GO 944/4 and the German Excellence Initiative via the ``Nanosystems Initiative Munich''(NIM) is gratefully acknowledged. Moreover, I would like to thank the following people for fruitful discussions, helpful input while writing this article and dedicating their resources to our results discussed here: S.~Gepr\"{a}gs, S.T.B.~Goennenwein, R.~Gross, H.~Huebl, M.~Opel, M.~Weiler. In addition, I would like to thank F.D.~Coletta, J.~Fischer, K.~Ganzhorn, A.~Kamra, S.~Meyer, R.~Schlitz, M.~Schreier, N.~Vlietstra, M.~Wagner, T.~Wimmer from the Walther-Mei{\ss}ner-Institut, and C.~Back, J.~Barker, G.E.W.~Bauer, Y.-T.~Chen, H.~Ebert, A.~Gupta, M.~Kl\"{a}ui, T.~Kuschel, G.~Reiss, E.~Saitoh, G.~Wolthersdorf, and many more unsung heroes for fruitful collaborations over the past decade dedicated to pure spin current physics in MOIs.
\end{acknowledgments}
\bibliographystyle{iopart-num}
\bibliography{althammer_PureSpinCurrents}

\providecommand{\newblock}{}
\begin{thebibliography}{100}
\expandafter\ifx\csname url\endcsname\relax
  \def\url#1{{\tt #1}}\fi
\expandafter\ifx\csname urlprefix\endcsname\relax\def\urlprefix{URL }\fi
\providecommand{\eprint}[2][]{\url{#2}}

\bibitem{waldrop_chips_2016}
Waldrop M~M 2016 {\em Nature\/} {\bf 530} 144--147 ISSN 0028-0836, 1476-4687
  \urlprefix\url{http://www.nature.com/doifinder/10.1038/530144a}

\bibitem{bourianoff_nanoelectronics_2010}
Bourianoff G, Brillouet M, Cavin R~K, Hiramoto T, Hutchby J~A, Ionescu A~M and
  Uchida K 2010 {\em Proceedings of the IEEE\/} {\bf 98} 1986--1992 ISSN
  0018-9219, 1558-2256
  \urlprefix\url{http://ieeexplore.ieee.org/document/5628289/}

\bibitem{sander_2017_2017}
Sander D, Valenzuela S~O, Makarov D, Marrows C~H, Fullerton E~E, Fischer P,
  McCord J, Vavassori P, Mangin S, Pirro P, Hillebrands B, Kent A~D, Jungwirth
  T, Gutfleisch O, Kim C~G and Berger A 2017 {\em Journal of Physics D: Applied
  Physics\/} {\bf 50} 363001 ISSN 0022-3727, 1361-6463
  \urlprefix\url{http://stacks.iop.org/0022-3727/50/i=36/a=363001?key=crossref.d604d760857c305b1bff22db68f4c139}

\bibitem{hoffmann_pure_2007}
Hoffmann A 2007 {\em physica status solidi (c)\/} {\bf 4} 4236--4241 ISSN
  1610-1642
  \urlprefix\url{http://onlinelibrary.wiley.com/doi/10.1002/pssc.200775942/abstract;jsessionid=93C3FD37B2788B0E804160AD2C2B8A48.d01t03}

\bibitem{Chumak2015}
Chumak A~V, Vasyuchka V~I, Serga A~A and Hillebrands B 2015 {\em Nature
  Physics\/} {\bf 11} 453--461
  \urlprefix\url{https://doi.org/10.1038/nphys3347}

\bibitem{Chumak2017}
Chumak A~V, Serga A~A and Hillebrands B 2017 {\em Journal of Physics D: Applied
  Physics\/} {\bf 50} 244001
  \urlprefix\url{https://doi.org/10.1088/1361-6463/aa6a65}

\bibitem{nakata_spin_2017}
Nakata K, Simon P and Loss D 2017 {\em Journal of Physics D: Applied Physics\/}
  {\bf 50} 114004 ISSN 0022-3727, 1361-6463
  \urlprefix\url{http://stacks.iop.org/0022-3727/50/i=11/a=114004?key=crossref.ceca069820e6a4325424815f4abdf6bb}

\bibitem{tserkovnyak_nonlocal_2005}
Tserkovnyak Y, Brataas A, Bauer G~E~W and Halperin B~I 2005 {\em Reviews of
  Modern Physics\/} {\bf 77} 1375--1421 ISSN 0034-6861, 1539-0756
  \urlprefix\url{https://link.aps.org/doi/10.1103/RevModPhys.77.1375}

\bibitem{Bauer2012}
Bauer G~E~W, Saitoh E and van Wees B~J 2012 {\em Nature Materials\/} {\bf 11}
  391--399 \urlprefix\url{https://doi.org/10.1038/nmat3301}

\bibitem{Dyakonov1971}
Dyakonov M and Perel V 1971 {\em Physics Letters A\/} {\bf 35} 459--460
  \urlprefix\url{https://doi.org/10.1016%2F0375-9601%2871%2990196-4}

\bibitem{SHE:Hirsch:PRL:1999}
Hirsch J~E 1999 {\em Physical Review Letters\/} {\bf 83} 1834--1837

\bibitem{Hoffmann2013}
Hoffmann A 2013 {\em {IEEE} Transactions on Magnetics\/} {\bf 49} 5172--5193

\bibitem{Sinova2015}
Sinova J, Valenzuela S~O, Wunderlich J, Back C and Jungwirth T 2015 {\em
  Reviews of Modern Physics\/} {\bf 87} 1213--1260

\bibitem{brataas_finite-element_2000}
Brataas A, Nazarov Y~V and Bauer G~E~W 2000 {\em Physical Review Letters\/}
  {\bf 84} 2481--2484 ISSN 0031-9007, 1079-7114
  \urlprefix\url{https://link.aps.org/doi/10.1103/PhysRevLett.84.2481}

\bibitem{Tserkovnyak2002}
Tserkovnyak Y, Brataas A and Bauer G~E~W 2002 {\em Physical Review B\/} {\bf
  66} 224403 \urlprefix\url{https://doi.org/10.1103/physrevb.66.224403}

\bibitem{Adachi2013}
Adachi H, ichi Uchida K, Saitoh E and Maekawa S 2013 {\em Reports on Progress
  in Physics\/} {\bf 76} 036501
  \urlprefix\url{https://doi.org/10.1088/0034-4885/76/3/036501}

\bibitem{Bender2015}
Bender S~A and Tserkovnyak Y 2015 {\em Physical Review B\/} {\bf 91} 140402
  \urlprefix\url{https://doi.org/10.1103/physrevb.91.140402}

\bibitem{brataas_spin-pumping_2004}
Brataas A, Tserkovnyak Y and Bauer G~E 2004 {\em Journal of Magnetism and
  Magnetic Materials\/} {\bf 272-276} 1981--1982 ISSN 03048853
  \urlprefix\url{http://linkinghub.elsevier.com/retrieve/pii/S030488530301672X}

\bibitem{Costache:vanWees:spin-pumping:experiment:PRL2006}
Costache M~V, Sladkov M, Watts S~M, van~der Wal C~H and van Wees B~J 2006 {\em
  Physical Review Letters\/} {\bf 97}(21) 216603
  \urlprefix\url{http://link.aps.org/doi/10.1103/PhysRevLett.97.216603}

\bibitem{spin-pumping:saitoh:APL:2006}
Saitoh E, Ueda M, Miyajima H and Tatara G 2006 {\em Applied Physics Letters\/}
  {\bf 88} 182509 (pages~3)
  \urlprefix\url{http://link.aip.org/link/?APL/88/182509/1}

\bibitem{woltersdorf_magnetization_2007}
Woltersdorf G, Mosendz O, Heinrich B and Back C~H 2007 {\em Physical Review
  Letters\/} {\bf 99} 246603
  \urlprefix\url{https://link.aps.org/doi/10.1103/PhysRevLett.99.246603}

\bibitem{Mosendz2010}
Mosendz O, Pearson J~E, Fradin F~Y, Bauer G~E~W, Bader S~D and Hoffmann A 2010
  {\em Physical Review Letters\/} {\bf 104} 046601
  \urlprefix\url{https://doi.org/10.1103/physrevlett.104.046601}

\bibitem{Ando2010}
Ando K, Kajiwara Y, Sasage K, Uchida K and Saitoh E 2010 {\em {IEEE} Trans.
  Magn.\/} {\bf 46} 3694--3696
  \urlprefix\url{http://dx.doi.org/10.1109/TMAG.2010.2060382}

\bibitem{ando_inverse_2011}
Ando K, Takahashi S, Ieda J, Kajiwara Y, Nakayama H, Yoshino T, Harii K,
  Fujikawa Y, Matsuo M, Maekawa S and Saitoh E 2011 {\em Journal of Applied
  Physics\/} {\bf 109} 103913 ISSN 00218979
  \urlprefix\url{http://link.aip.org/link/JAPIAU/v109/i10/p103913/s1&Agg=doi}

\bibitem{Czeschka2011}
Czeschka F~D, Dreher L, Brandt M~S, Weiler M, Althammer M, Imort I~M, Reiss G,
  Thomas A, Schoch W, Limmer W, Huebl H, Gross R and Goennenwein S~T~B 2011
  {\em Phys. Rev. Lett.\/} {\bf 107}
  \urlprefix\url{http://dx.doi.org/10.1103/PhysRevLett.107.046601}

\bibitem{ando_observation_2012}
Ando K and Saitoh E 2012 {\em Nature Communications\/} {\bf 3} 629 ISSN
  2041-1723 \urlprefix\url{http://www.nature.com/doifinder/10.1038/ncomms1640}

\bibitem{kapelrud_spin_2013}
Kapelrud A and Brataas A 2013 {\em Physical Review Letters\/} {\bf 111} ISSN
  0031-9007, 1079-7114
  \urlprefix\url{https://link.aps.org/doi/10.1103/PhysRevLett.111.097602}

\bibitem{Hahn2013}
Hahn C, de~Loubens G, Viret M, Klein O, Naletov V~V and Youssef J~B 2013 {\em
  Physical Review Letters\/} {\bf 111} 217204
  \urlprefix\url{https://doi.org/10.1103/physrevlett.111.217204}

\bibitem{Kampfrath2013}
Kampfrath T, Battiato M, Maldonado P, Eilers G, N\"{o}tzold J, M\"{a}hrlein S,
  Zbarsky V, Freimuth F, Mokrousov Y, Bl\"{u}gel S, Wolf M, Radu I, Oppeneer
  P~M and M\"{u}nzenberg M 2013 {\em Nature Nanotechnology\/} {\bf 8} 256--260
  \urlprefix\url{https://doi.org/10.1038/nnano.2013.43}

\bibitem{Cheng2014}
Cheng R, Xiao J, Niu Q and Brataas A 2014 {\em Physical Review Letters\/} {\bf
  113} 057601 \urlprefix\url{https://doi.org/10.1103/physrevlett.113.057601}

\bibitem{Wei2014}
Wei D, Obstbaum M, Ribow M, Back C~H and Woltersdorf G 2014 {\em Nature
  Communications\/} {\bf 5} \urlprefix\url{https://doi.org/10.1038/Fncomms4768}

\bibitem{Weiler2014}
Weiler M, Shaw J~M, Nembach H~T and Silva T~J 2014 {\em Physical Review
  Letters\/} {\bf 113} 157204
  \urlprefix\url{https://doi.org/10.1103/physrevlett.113.157204}

\bibitem{huisman_femtosecond_2016}
Huisman T~J, Mikhaylovskiy R~V, Costa J~D, Freimuth F, Paz E, Ventura J,
  Freitas P~P, Blügel S, Mokrousov Y, Rasing T and Kimel A~V 2016 {\em Nature
  Nanotechnology\/} {\bf 11} 455--458 ISSN 1748-3387, 1748-3395
  \urlprefix\url{http://www.nature.com/doifinder/10.1038/nnano.2015.331}

\bibitem{khymyn_transformation_2016}
Khymyn R, Lisenkov I, Tiberkevich V~S, Slavin A~N and Ivanov B~A 2016 {\em
  Physical Review B\/} {\bf 93} ISSN 2469-9950, 2469-9969
  \urlprefix\url{https://link.aps.org/doi/10.1103/PhysRevB.93.224421}

\bibitem{li_direct_2016}
Li J, Shelford L, Shafer P, Tan A, Deng J, Keatley P, Hwang C, Arenholz E,
  van~der Laan G, Hicken R and Qiu Z 2016 {\em Physical Review Letters\/} {\bf
  117} ISSN 0031-9007, 1079-7114
  \urlprefix\url{https://link.aps.org/doi/10.1103/PhysRevLett.117.076602}

\bibitem{Seifert2016}
Seifert T, Jaiswal S, Martens U, Hannegan J, Braun L, Maldonado P, Freimuth F,
  Kronenberg A, Henrizi J, Radu I, Beaurepaire E, Mokrousov Y, Oppeneer P~M,
  Jourdan M, Jakob G, Turchinovich D, Hayden L~M, Wolf M, M\"{u}nzenberg M,
  Kl\"{a}ui M and Kampfrath T 2016 {\em Nature Photonics\/} {\bf 10} 483--488
  \urlprefix\url{https://doi.org/10.1038/Fnphoton.2016.91}

\bibitem{bocklage_coherent_2017}
Bocklage L 2017 {\em Physical Review Letters\/} {\bf 118} ISSN 0031-9007,
  1079-7114
  \urlprefix\url{http://link.aps.org/doi/10.1103/PhysRevLett.118.257202}

\bibitem{huisman_spin-photo-currents_2017}
Huisman T~J, Ciccarelli C, Tsukamoto A, Mikhaylovskiy R~V, Rasing T and Kimel
  A~V 2017 {\em Applied Physics Letters\/} {\bf 110} 072402 ISSN 0003-6951,
  1077-3118 \urlprefix\url{http://aip.scitation.org/doi/10.1063/1.4976202}

\bibitem{Johansen2017}
Johansen {\O} and Brataas A 2017 {\em Physical Review B\/} {\bf 95} 220408
  \urlprefix\url{https://doi.org/10.1103/physrevb.95.220408}

\bibitem{kapelrud_spin_2017}
Kapelrud A and Brataas A 2017 {\em Physical Review B\/} {\bf 95} ISSN
  2469-9950, 2469-9969
  \urlprefix\url{http://link.aps.org/doi/10.1103/PhysRevB.95.214413}

\bibitem{semenov_spin_2017}
Semenov Y~G and Kim K~W 2017 {\em Applied Physics Letters\/} {\bf 110} 192405
  ISSN 0003-6951, 1077-3118
  \urlprefix\url{http://aip.scitation.org/doi/10.1063/1.4983196}

\bibitem{seifert_ultrabroadband_2017}
Seifert T, Jaiswal S, Sajadi M, Jakob G, Winnerl S, Wolf M, Kläui M and
  Kampfrath T 2017 {\em Applied Physics Letters\/} {\bf 110} 252402 ISSN
  0003-6951, 1077-3118
  \urlprefix\url{http://aip.scitation.org/doi/10.1063/1.4986755}

\bibitem{Collet2016}
Collet M, de~Milly X, d'Allivy Kelly O, Naletov V~V, Bernard R, Bortolotti P,
  Youssef J~B, Demidov V~E, Demokritov S~O, Prieto J~L, Mu{\~{n}}oz M, Cros V,
  Anane A, de~Loubens G and Klein O 2016 {\em Nature Communications\/} {\bf 7}
  10377 \urlprefix\url{https://doi.org/10.1038/Fncomms10377}

\bibitem{chen_spin-torque_2016}
Chen T, Dumas R~K, Eklund A, Muduli P~K, Houshang A, Awad A~A, Durrenfeld P,
  Malm B~G, Rusu A and Akerman J 2016 {\em Proceedings of the IEEE\/} {\bf 104}
  1919--1945 ISSN 0018-9219, 1558-2256
  \urlprefix\url{http://ieeexplore.ieee.org/document/7505988/}

\bibitem{jungfleisch_insulating_2017}
Jungfleisch M~B, Ding J, Zhang W, Jiang W, Pearson J~E, Novosad V and Hoffmann
  A 2017 {\em Nano Letters\/} {\bf 17} 8--14 ISSN 1530-6984, 1530-6992
  \urlprefix\url{http://pubs.acs.org/doi/10.1021/acs.nanolett.6b02794}

\bibitem{evelt_spin_2018}
Evelt M, Safranski C, Aldosary M, Demidov V~E, Barsukov I, Nosov A~P, Rinkevich
  A~B, Sobotkiewich K, Li X, Shi J, Krivorotov I~N and Demokritov S~O 2018 {\em
  Scientific Reports\/} {\bf 8} ISSN 2045-2322
  \urlprefix\url{http://www.nature.com/articles/s41598-018-19606-5}

\bibitem{Uchida:2010}
Uchida K, Xiao J, Adachi H, Ohe J, Takahashi S, Ieda J, Ota T, Kajiwara Y,
  Umezawa H, Kawai H, Bauer G~E~W, Maekawa S and Saitoh E 2010 {\em Nat
  Mater\/} {\bf 9} 894--897 \urlprefix\url{http://dx.doi.org/10.1038/nmat2856}

\bibitem{Uchida2010}
ichi Uchida K, Adachi H, Ota T, Nakayama H, Maekawa S and Saitoh E 2010 {\em
  Appl. Phys. Lett.\/} {\bf 97} 172505
  \urlprefix\url{http://dx.doi.org/10.1063/1.3507386}

\bibitem{Xiao2010}
Xiao J, Bauer G~E~W, chi Uchida K, Saitoh E and Maekawa S 2010 {\em Physical
  Review B\/} {\bf 81} 214418
  \urlprefix\url{https://doi.org/10.1103/physrevb.81.214418}

\bibitem{Uchida2014}
Uchida K, Ishida M, Kikkawa T, Kirihara A, Murakami T and Saitoh E 2014 {\em
  Journal of Physics: Condensed Matter\/} {\bf 26} 343202
  \urlprefix\url{https://doi.org/10.1088/0953-8984/26/34/343202}

\bibitem{Rezende2014}
Rezende S~M, Rodr{\'{\i}}guez-Su{\'{a}}rez R~L, Cunha R~O, Rodrigues A~R,
  Machado F~L~A, Guerra G~A~F, Ortiz J~C~L and Azevedo A 2014 {\em Physical
  Review B\/} {\bf 89} 014416
  \urlprefix\url{https://doi.org/10.1103/physrevb.89.014416}

\bibitem{Cahaya2015}
Cahaya A~B, Tretiakov O~A and Bauer G~E~W 2015 {\em {IEEE} Transactions on
  Magnetics\/} {\bf 51} 1--14
  \urlprefix\url{https://doi.org/10.1109/tmag.2015.2436362}

\bibitem{Geprgs2016}
Gepr\"{a}gs S, Kehlberger A, Coletta F~D, Qiu Z, Guo E~J, Schulz T, Mix C,
  Meyer S, Kamra A, Althammer M, Huebl H, Jakob G, Ohnuma Y, Adachi H, Barker
  J, Maekawa S, Bauer G~E~W, Saitoh E, Gross R, Goennenwein S~T~B and Kl\"{a}ui
  M 2016 {\em Nature Communications\/} {\bf 7} 10452
  \urlprefix\url{https://doi.org/10.1038/ncomms10452}

\bibitem{althammer_quantitative_2013}
Althammer M, Meyer S, Nakayama H, Schreier M, Altmannshofer S, Weiler M, Huebl
  H, Gepr\"{a}gs S, Opel M, Gross R, Meier D, Klewe C, Kuschel T, Schmalhorst
  J~M, Reiss G, Shen L, Gupta A, Chen Y~T, Bauer G~E~W, Saitoh E and
  Goennenwein S~T~B 2013 {\em Phys. Rev. B\/} {\bf 87} 224401
  \urlprefix\url{http://link.aps.org/doi/10.1103/PhysRevB.87.224401}

\bibitem{chen_theory_2013}
Chen Y~T, Takahashi S, Nakayama H, Althammer M, Goennenwein S~T~B, Saitoh E and
  Bauer G~E~W 2013 {\em Phys. Rev. B\/} {\bf 87} 144411
  \urlprefix\url{http://link.aps.org/doi/10.1103/PhysRevB.87.144411}

\bibitem{Nakayama2013}
Nakayama H, Althammer M, Chen Y~T, Uchida K, Kajiwara Y, Kikuchi D, Ohtani T,
  Gepr\"{a}gs S, Opel M, Takahashi S, Gross R, Bauer G~E~W, Goennenwein S~T~B
  and Saitoh E 2013 {\em Physical Review Letters\/} {\bf 110} 206601
  \urlprefix\url{https://doi.org/10.1103/Fphysrevlett.110.206601}

\bibitem{Hahn2013SMR}
Hahn C, de~Loubens G, Klein O, Viret M, Naletov V~V and Youssef J~B 2013 {\em
  Phys. Rev. B\/} {\bf 87} 174417
  \urlprefix\url{http://dx.doi.org/10.1103/PhysRevB.87.174417}

\bibitem{Vlietstra2013}
Vlietstra N, Shan J, Castel V, van Wees B~J and Youssef J~B 2013 {\em Physical
  Review B\/} {\bf 87} 184421
  \urlprefix\url{https://doi.org/10.1103/Fphysrevb.87.184421}

\bibitem{Chen2016SMRReview}
Chen Y~T, Takahashi S, Nakayama H, Althammer M, Goennenwein S~T~B, Saitoh E and
  Bauer G~E~W 2016 {\em Journal of Physics: Condensed Matter\/} {\bf 28} 103004
  \urlprefix\url{https://doi.org/10.1088/F0953-8984/F28/F10/F103004}

\bibitem{Aqeel2015}
Aqeel A, Vlietstra N, Heuver J~A, Bauer G~E~W, Noheda B, van Wees B~J and
  Palstra T~T~M 2015 {\em Physical Review B\/} {\bf 92} 224410
  \urlprefix\url{https://doi.org/10.1103/Fphysrevb.92.224410}

\bibitem{aqeel_electrical_2016}
Aqeel A, Vlietstra N, Roy A, Mostovoy M, van Wees B~J and Palstra T~T~M 2016
  {\em Physical Review B\/} {\bf 94} ISSN 2469-9950, 2469-9969
  \urlprefix\url{https://link.aps.org/doi/10.1103/PhysRevB.94.134418}

\bibitem{Ganzhorn2016}
Ganzhorn K, Barker J, Schlitz R, Piot B~A, Ollefs K, Guillou F, Wilhelm F,
  Rogalev A, Opel M, Althammer M, Gepr\"{a}gs S, Huebl H, Gross R, Bauer G~E~W
  and Goennenwein S~T~B 2016 {\em Physical Review B\/} {\bf 94} 094401
  \urlprefix\url{https://doi.org/10.1103/Fphysrevb.94.094401}

\bibitem{Han2014}
Han J~H, Song C, Li F, Wang Y~Y, Wang G~Y, Yang Q~H and Pan F 2014 {\em
  Physical Review B\/} {\bf 90} 144431
  \urlprefix\url{https://doi.org/10.1103/Fphysrevb.90.144431}

\bibitem{hoogeboom_negative_2017}
Hoogeboom G~R, Aqeel A, Kuschel T, Palstra T~T~M and van Wees B~J 2017 {\em
  Applied Physics Letters\/} {\bf 111} 052409 ISSN 0003-6951, 1077-3118
  \urlprefix\url{http://aip.scitation.org/doi/10.1063/1.4997588}

\bibitem{ji_spin_2017}
Ji Y, Miao J, Meng K~K, Ren Z~Y, Dong B~W, Xu X~G, Wu Y and Jiang Y 2017 {\em
  Applied Physics Letters\/} {\bf 110} 262401 ISSN 0003-6951, 1077-3118
  \urlprefix\url{http://aip.scitation.org/doi/10.1063/1.4989680}

\bibitem{hou_tunable_2017}
Hou D, Qiu Z, Barker J, Sato K, Yamamoto K, Vélez S, Gomez-Perez J~M, Hueso
  L~E, Casanova F and Saitoh E 2017 {\em Physical Review Letters\/} {\bf 118}
  ISSN 0031-9007, 1079-7114
  \urlprefix\url{http://link.aps.org/doi/10.1103/PhysRevLett.118.147202}

\bibitem{manchon_spin_2017}
Manchon A 2017 {\em physica status solidi (RRL) - Rapid Research Letters\/}
  {\bf 11} 1600409 ISSN 18626254
  \urlprefix\url{http://doi.wiley.com/10.1002/pssr.201600409}

\bibitem{fischer_spin_2018}
Fischer J, Gomonay O, Schlitz R, Ganzhorn K, Vlietstra N, Althammer M, Huebl H,
  Opel M, Gross R, Goennenwein S~T~B and Geprägs S 2018 {\em Physical Review
  B\/} {\bf 97} ISSN 2469-9950, 2469-9969
  \urlprefix\url{https://link.aps.org/doi/10.1103/PhysRevB.97.014417}

\bibitem{Zhang2012}
Zhang S~S~L and Zhang S 2012 {\em Physical Review Letters\/} {\bf 109} 096603
  \urlprefix\url{https://doi.org/10.1103/physrevlett.109.096603}

\bibitem{Zhang2012_PRB}
Zhang S~S~L and Zhang S 2012 {\em Physical Review B\/} {\bf 86} 214424
  \urlprefix\url{https://doi.org/10.1103/physrevb.86.214424}

\bibitem{Cornelissen2015}
Cornelissen L~J, Liu J, Duine R~A, Youssef J~B and van Wees B~J 2015 {\em
  Nature Physics\/} {\bf 11} 1022--1026
  \urlprefix\url{https://doi.org/10.1038/Fnphys3465}

\bibitem{Goennenwein2015}
Goennenwein S~T~B, Schlitz R, Pernpeintner M, Ganzhorn K, Althammer M, Gross R
  and Huebl H 2015 {\em Applied Physics Letters\/} {\bf 107} 172405
  \urlprefix\url{https://doi.org/10.1063/1.4935074}

\bibitem{julliere_tunneling_1975}
Julliere M 1975 {\em Physics Letters A\/} {\bf 54} 225--226 ISSN 03759601
  \urlprefix\url{http://linkinghub.elsevier.com/retrieve/pii/0375960175901747}

\bibitem{Hall1881}
Hall E~H 1881 {\em Philos. Mag.\/} {\bf 12}

\bibitem{Onoda2008}
Onoda S, Sugimoto N and Nagaosa N 2008 {\em Physical Review B\/} {\bf 77}
  165103

\bibitem{nagaosa_anomalous_2010}
Nagaosa N, Sinova J, Onoda S, {MacDonald} A~H and Ong N~P 2010 {\em Reviews of
  Modern Physics\/} {\bf 82} 1539--1592
  \urlprefix\url{http://link.aps.org/doi/10.1103/RevModPhys.82.1539}

\bibitem{Smit1958}
Smit J 1958 {\em Physica\/} {\bf 24} 39--51
  \urlprefix\url{https://doi.org/10.1016/Fs0031-8914/858/993541-9}

\bibitem{Berger1970}
Berger L 1970 {\em Physical Review B\/} {\bf 2} 4559--4566
  \urlprefix\url{https://doi.org/10.1103/Fphysrevb.2.4559}

\bibitem{xiao_berry_2010}
Xiao D, Chang M~C and Niu Q 2010 {\em Reviews of Modern Physics\/} {\bf 82}
  1959--2007 ISSN 0034-6861, 1539-0756
  \urlprefix\url{https://link.aps.org/doi/10.1103/RevModPhys.82.1959}

\bibitem{spin-currents:Kato:Science:2004}
Kato Y~K, Myers R~C, Gossard A~C and Awschalom D~D 2004 {\em Science\/} {\bf
  306} 1910--1913

\bibitem{Wunderlich2005}
Wunderlich J, Kaestner B, Sinova J and Jungwirth T 2005 {\em Physical Review
  Letters\/} {\bf 94} 047204
  \urlprefix\url{https://doi.org/10.1103/Fphysrevlett.94.047204}

\bibitem{Valenzuela:2006}
Valenzuela S~O and Tinkham M 2006 {\em Nature\/} {\bf 442} 176--179 ISSN
  0028-0836 \urlprefix\url{http://dx.doi.org/10.1038/nature04937}

\bibitem{Morota:2011}
Morota M, Niimi Y, Ohnishi K, Wei D~H, Tanaka T, Kontani H, Kimura T and Otani
  Y 2011 {\em Phys. Rev. B\/} {\bf 83}(17) 174405
  \urlprefix\url{http://link.aps.org/doi/10.1103/PhysRevB.83.174405}

\bibitem{Liu2012Pt}
Liu L, Lee O~J, Gudmundsen T~J, Ralph D~C and Buhrman R~A 2012 {\em Physical
  Review Letters\/} {\bf 109} 096602
  \urlprefix\url{https://doi.org/10.1103/physrevlett.109.096602}

\bibitem{Liu2012}
Liu L, Pai C~F, Li Y, Tseng H~W, Ralph D~C and Buhrman R~A 2012 {\em Science\/}
  {\bf 336} 555--558 \urlprefix\url{https://doi.org/10.1126/science.1218197}

\bibitem{Pai2012}
Pai C~F, Liu L, Li Y, Tseng H~W, Ralph D~C and Buhrman R~A 2012 {\em Applied
  Physics Letters\/} {\bf 101} 122404
  \urlprefix\url{https://doi.org/10.1063/1.4753947}

\bibitem{Niimi:2012}
Niimi Y, Kawanishi Y, Wei D~H, Deranlot C, Yang H~X, Chshiev M, Valet T, Fert A
  and Otani Y 2012 {\em Physical Review Letters\/} {\bf 109} 156602

\bibitem{Garello2013}
Garello K, Miron I~M, Avci C~O, Freimuth F, Mokrousov Y, Bl\"{u}gel S, Auffret
  S, Boulle O, Gaudin G and Gambardella P 2013 {\em Nature Nanotechnology\/}
  {\bf 8} 587--593 \urlprefix\url{https://doi.org/10.1038/Fnnano.2013.145}

\bibitem{Weiler:Solid-state-physics-64:2013}
Weiler M, Woltersdorf G, Althammer M, Huebl H and Goennenwein S~T~B 2013 Spin
  pumping and spin currents in magnetic insulators {\em Recent Advances in
  Magnetic Insulators -- From Spintronics to Microwave Applications\/} ({\em
  Solid State Physics\/} vol~64) ed Wu M and Hoffmann A (Academic Press)
  chap~5, pp 123--156

\bibitem{cheng_spin_2008}
Cheng S~g, Xing Y, Sun Q~f and Xie X~C 2008 {\em Physical Review B\/} {\bf 78}
  ISSN 1098-0121, 1550-235X
  \urlprefix\url{https://link.aps.org/doi/10.1103/PhysRevB.78.045302}

\bibitem{liu_spin_2010}
Liu X and Xie X 2010 {\em Solid State Communications\/} {\bf 150} 471--474 ISSN
  00381098
  \urlprefix\url{http://linkinghub.elsevier.com/retrieve/pii/S0038109809007625}

\bibitem{meyer_observation_2017}
Meyer S, Chen Y~T, Wimmer S, Althammer M, Wimmer T, Schlitz R, Geprägs S,
  Huebl H, Ködderitzsch D, Ebert H, Bauer G~E~W, Gross R and Goennenwein S~T~B
  2017 {\em Nature Materials\/} {\bf 16} 977--981 ISSN 1476-1122, 1476-4660
  \urlprefix\url{http://www.nature.com/doifinder/10.1038/nmat4964}

\bibitem{edelstein_spin_1990}
Edelstein V 1990 {\em Solid State Communications\/} {\bf 73} 233--235 ISSN
  00381098
  \urlprefix\url{http://linkinghub.elsevier.com/retrieve/pii/003810989090963C}

\bibitem{ganichev_spin-galvanic_2002}
Ganichev S~D, Ivchenko E~L, Bel'kov V~V, Tarasenko S~A, Sollinger M, Weiss D,
  Wegscheider W and Prettl W 2002 {\em Nature\/} {\bf 417} 153--156 ISSN
  0028-0836 \urlprefix\url{http://www.nature.com/doifinder/10.1038/417153a}

\bibitem{sanchez_spin--charge_2013}
Sánchez J~C~R, Vila L, Desfonds G, Gambarelli S, Attané J~P, De~Teresa J~M,
  Magén C and Fert A 2013 {\em Nature Communications\/} {\bf 4} ISSN 2041-1723
  \urlprefix\url{http://www.nature.com/doifinder/10.1038/ncomms3944}

\bibitem{jungfleisch_interface-driven_2016}
Jungfleisch M~B, Zhang W, Sklenar J, Jiang W, Pearson J~E, Ketterson J~B and
  Hoffmann A 2016 {\em Physical Review B\/} {\bf 93} ISSN 2469-9950, 2469-9969
  \urlprefix\url{https://link.aps.org/doi/10.1103/PhysRevB.93.224419}

\bibitem{seibold_theory_2017}
Seibold G, Caprara S, Grilli M and Raimondi R 2017 {\em Physical Review
  Letters\/} {\bf 119} ISSN 0031-9007, 1079-7114
  \urlprefix\url{https://link.aps.org/doi/10.1103/PhysRevLett.119.256801}

\bibitem{jia_spin_2011}
Jia X, Liu K, Xia K and Bauer G~E~W 2011 {\em EPL (Europhysics Letters)\/} {\bf
  96} 17005 ISSN 0295-5075
  \urlprefix\url{http://stacks.iop.org/0295-5075/96/i=1/a=17005}

\bibitem{Cornelissen2016}
Cornelissen L~J, Peters K~J~H, Bauer G~E~W, Duine R~A and van Wees B~J 2016
  {\em Physical Review B\/} {\bf 94} 014412
  \urlprefix\url{https://doi.org/10.1103/physrevb.94.014412}

\bibitem{niimi_reciprocal_2015}
Niimi Y and Otani Y 2015 {\em Reports on Progress in Physics\/} {\bf 78} 124501
  ISSN 0034-4885, 1361-6633
  \urlprefix\url{http://stacks.iop.org/0034-4885/78/i=12/a=124501?key=crossref.36b47a051f9426c7efbfcbeb67bd2637}

\bibitem{coey_magnetism_2010}
Coey J~M~D 2010 {\em Magnetism and {Magnetic} {Materials}\/} (Cambridge
  University Press) ISBN 0-521-81614-9

\bibitem{gilleo_magnetic_1958}
Gilleo M~A and Geller S 1958 {\em Physical Review\/} {\bf 110} 73
  \urlprefix\url{http://link.aps.org/doi/10.1103/PhysRev.110.73}

\bibitem{geller_structure_1957}
Geller S and Gilleo M~A 1957 {\em Acta Crystallographica\/} {\bf 10} 239--239
  ISSN 0365110X
  \urlprefix\url{http://scripts.iucr.org/cgi-bin/paper?S0365110X57000729}

\bibitem{helszajn_yig_1985}
Helszajn J 1985 {\em {YIG} resonators and filters\/} (Wiley) ISBN
  978-0-471-90516-5

\bibitem{winkler_magnetic_1981}
Winkler G 1981 {\em Magnetic garnets\/} (Braunschweig; Wiesbaden: Vieweg) ISBN
  3-528-08487-1 978-3-528-08487-5

\bibitem{elezzabi_ultrafast_1996}
Elezzabi A~Y and Freeman M~R 1996 {\em Applied Physics Letters\/} {\bf 68}
  3546--3548 ISSN 00036951
  \urlprefix\url{http://apl.aip.org/resource/1/applab/v68/i25/p3546_s1}

\bibitem{geller_crystal_1957}
Geller S and Gilleo M 1957 {\em Journal of Physics and Chemistry of Solids\/}
  {\bf 3} 30--36 ISSN 0022-3697
  \urlprefix\url{http://www.sciencedirect.com/science/article/B6TXR-46MF5CJ-9J/2/b935d08036890429e8fa9b24b750d051}

\bibitem{anderson_molecular_1964}
Anderson E~E 1964 {\em Physical Review\/} {\bf 134} A1581--A1585
  \urlprefix\url{http://link.aps.org/doi/10.1103/PhysRev.134.A1581}

\bibitem{hansen_saturation_1974}
Hansen P, Röschmann P and Tolksdorf W 1974 {\em Journal of Applied Physics\/}
  {\bf 45} 2728--2732 ISSN 00218979
  \urlprefix\url{http://jap.aip.org/resource/1/japiau/v45/i6/p2728_s1}

\bibitem{Cherepanov1993}
Cherepanov V, Kolokolov I and L{'}vov V 1993 {\em Physics Reports\/} {\bf 229}
  81--144 \urlprefix\url{https://doi.org/10.1016/0370-1573(93)90107-o}

\bibitem{dionne_molecularfield_1976}
Dionne G~F 1976 {\em Journal of Applied Physics\/} {\bf 47} 4220--4221 ISSN
  0021-8979, 1089-7550
  \urlprefix\url{http://scitation.aip.org/content/aip/journal/jap/47/9/10.1063/1.323204}

\bibitem{SchlitzMaster2015}
Schlitz R 2015 {\em Spin Transport Experiments in Hybrid Nanostructures\/}
  Master's thesis Technical University Munich
  \urlprefix\url{http://www.wmi.badw.de/publications/theses/Schlitz,Richard_Masterarbeit_2015.pdf}

\bibitem{bernasconi_canted_1971}
Bernasconi J and Kuse D 1971 {\em Physical Review B\/} {\bf 3} 811--815
  \urlprefix\url{http://link.aps.org/doi/10.1103/PhysRevB.3.811}

\bibitem{eremenko_field-induced_1979}
Eremenko V~V and Kharchenko N~F 1979 {\em Phase Transitions\/} {\bf 1} 61--98
  ISSN 0141-1594 \urlprefix\url{http://dx.doi.org/10.1080/01411597908213185}

\bibitem{srivastava_exchange_1982}
Srivastava C~M, Srinivasan C and Aiyar R 1982 {\em Journal of Applied
  Physics\/} {\bf 53} 781--783 ISSN 0021-8979, 1089-7550
  \urlprefix\url{http://scitation.aip.org/content/aip/journal/jap/53/1/10.1063/1.329990}

\bibitem{clark_neferrimagnets_1968}
Clark A~E and Callen E 1968 {\em Journal of Applied Physics\/} {\bf 39}
  5972--5982 ISSN 0021-8979, 1089-7550
  \urlprefix\url{http://scitation.aip.org/content/aip/journal/jap/39/13/10.1063/1.1656100}

\bibitem{alben_phase_1970}
Alben R 1970 {\em Physical Review B\/} {\bf 2} 2767--2784
  \urlprefix\url{http://link.aps.org/doi/10.1103/PhysRevB.2.2767}

\bibitem{seul_domain_1995}
Seul M and Andelman D 1995 {\em Science\/} {\bf 267} 476--483 ISSN 0036-8075,
  1095-9203 \urlprefix\url{http://www.sciencemag.org/content/267/5197/476}

\bibitem{hansen_magnetic_1983}
Hansen P, Witter K and Tolksdorf W 1983 {\em Physical Review B\/} {\bf 27} 6608
  \urlprefix\url{http://link.aps.org/doi/10.1103/PhysRevB.27.6608}

\bibitem{blank_growth_1972}
Blank S and Nielsen J 1972 {\em Journal of Crystal Growth\/} {\bf 17} 302--311
  ISSN 0022-0248
  \urlprefix\url{http://www.sciencedirect.com/science/article/pii/0022024872902618}

\bibitem{aichele_garnet_2003}
Aichele T, Lorenz A, Hergt R and Görnert P 2003 {\em Crystal Research and
  Technology\/} {\bf 38} 575--587 ISSN 1521-4079
  \urlprefix\url{http://onlinelibrary.wiley.com/doi/10.1002/crat.200310071/abstract;jsessionid=C044891ACF094D6F26B5B4880B4B287F.d02t02?systemMessage=Wiley+Online+Library+will+be+disrupted+17+March+from+10-14+GMT+%2806-10+EDT%29+for+essential+maintenance}

\bibitem{krockenberger_solid_2008}
Krockenberger Y, Matsui H, Hasegawa T, Kawasaki M and Tokura Y 2008 {\em
  Applied Physics Letters\/} {\bf 93} 092505 ISSN 00036951
  \urlprefix\url{http://link.aip.org/link/APPLAB/v93/i9/p092505/s1&Agg=doi}

\bibitem{krockenberger_layer-by-layer_2009}
Krockenberger Y, Yun K~S, Hatano T, Arisawa S, Kawasaki M and Tokura Y 2009
  {\em Journal of Applied Physics\/} {\bf 106} 123911 ISSN 00218979
  \urlprefix\url{http://link.aip.org/link/JAPIAU/v106/i12/p123911/s1&Agg=doi}

\bibitem{kahl_pulsed_2003}
Kahl S and Grishin A~M 2003 {\em Journal of Applied Physics\/} {\bf 93} 6945
  ISSN 00218979
  \urlprefix\url{http://link.aip.org/link/JAPIAU/v93/i10/p6945/s1&Agg=doi}

\bibitem{manuilov_submicron_2009}
Manuilov S~A, Fors R, Khartsev S~I and Grishin A~M 2009 {\em Journal of Applied
  Physics\/} {\bf 105} 033917--033917--9 ISSN 00218979
  \urlprefix\url{http://jap.aip.org/resource/1/japiau/v105/i3/p033917_s1}

\bibitem{manuilov_pulsed_2010}
Manuilov S~A and Grishin A~M 2010 {\em Journal of Applied Physics\/} {\bf 108}
  013902 ISSN 00218979
  \urlprefix\url{http://link.aip.org/link/JAPIAU/v108/i1/p013902/s1&Agg=doi}

\bibitem{dorsey_epitaxial_1993}
Dorsey P~C, Bushnell S~E, Seed R~G and Vittoria C 1993 {\em Journal of Applied
  Physics\/} {\bf 74} 1242 ISSN 00218979
  \urlprefix\url{http://link.aip.org/link/JAPIAU/v74/i2/p1242/s1&Agg=doi}

\bibitem{manuilov_pulsed_2009}
Manuilov S~A, Khartsev S~I and Grishin A~M 2009 {\em Journal of Applied
  Physics\/} {\bf 106} 123917 ISSN 00218979
  \urlprefix\url{http://link.aip.org/link/JAPIAU/v106/i12/p123917/s1&Agg=doi}

\bibitem{onbasli_pulsed_2014}
Onbasli M~C, Kehlberger A, Kim D~H, Jakob G, Kläui M, Chumak A~V, Hillebrands
  B and Ross C~A 2014 {\em APL Materials\/} {\bf 2} 106102 ISSN 2166-532X
  \urlprefix\url{http://aip.scitation.org/doi/10.1063/1.4896936}

\bibitem{hauser_yttrium_2016}
Hauser C, Richter T, Homonnay N, Eisenschmidt C, Qaid M, Deniz H, Hesse D,
  Sawicki M, Ebbinghaus S~G and Schmidt G 2016 {\em Scientific Reports\/} {\bf
  6} ISSN 2045-2322 \urlprefix\url{http://www.nature.com/articles/srep20827}

\bibitem{zaki_growth_2017}
Zaki A~M, Blythe H~J, Heald S~M, Fox A~M and Gehring G~A 2017 {\em Journal of
  Magnetism and Magnetic Materials\/} ISSN 03048853
  \urlprefix\url{http://linkinghub.elsevier.com/retrieve/pii/S0304885317323703}

\bibitem{houchen_chang_nanometer-thick_2014}
{Houchen Chang}, {Peng Li}, {Wei Zhang}, {Tao Liu}, Hoffmann A, {Longjiang
  Deng} and {Mingzhong Wu} 2014 {\em IEEE Magnetics Letters\/} {\bf 5} 1--4
  ISSN 1949-307X, 1949-3088
  \urlprefix\url{http://ieeexplore.ieee.org/lpdocs/epic03/wrapper.htm?arnumber=6882836}

\bibitem{liu_ferromagnetic_2014}
Liu T, Chang H, Vlaminck V, Sun Y, Kabatek M, Hoffmann A, Deng L and Wu M 2014
  {\em Journal of Applied Physics\/} {\bf 115} 17A501 ISSN 0021-8979, 1089-7550
  \urlprefix\url{http://aip.scitation.org/doi/10.1063/1.4852135}

\bibitem{lustikova_spin_2014}
Lustikova J, Shiomi Y, Qiu Z, Kikkawa T, Iguchi R, Uchida K and Saitoh E 2014
  {\em Journal of Applied Physics\/} {\bf 116} 153902 ISSN 0021-8979, 1089-7550
  \urlprefix\url{http://aip.scitation.org/doi/10.1063/1.4898161}

\bibitem{du_y_2015}
Du C, Wang H, Hammel P~C and Yang F 2015 {\em Journal of Applied Physics\/}
  {\bf 117} 172603 ISSN 0021-8979, 1089-7550
  \urlprefix\url{http://aip.scitation.org/doi/10.1063/1.4913813}

\bibitem{cao_van_effect_2018}
Cao~Van P, Surabhi S, Dongquoc V, Kuchi R, Yoon S~G and Jeong J~R 2018 {\em
  Applied Surface Science\/} {\bf 435} 377--383 ISSN 01694332
  \urlprefix\url{http://linkinghub.elsevier.com/retrieve/pii/S0169433217334116}

\bibitem{knorr_lattice_1984}
Knörr B and Tolksdorf W 1984 {\em Materials Research Bulletin\/} {\bf 19}
  1507--1513 ISSN 0025-5408
  \urlprefix\url{http://www.sciencedirect.com/science/article/pii/0025540884902654}

\bibitem{Avci2016}
Avci C~O, Quindeau A, Pai C~F, Mann M, Caretta L, Tang A~S, Onbasli M~C, Ross
  C~A and Beach G~S~D 2016 {\em Nature Materials\/} {\bf 16} 309--314
  \urlprefix\url{https://doi.org/10.1038/nmat4812}

\bibitem{tang_anomalous_2016}
Tang C, Sellappan P, Liu Y, Xu Y, Garay J~E and Shi J 2016 {\em Physical Review
  B\/} {\bf 94} ISSN 2469-9950, 2469-9969
  \urlprefix\url{https://link.aps.org/doi/10.1103/PhysRevB.94.140403}

\bibitem{quindeau_tm_2017}
Quindeau A, Avci C~O, Liu W, Sun C, Mann M, Tang A~S, Onbasli M~C, Bono D,
  Voyles P~M, Xu Y, Robinson J, Beach G~S~D and Ross C~A 2017 {\em Advanced
  Electronic Materials\/} {\bf 3} 1600376 ISSN 2199160X
  \urlprefix\url{http://doi.wiley.com/10.1002/aelm.201600376}

\bibitem{fu_epitaxial_2017}
Fu J, Hua M, Wen X, Xue M, Ding S, Wang M, Yu P, Liu S, Han J, Wang C, Du H,
  Yang Y and Yang J 2017 {\em Applied Physics Letters\/} {\bf 110} 202403 ISSN
  0003-6951, 1077-3118
  \urlprefix\url{http://aip.scitation.org/doi/10.1063/1.4983783}

\bibitem{fujiwara_5d_2013}
Fujiwara K, Fukuma Y, Matsuno J, Idzuchi H, Niimi Y, Otani Y and Takagi H 2013
  {\em Nature Communications\/} {\bf 4} ISSN 2041-1723
  \urlprefix\url{http://www.nature.com/doifinder/10.1038/ncomms3893}

\bibitem{qiu_experimental_2013}
Qiu Z, An T, Uchida K, Hou D, Shiomi Y, Fujikawa Y and Saitoh E 2013 {\em
  Applied Physics Letters\/} {\bf 103} 182404 ISSN 0003-6951, 1077-3118
  \urlprefix\url{http://aip.scitation.org/doi/10.1063/1.4827808}

\bibitem{qiu_all-oxide_2015}
Qiu Z, Hou D, Kikkawa T, Uchida K~i and Saitoh E 2015 {\em Applied Physics
  Express\/} {\bf 8} 083001 ISSN 1882-0778, 1882-0786
  \urlprefix\url{http://stacks.iop.org/1882-0786/8/i=8/a=083001?key=crossref.55d210b01009c2d1684cc31daf99079f}

\bibitem{demasius_enhanced_2016}
Demasius K~U, Phung T, Zhang W, Hughes B~P, Yang S~H, Kellock A, Han W, Pushp A
  and Parkin S~S~P 2016 {\em Nature Communications\/} {\bf 7} 10644 ISSN
  2041-1723 \urlprefix\url{http://www.nature.com/doifinder/10.1038/ncomms10644}

\bibitem{zeb_interplay_2012}
Zeb M~A and Kee H~Y 2012 {\em Physical Review B\/} {\bf 86} ISSN 1098-0121,
  1550-235X \urlprefix\url{https://link.aps.org/doi/10.1103/PhysRevB.86.085149}

\bibitem{chen_topological_2015}
Chen Y, Lu Y~M and Kee H~Y 2015 {\em Nature Communications\/} {\bf 6} ISSN
  2041-1723 \urlprefix\url{http://www.nature.com/articles/ncomms7593}

\bibitem{nie_interplay_2015}
Nie Y, King P, Kim C, Uchida M, Wei H, Faeth B, Ruf J, Ruff J, Xie L, Pan X,
  Fennie C, Schlom D and Shen K 2015 {\em Physical Review Letters\/} {\bf 114}
  ISSN 0031-9007, 1079-7114
  \urlprefix\url{https://link.aps.org/doi/10.1103/PhysRevLett.114.016401}

\bibitem{martins_coulomb_2017}
Martins C, Aichhorn M and Biermann S 2017 {\em Journal of Physics: Condensed
  Matter\/} {\bf 29} 263001 ISSN 0953-8984, 1361-648X
  \urlprefix\url{http://stacks.iop.org/0953-8984/29/i=26/a=263001?key=crossref.2aec7e4c6cf0f6377b3759470888f2d9}

\bibitem{feng_intrinsic_2012}
Feng W, Yao Y, Zhu W, Zhou J, Yao W and Xiao D 2012 {\em Physical Review B\/}
  {\bf 86} ISSN 1098-0121, 1550-235X
  \urlprefix\url{https://link.aps.org/doi/10.1103/PhysRevB.86.165108}

\bibitem{qian_quantum_2014}
Qian X, Liu J, Fu L and Li J 2014 {\em Science\/} {\bf 346} 1344--1347 ISSN
  0036-8075, 1095-9203
  \urlprefix\url{http://www.sciencemag.org/cgi/doi/10.1126/science.1256815}

\bibitem{cazalilla_quantum_2014}
Cazalilla M, Ochoa H and Guinea F 2014 {\em Physical Review Letters\/} {\bf
  113} ISSN 0031-9007, 1079-7114
  \urlprefix\url{https://link.aps.org/doi/10.1103/PhysRevLett.113.077201}

\bibitem{shao_strong_2016}
Shao Q, Yu G, Lan Y~W, Shi Y, Li M~Y, Zheng C, Zhu X, Li L~J, Amiri P~K and
  Wang K~L 2016 {\em Nano Letters\/} {\bf 16} 7514--7520 ISSN 1530-6984,
  1530-6992
  \urlprefix\url{http://pubs.acs.org/doi/10.1021/acs.nanolett.6b03300}

\bibitem{Jamali2015}
Jamali M, Lee J~S, Jeong J~S, Mahfouzi F, Lv Y, Zhao Z, Nikoli{\'{c}} B~K,
  Mkhoyan K~A, Samarth N and Wang J~P 2015 {\em Nano Letters\/} {\bf 15}
  7126--7132 \urlprefix\url{https://doi.org/10.1021/acs.nanolett.5b03274}

\bibitem{Miao2013}
Miao B~F, Huang S~Y, Qu D and Chien C~L 2013 {\em Physical Review Letters\/}
  {\bf 111} 066602
  \urlprefix\url{https://doi.org/10.1103/Fphysrevlett.111.066602}

\bibitem{zhang_spin_2014}
Zhang W, Jungfleisch M~B, Jiang W, Pearson J~E, Hoffmann A, Freimuth F and
  Mokrousov Y 2014 {\em Physical Review Letters\/} {\bf 113} ISSN 0031-9007,
  1079-7114
  \urlprefix\url{https://link.aps.org/doi/10.1103/PhysRevLett.113.196602}

\bibitem{Wahler2016}
Wahler M, Homonnay N, Richter T, M\"{u}ller A, Eisenschmidt C, Fuhrmann B and
  Schmidt G 2016 {\em Scientific Reports\/} {\bf 6} 28727
  \urlprefix\url{https://doi.org/10.1038/Fsrep28727}

\bibitem{sangiao_control_2015}
Sangiao S, De~Teresa J~M, Morellon L, Lucas I, Martinez-Velarte M~C and Viret M
  2015 {\em Applied Physics Letters\/} {\bf 106} 172403 ISSN 0003-6951,
  1077-3118 \urlprefix\url{http://aip.scitation.org/doi/10.1063/1.4919129}

\bibitem{nakayama_rashba-edelstein_2016}
Nakayama H, Kanno Y, An H, Tashiro T, Haku S, Nomura A and Ando K 2016 {\em
  Physical Review Letters\/} {\bf 117} ISSN 0031-9007, 1079-7114
  \urlprefix\url{https://link.aps.org/doi/10.1103/PhysRevLett.117.116602}

\bibitem{ohshima_strong_2017}
Ohshima R, Ando Y, Matsuzaki K, Susaki T, Weiler M, Klingler S, Huebl H, Shikoh
  E, Shinjo T, Goennenwein S~T~B and Shiraishi M 2017 {\em Nature Materials\/}
  {\bf 16} 609--614 ISSN 1476-1122, 1476-4660
  \urlprefix\url{http://www.nature.com/doifinder/10.1038/nmat4857}

\bibitem{singh_bulk_2017}
Singh A~V, Khodadadi B, Mohammadi J~B, Keshavarz S, Mewes T, Negi D~S, Datta R,
  Galazka Z, Uecker R and Gupta A 2017 {\em Advanced Materials\/} {\bf 29}
  1701222 ISSN 09359648
  \urlprefix\url{http://doi.wiley.com/10.1002/adma.201701222}

\bibitem{qin_ultra-low_2017}
Qin Q, He S, Song W, Yang P, Wu Q, Feng Y~P and Chen J 2017 {\em Applied
  Physics Letters\/} {\bf 110} 112401 ISSN 0003-6951, 1077-3118
  \urlprefix\url{http://aip.scitation.org/doi/10.1063/1.4978431}

\bibitem{hahn_conduction_2014}
Hahn C, de~Loubens G, Naletov V~V, Ben~Youssef J, Klein O and Viret M 2014 {\em
  EPL (Europhysics Letters)\/} {\bf 108} 57005 ISSN 0295-5075, 1286-4854
  \urlprefix\url{http://stacks.iop.org/0295-5075/108/i=5/a=57005?key=crossref.b9777bc216ac1f03cd966909d7c77025}

\bibitem{wang_antiferromagnonic_2014}
Wang H, Du C, Hammel P~C and Yang F 2014 {\em Physical Review Letters\/} {\bf
  113} ISSN 0031-9007, 1079-7114
  \urlprefix\url{https://link.aps.org/doi/10.1103/PhysRevLett.113.097202}

\bibitem{wang_spin_2015}
Wang H, Du C, Hammel P~C and Yang F 2015 {\em Physical Review B\/} {\bf 91}
  ISSN 1098-0121, 1550-235X
  \urlprefix\url{https://link.aps.org/doi/10.1103/PhysRevB.91.220410}

\bibitem{seki_thermal_2015}
Seki S, Ideue T, Kubota M, Kozuka Y, Takagi R, Nakamura M, Kaneko Y, Kawasaki M
  and Tokura Y 2015 {\em Physical Review Letters\/} {\bf 115} ISSN 0031-9007,
  1079-7114
  \urlprefix\url{https://link.aps.org/doi/10.1103/PhysRevLett.115.266601}

\bibitem{moriyama_anti-damping_2015}
Moriyama T, Takei S, Nagata M, Yoshimura Y, Matsuzaki N, Terashima T,
  Tserkovnyak Y and Ono T 2015 {\em Applied Physics Letters\/} {\bf 106} 162406
  ISSN 0003-6951, 1077-3118
  \urlprefix\url{http://aip.scitation.org/doi/10.1063/1.4918990}

\bibitem{jungwirth_antiferromagnetic_2016}
Jungwirth T, Marti X, Wadley P and Wunderlich J 2016 {\em Nature
  Nanotechnology\/} {\bf 11} 231--241 ISSN 1748-3387, 1748-3395
  \urlprefix\url{http://www.nature.com/articles/nnano.2016.18}

\bibitem{rezende_theory_2016}
Rezende S~M, Rodríguez-Suárez R~L and Azevedo A 2016 {\em Physical Review
  B\/} {\bf 93} ISSN 2469-9950, 2469-9969
  \urlprefix\url{https://link.aps.org/doi/10.1103/PhysRevB.93.014425}

\bibitem{wu_antiferromagnetic_2016}
Wu S~M, Zhang W, Kc A, Borisov P, Pearson J~E, Jiang J~S, Lederman D, Hoffmann
  A and Bhattacharya A 2016 {\em Physical Review Letters\/} {\bf 116} ISSN
  0031-9007, 1079-7114
  \urlprefix\url{https://link.aps.org/doi/10.1103/PhysRevLett.116.097204}

\bibitem{holanda_spin_2017}
Holanda J, Maior D~S, Alves~Santos O, Vilela-Leão L~H, Mendes J~B~S, Azevedo
  A, Rodríguez-Suárez R~L and Rezende S~M 2017 {\em Applied Physics
  Letters\/} {\bf 111} 172405 ISSN 0003-6951, 1077-3118
  \urlprefix\url{http://aip.scitation.org/doi/10.1063/1.5001694}

\bibitem{Gayles2015}
Gayles J, Freimuth F, Schena T, Lani G, Mavropoulos P, Duine R, Bl\"{u}gel S,
  Sinova J and Mokrousov Y 2015 {\em Physical Review Letters\/} {\bf 115}
  036602 \urlprefix\url{https://doi.org/10.1103/Fphysrevlett.115.036602}

\bibitem{jiang_blowing_2015}
Jiang W, Upadhyaya P, Zhang W, Yu G, Jungfleisch M~B, Fradin F~Y, Pearson J~E,
  Tserkovnyak Y, Wang K~L, Heinonen O, te~Velthuis S~G~E and Hoffmann A 2015
  {\em Science\/} {\bf 349} 283--286 ISSN 0036-8075, 1095-9203
  \urlprefix\url{http://www.sciencemag.org/cgi/doi/10.1126/science.aaa1442}

\bibitem{Woo2016}
Woo S, Litzius K, Kr\"{u}ger B, Im M~Y, Caretta L, Richter K, Mann M, Krone A,
  Reeve R~M, Weigand M, Agrawal P, Lemesh I, Mawass M~A, Fischer P, Kl\"{a}ui M
  and Beach G~S~D 2016 {\em Nature Materials\/} {\bf 15} 501--506
  \urlprefix\url{https://doi.org/10.1038/Fnmat4593}

\bibitem{fert_magnetic_2017}
Fert A, Reyren N and Cros V 2017 {\em Nature Reviews Materials\/} {\bf 2} 17031
  ISSN 2058-8437
  \urlprefix\url{http://www.nature.com/articles/natrevmats201731}

\bibitem{chen_all-oxidebased_2017}
Chen B, Xu H, Ma C, Mattauch S, Lan D, Jin F, Guo Z, Wan S, Chen P, Gao G, Chen
  F, Su Y and Wu W 2017 {\em Science\/} {\bf 357} 191--194 ISSN 0036-8075,
  1095-9203
  \urlprefix\url{http://www.sciencemag.org/lookup/doi/10.1126/science.aak9717}

\bibitem{hao_two-dimensional_2017}
Hao L, Meyers D, Frederick C, Fabbris G, Yang J, Traynor N, Horak L, Kriegner
  D, Choi Y, Kim J~W, Haskel D, Ryan P~J, Dean M and Liu J 2017 {\em Physical
  Review Letters\/} {\bf 119} ISSN 0031-9007, 1079-7114
  \urlprefix\url{http://link.aps.org/doi/10.1103/PhysRevLett.119.027204}

\bibitem{Vonsovskii:FMR:1960}
Vonsovskii S~V 1960 {\em Ferromagnetic Resonance\/} (New York: Pergamon Press)

\bibitem{MorrishBuch}
Morrish A 2001 {\em The Physical Principles of Magnetism\/} (New York: IEEE
  Press)

\bibitem{Mosendz:2010:PRB}
Mosendz O, Vlaminck V, Pearson J~E, Fradin F~Y, Bauer G~E~W, Bader S~D and
  Hoffmann A 2010 {\em Phys. Rev. B\/} {\bf 82}(21) 214403
  \urlprefix\url{http://link.aps.org/doi/10.1103/PhysRevB.82.214403}

\bibitem{Jiao2013}
Jiao H and Bauer G~E~W 2013 {\em Physical Review Letters\/} {\bf 110} 217602
  \urlprefix\url{https://doi.org/10.1103/Fphysrevlett.110.217602}

\bibitem{Chen:SMR:theory:PRB:2013}
Chen Y~T, Takahashi S, Nakayama H, Althammer M, Goennenwein S~T~B, Saitoh E and
  Bauer G~E~W 2013 {\em Phys. Rev. B\/} {\bf 87}(14) 144411
  \urlprefix\url{http://link.aps.org/doi/10.1103/PhysRevB.87.144411}

\bibitem{haertinger_spin_2015}
Haertinger M, Back C~H, Lotze J, Weiler M, Geprägs S, Huebl H, Goennenwein
  S~T~B and Woltersdorf G 2015 {\em Physical Review B\/} {\bf 92} ISSN
  1098-0121, 1550-235X
  \urlprefix\url{https://link.aps.org/doi/10.1103/PhysRevB.92.054437}

\bibitem{nembach_perpendicular_2011}
Nembach H~T, Silva T~J, Shaw J~M, Schneider M~L, Carey M~J, Maat S and
  Childress J~R 2011 {\em Physical Review B\/} {\bf 84} ISSN 1098-0121,
  1550-235X \urlprefix\url{https://link.aps.org/doi/10.1103/PhysRevB.84.054424}

\bibitem{shaw_determination_2012}
Shaw J~M, Nembach H~T and Silva T~J 2012 {\em Physical Review B\/} {\bf 85}
  ISSN 1098-0121, 1550-235X
  \urlprefix\url{https://link.aps.org/doi/10.1103/PhysRevB.85.054412}

\bibitem{shaw_precise_2013}
Shaw J~M, Nembach H~T, Silva T~J and Boone C~T 2013 {\em Journal of Applied
  Physics\/} {\bf 114} 243906 ISSN 0021-8979, 1089-7550
  \urlprefix\url{http://aip.scitation.org/doi/10.1063/1.4852415}

\bibitem{Heinrich2011}
Heinrich B, Burrowes C, Montoya E, Kardasz B, Girt E, Song Y~Y, Sun Y and Wu M
  2011 {\em Physical Review Letters\/} {\bf 107} 066604
  \urlprefix\url{https://doi.org/10.1103/physrevlett.107.066604}

\bibitem{Burrowes2012}
Burrowes C, Heinrich B, Kardasz B, Montoya E~A, Girt E, Sun Y, Song Y~Y and Wu
  M 2012 {\em Applied Physics Letters\/} {\bf 100} 092403
  \urlprefix\url{https://doi.org/10.1063/1.3690918}

\bibitem{Sun:2013go}
Sun Y, Chang H, Kabatek M, Song Y~Y, Wang Z, Jantz M, Schneider W, Wu M,
  Montoya E, Kardasz B, Heinrich B, te~Velthuis S~G~E, Schultheiss H and
  Hoffmann A 2013 {\em Physical Review Letters\/} {\bf 111} 106601
  \urlprefix\url{http://link.aps.org/doi/10.1103/PhysRevLett.111.106601}

\bibitem{rezende_enhanced_2013}
Rezende S~M, Rodríguez-Suárez R~L, Soares M~M, Vilela-Leão L~H,
  Ley~Domínguez D and Azevedo A 2013 {\em Applied Physics Letters\/} {\bf 102}
  012402 ISSN 0003-6951, 1077-3118
  \urlprefix\url{http://aip.scitation.org/doi/10.1063/1.4773993}

\bibitem{weiler_experimental_2013}
Weiler M, Althammer M, Schreier M, Lotze J, Pernpeintner M, Meyer S, Huebl H,
  Gross R, Kamra A, Xiao J, Chen Y~T, Jiao H, Bauer G~E~W and Goennenwein S~T~B
  2013 {\em Phys. Rev. Lett.\/} {\bf 111} 176601
  \urlprefix\url{http://link.aps.org/doi/10.1103/PhysRevLett.111.176601}

\bibitem{jungfleisch_temporal_2011}
Jungfleisch M~B, Chumak A~V, Vasyuchka V~I, Serga A~A, Obry B, Schultheiss H,
  Beck P~A, Karenowska A~D, Saitoh E and Hillebrands B 2011 {\em Applied
  Physics Letters\/} {\bf 99} 182512 ISSN 0003-6951, 1077-3118
  \urlprefix\url{http://aip.scitation.org/doi/10.1063/1.3658398}

\bibitem{takahashi_electrical_2012}
Takahashi R, Iguchi R, Ando K, Nakayama H, Yoshino T and Saitoh E 2012 {\em
  Journal of Applied Physics\/} {\bf 111} 07C307 ISSN 0021-8979, 1089-7550
  \urlprefix\url{http://aip.scitation.org/doi/10.1063/1.3673429}

\bibitem{dallivy_kelly_inverse_2013}
d'Allivy Kelly O, Anane A, Bernard R, Ben~Youssef J, Hahn C, Molpeceres A~H,
  Carrétéro C, Jacquet E, Deranlot C, Bortolotti P, Lebourgeois R, Mage J~C,
  de~Loubens G, Klein O, Cros V and Fert A 2013 {\em Applied Physics Letters\/}
  {\bf 103} 082408 ISSN 0003-6951, 1077-3118
  \urlprefix\url{http://aip.scitation.org/doi/10.1063/1.4819157}

\bibitem{jungfleisch_thickness_2015}
Jungfleisch M~B, Chumak A~V, Kehlberger A, Lauer V, Kim D~H, Onbasli M~C, Ross
  C~A, Kläui M and Hillebrands B 2015 {\em Physical Review B\/} {\bf 91} ISSN
  1098-0121, 1550-235X
  \urlprefix\url{https://link.aps.org/doi/10.1103/PhysRevB.91.134407}

\bibitem{takemasa_spatial_2017}
Takemasa R, Tateno Y and Ando K 2017 {\em Applied Physics Letters\/} {\bf 110}
  042404 ISSN 0003-6951, 1077-3118
  \urlprefix\url{http://aip.scitation.org/doi/10.1063/1.4974823}

\bibitem{wang_scaling_2014}
Wang H, Du C, Pu Y, Adur R, Hammel P and Yang F 2014 {\em Physical Review
  Letters\/} {\bf 112} ISSN 0031-9007, 1079-7114
  \urlprefix\url{https://link.aps.org/doi/10.1103/PhysRevLett.112.197201}

\bibitem{Sandweg:2011ig}
Sandweg C, Kajiwara Y, Chumak A, Serga A, Vasyuchka V, Jungfleisch M, Saitoh E
  and Hillebrands B 2011 {\em Physical Review Letters\/} {\bf 106} 216601
  \urlprefix\url{http://link.aps.org/doi/10.1103/PhysRevLett.106.216601}

\bibitem{chiba_current-induced_2014}
Chiba T, Bauer G~E and Takahashi S 2014 {\em Physical Review Applied\/} {\bf 2}
  ISSN 2331-7019
  \urlprefix\url{https://link.aps.org/doi/10.1103/PhysRevApplied.2.034003}

\bibitem{klingler_spin_2017}
Klingler S, Amin V, Geprägs S, Ganzhorn K, Maier-Flaig H, Althammer M, Huebl
  H, Gross R, McMichael R~D, Stiles M~D, Goennenwein S~T~B and Weiler M 2017
  {\em arXiv:1712.02561 [cond-mat]\/} ArXiv: 1712.02561
  \urlprefix\url{http://arxiv.org/abs/1712.02561}

\bibitem{skarsvag_spin_2014}
Skarsvåg H, Kapelrud A and Brataas A 2014 {\em Physical Review B\/} {\bf 90}
  ISSN 1098-0121, 1550-235X
  \urlprefix\url{https://link.aps.org/doi/10.1103/PhysRevB.90.094418}

\bibitem{Liu2011}
Liu L, Moriyama T, Ralph D~C and Buhrman R~A 2011 {\em Physical Review
  Letters\/} {\bf 106} 036601
  \urlprefix\url{https://doi.org/10.1103/physrevlett.106.036601}

\bibitem{Jungwirth:2012em}
Jungwirth T, Wunderlich J and Olejn{\'\i}k K 2012 {\em Nature Materials\/} {\bf
  11} 382 \urlprefix\url{http://www.nature.com/doifinder/10.1038/nmat3279}

\bibitem{spin-torque:Miron:NatMat:2010}
Mihai~Miron I, Gaudin G, Auffret S, Rodmacq B, Schuhl A, Pizzini S, Vogel J and
  Gambardella P 2010 {\em Nat. Mater.\/} {\bf 9}(3) 230--234

\bibitem{Miron2011}
Miron I~M, Garello K, Gaudin G, Zermatten P~J, Costache M~V, Auffret S,
  Bandiera S, Rodmacq B, Schuhl A and Gambardella P 2011 {\em Nature\/} {\bf
  476} 189--193 \urlprefix\url{https://doi.org/10.1038/Fnature10309}

\bibitem{Hamadeh2014}
Hamadeh A, d'Allivy Kelly O, Hahn C, Meley H, Bernard R, Molpeceres A, Naletov
  V, Viret M, Anane A, Cros V, Demokritov S, Prieto J, Mu{\~{n}}oz M,
  de~Loubens G and Klein O 2014 {\em Physical Review Letters\/} {\bf 113}
  197203 \urlprefix\url{https://doi.org/10.1103/Fphysrevlett.113.197203}

\bibitem{Schreier2015}
Schreier M, Chiba T, Niedermayr A, Lotze J, Huebl H, Gepr\"{a}gs S, Takahashi
  S, Bauer G~E~W, Gross R and Goennenwein S~T~B 2015 {\em Physical Review B\/}
  {\bf 92} 144411 \urlprefix\url{https://doi.org/10.1103/Fphysrevb.92.144411}

\bibitem{baumgartner_spatially_2017}
Baumgartner M, Garello K, Mendil J, Avci C~O, Grimaldi E, Murer C, Feng J,
  Gabureac M, Stamm C, Acremann Y, Finizio S, Wintz S, Raabe J and Gambardella
  P 2017 {\em Nature Nanotechnology\/} {\bf 12} 980--986 ISSN 1748-3387,
  1748-3395
  \urlprefix\url{http://www.nature.com/doifinder/10.1038/nnano.2017.151}

\bibitem{Duan2014}
Duan Z, Smith A, Yang L, Youngblood B, Lindner J, Demidov V~E, Demokritov S~O
  and Krivorotov I~N 2014 {\em Nature Communications\/} {\bf 5} 5616
  \urlprefix\url{https://doi.org/10.1038/Fncomms6616}

\bibitem{Cheng2016}
Cheng R, Xiao D and Brataas A 2016 {\em Physical Review Letters\/} {\bf 116}
  207603 \urlprefix\url{https://doi.org/10.1103/Fphysrevlett.116.207603}

\bibitem{Manchon2015}
Manchon A, Koo H~C, Nitta J, Frolov S~M and Duine R~A 2015 {\em Nature
  Materials\/} {\bf 14} 871--882
  \urlprefix\url{https://doi.org/10.1038/Fnmat4360}

\bibitem{Miron2011:domainwall}
Miron I~M, Moore T, Szambolics H, Buda-Prejbeanu L~D, Auffret S, Rodmacq B,
  Pizzini S, Vogel J, Bonfim M, Schuhl A and Gaudin G 2011 {\em Nature
  Materials\/} {\bf 10} 419--423
  \urlprefix\url{https://doi.org/10.1038/Fnmat3020}

\bibitem{Emori2013}
Emori S, Bauer U, Ahn S~M, Martinez E and Beach G~S~D 2013 {\em Nature
  Materials\/} {\bf 12} 611--616
  \urlprefix\url{https://doi.org/10.1038/Fnmat3675}

\bibitem{Ryu2013}
Ryu K~S, Thomas L, Yang S~H and Parkin S 2013 {\em Nature Nanotechnology\/}
  {\bf 8} 527--533 \urlprefix\url{https://doi.org/10.1038/Fnnano.2013.102}

\bibitem{Ryu2016}
Ryu K~S, Yang S~H and Parkin S 2016 {\em New Journal of Physics\/} {\bf 18}
  053027 \urlprefix\url{https://doi.org/10.1088/F1367-2630/F18/F5/F053027}

\bibitem{Muhlbauer2009}
Muhlbauer S, Binz B, Jonietz F, Pfleiderer C, Rosch A, Neubauer A, Georgii R
  and Boni P 2009 {\em Science\/} {\bf 323} 915--919
  \urlprefix\url{https://doi.org/10.1126/Fscience.1166767}

\bibitem{sampaio_nucleation_2013}
Sampaio J, Cros V, Rohart S, Thiaville A and Fert A 2013 {\em Nature
  Nanotechnology\/} {\bf 8} 839--844 ISSN 1748-3387, 1748-3395
  \urlprefix\url{http://www.nature.com/doifinder/10.1038/nnano.2013.210}

\bibitem{Bttner2015}
B\"{u}ttner F, Moutafis C, Schneider M, Kr\"{u}ger B, G\"{u}nther C~M, Geilhufe
  J, v~Korff~Schmising C, Mohanty J, Pfau B, Schaffert S, Bisig A, Foerster M,
  Schulz T, Vaz C~A~F, Franken J~H, Swagten H~J~M, Kl\"{a}ui M and Eisebitt S
  2015 {\em Nature Physics\/} {\bf 11} 225--228
  \urlprefix\url{https://doi.org/10.1038/Fnphys3234}

\bibitem{kikkawa_longitudinal_2013}
Kikkawa T, Uchida K, Shiomi Y, Qiu Z, Hou D, Tian D, Nakayama H, Jin X~F and
  Saitoh E 2013 {\em Physical Review Letters\/} {\bf 110} ISSN 0031-9007,
  1079-7114
  \urlprefix\url{https://link.aps.org/doi/10.1103/PhysRevLett.110.067207}

\bibitem{meier_longitudinal_2015}
Meier D, Reinhardt D, van Straaten M, Klewe C, Althammer M, Schreier M,
  Goennenwein S~T~B, Gupta A, Schmid M, Back C~H, Schmalhorst J~M, Kuschel T
  and Reiss G 2015 {\em Nature Communications\/} {\bf 6} ISSN 2041-1723
  \urlprefix\url{http://www.nature.com/articles/ncomms9211}

\bibitem{sola_longitudinal_2017}
Sola A, Bougiatioti P, Kuepferling M, Meier D, Reiss G, Pasquale M, Kuschel T
  and Basso V 2017 {\em Scientific Reports\/} {\bf 7} 46752 ISSN 2045-2322
  \urlprefix\url{http://www.nature.com/articles/srep46752}

\bibitem{sultan_thermal_2009}
Sultan R, Avery A~D, Stiehl G and Zink B~L 2009 {\em Journal of Applied
  Physics\/} {\bf 105} 043501 ISSN 0021-8979, 1089-7550
  \urlprefix\url{http://aip.scitation.org/doi/10.1063/1.3078025}

\bibitem{zink_exploring_2010}
Zink B, Avery A, Sultan R, Bassett D and Pufall M 2010 {\em Solid State
  Communications\/} {\bf 150} 514--518 ISSN 00381098
  \urlprefix\url{http://linkinghub.elsevier.com/retrieve/pii/S0038109809006759}

\bibitem{avery_thermopower_2011}
Avery A~D, Sultan R, Bassett D, Wei D and Zink B~L 2011 {\em Physical Review
  B\/} {\bf 83} ISSN 1098-0121, 1550-235X
  \urlprefix\url{https://link.aps.org/doi/10.1103/PhysRevB.83.100401}

\bibitem{Weiler2012}
Weiler M, Althammer M, Czeschka F~D, Huebl H, Wagner M~S, Opel M, Imort I~M,
  Reiss G, Thomas A, Gross R and Goennenwein S~T~B 2012 {\em Phys. Rev.
  Lett.\/} {\bf 108} 106602
  \urlprefix\url{http://dx.doi.org/10.1103/PhysRevLett.108.106602}

\bibitem{schreier_magnon_2013}
Schreier M, Kamra A, Weiler M, Xiao J, Bauer G~E~W, Gross R and Goennenwein
  S~T~B 2013 {\em Physical Review B\/} {\bf 88} ISSN 1098-0121, 1550-235X
  \urlprefix\url{https://link.aps.org/doi/10.1103/PhysRevB.88.094410}

\bibitem{Schreier2013}
Schreier M, Roschewsky N, Dobler E, Meyer S, Huebl H, Gross R and Goennenwein
  S~T~B 2013 {\em Appl. Phys. Lett.\/} {\bf 103} 242404
  \urlprefix\url{http://dx.doi.org/10.1063/1.4839395}

\bibitem{huang_transport_2012}
Huang S~Y, Fan X, Qu D, Chen Y~P, Wang W~G, Wu J, Chen T~Y, Xiao J~Q and Chien
  C~L 2012 {\em Physical Review Letters\/} {\bf 109} 107204
  \urlprefix\url{http://link.aps.org/doi/10.1103/PhysRevLett.109.107204}

\bibitem{qu_intrinsic_2013}
Qu D, Huang S~Y, Hu J, Wu R and Chien C~L 2013 {\em Physical Review Letters\/}
  {\bf 110} 067206
  \urlprefix\url{http://link.aps.org/doi/10.1103/PhysRevLett.110.067206}

\bibitem{kikkawa_separation_2013}
Kikkawa T, Uchida K, Daimon S, Shiomi Y, Adachi H, Qiu Z, Hou D, Jin X~F,
  Maekawa S and Saitoh E 2013 {\em Physical Review B\/} {\bf 88} ISSN
  1098-0121, 1550-235X
  \urlprefix\url{https://link.aps.org/doi/10.1103/PhysRevB.88.214403}

\bibitem{bougiatioti_quantitative_2017}
Bougiatioti P, Klewe C, Meier D, Manos O, Kuschel O, Wollschläger J,
  Bouchenoire L, Brown S~D, Schmalhorst J~M, Reiss G and Kuschel T 2017 {\em
  Physical Review Letters\/} {\bf 119} ISSN 0031-9007, 1079-7114
  \urlprefix\url{https://link.aps.org/doi/10.1103/PhysRevLett.119.227205}

\bibitem{roschewsky_time_2014}
Roschewsky N, Schreier M, Kamra A, Schade F, Ganzhorn K, Meyer S, Huebl H,
  Geprägs S, Gross R and Goennenwein S~T~B 2014 {\em Applied Physics
  Letters\/} {\bf 104} 202410 ISSN 0003-6951, 1077-3118
  \urlprefix\url{http://aip.scitation.org/doi/10.1063/1.4879462}

\bibitem{agrawal_role_2014}
Agrawal M, Vasyuchka V~I, Serga A~A, Kirihara A, Pirro P, Langner T,
  Jungfleisch M~B, Chumak A~V, Papaioannou E~T and Hillebrands B 2014 {\em
  Physical Review B\/} {\bf 89} ISSN 1098-0121, 1550-235X
  \urlprefix\url{https://link.aps.org/doi/10.1103/PhysRevB.89.224414}

\bibitem{schreier_spin_2016}
Schreier M, Kramer F, Huebl H, Geprägs S, Gross R, Goennenwein S~T~B, Noack T,
  Langner T, Serga A~A, Hillebrands B and Vasyuchka V~I 2016 {\em Physical
  Review B\/} {\bf 93} ISSN 2469-9950, 2469-9969
  \urlprefix\url{https://link.aps.org/doi/10.1103/PhysRevB.93.224430}

\bibitem{seifert_launching_2017}
Seifert T, Jaiswal S, Barker J, Razdolski I, Cramer J, Gueckstock O, Maehrlein
  S, Nadvornik L, Watanabe S, Ciccarelli C, Melnikov A, Jakob G, Münzenberg M,
  Goennenwein S~T~B, Woltersdorf G, Brouwer P~W, Wolf M, Kläui M and Kampfrath
  T 2017 {\em arXiv:1709.00768 [cond-mat]\/} ArXiv: 1709.00768
  \urlprefix\url{http://arxiv.org/abs/1709.00768}

\bibitem{kimling_picosecond_2017}
Kimling J, Choi G~M, Brangham J~T, Matalla-Wagner T, Huebner T, Kuschel T, Yang
  F and Cahill D~G 2017 {\em Physical Review Letters\/} {\bf 118} ISSN
  0031-9007, 1079-7114
  \urlprefix\url{https://link.aps.org/doi/10.1103/PhysRevLett.118.057201}

\bibitem{kehlberger_length_2015}
Kehlberger A, Ritzmann U, Hinzke D, Guo E~J, Cramer J, Jakob G, Onbasli M~C,
  Kim D~H, Ross C~A, Jungfleisch M~B, Hillebrands B, Nowak U and Kläui M 2015
  {\em Physical Review Letters\/} {\bf 115} ISSN 0031-9007, 1079-7114
  \urlprefix\url{https://link.aps.org/doi/10.1103/PhysRevLett.115.096602}

\bibitem{jin_effect_2015}
Jin H, Boona S~R, Yang Z, Myers R~C and Heremans J~P 2015 {\em Physical Review
  B\/} {\bf 92} ISSN 1098-0121, 1550-235X
  \urlprefix\url{https://link.aps.org/doi/10.1103/PhysRevB.92.054436}

\bibitem{ritzmann_propagation_2014}
Ritzmann U, Hinzke D and Nowak U 2014 {\em Physical Review B\/} {\bf 89} ISSN
  1098-0121, 1550-235X
  \urlprefix\url{https://link.aps.org/doi/10.1103/PhysRevB.89.024409}

\bibitem{ritzmann_thermally_2017}
Ritzmann U, Hinzke D and Nowak U 2017 {\em Physical Review B\/} {\bf 95} ISSN
  2469-9950, 2469-9969
  \urlprefix\url{https://link.aps.org/doi/10.1103/PhysRevB.95.054411}

\bibitem{kikkawa_critical_2015}
Kikkawa T, Uchida K~i, Daimon S, Qiu Z, Shiomi Y and Saitoh E 2015 {\em
  Physical Review B\/} {\bf 92} ISSN 1098-0121, 1550-235X
  \urlprefix\url{https://link.aps.org/doi/10.1103/PhysRevB.92.064413}

\bibitem{ritzmann_magnetic_2015}
Ritzmann U, Hinzke D, Kehlberger A, Guo E~J, Kläui M and Nowak U 2015 {\em
  Physical Review B\/} {\bf 92} ISSN 1098-0121, 1550-235X
  \urlprefix\url{https://link.aps.org/doi/10.1103/PhysRevB.92.174411}

\bibitem{uchida_longitudinal_2013}
Uchida K, Nonaka T, Kikkawa T, Kajiwara Y and Saitoh E 2013 {\em Physical
  Review B\/} {\bf 87} ISSN 1098-0121, 1550-235X
  \urlprefix\url{https://link.aps.org/doi/10.1103/PhysRevB.87.104412}

\bibitem{guo_thermal_2016}
Guo E~J, Herklotz A, Kehlberger A, Cramer J, Jakob G and Kläui M 2016 {\em
  Applied Physics Letters\/} {\bf 108} 022403 ISSN 0003-6951, 1077-3118
  \urlprefix\url{http://aip.scitation.org/doi/10.1063/1.4939625}

\bibitem{anadon_characteristic_2016}
Anadón A, Ramos R, Lucas I, Algarabel P~A, Morellón L, Ibarra M~R and Aguirre
  M~H 2016 {\em Applied Physics Letters\/} {\bf 109} 012404 ISSN 0003-6951,
  1077-3118 \urlprefix\url{http://aip.scitation.org/doi/10.1063/1.4955031}

\bibitem{wu_longitudinal_2017}
Wu B~W, Luo G~Y, Lin J~G and Huang S~Y 2017 {\em Physical Review B\/} {\bf 96}
  ISSN 2469-9950, 2469-9969
  \urlprefix\url{http://link.aps.org/doi/10.1103/PhysRevB.96.060402}

\bibitem{lin_enhancement_2016}
Lin W, Chen K, Zhang S and Chien C 2016 {\em Physical Review Letters\/} {\bf
  116} ISSN 0031-9007, 1079-7114
  \urlprefix\url{https://link.aps.org/doi/10.1103/PhysRevLett.116.186601}

\bibitem{prakash_spin_2016}
Prakash A, Brangham J, Yang F and Heremans J~P 2016 {\em Physical Review B\/}
  {\bf 94} ISSN 2469-9950, 2469-9969
  \urlprefix\url{https://link.aps.org/doi/10.1103/PhysRevB.94.014427}

\bibitem{Giles2015}
Giles B~L, Yang Z, Jamison J~S and Myers R~C 2015 {\em Physical Review B\/}
  {\bf 92} \urlprefix\url{https://doi.org/10.1103/physrevb.92.224415}

\bibitem{cramer_magnon_2017}
Cramer J, Guo E~J, Geprägs S, Kehlberger A, Ivanov Y~P, Ganzhorn K,
  Della~Coletta F, Althammer M, Huebl H, Gross R, Kosel J, Kläui M and
  Goennenwein S~T~B 2017 {\em Nano Letters\/} {\bf 17} 3334--3340 ISSN
  1530-6984, 1530-6992
  \urlprefix\url{http://pubs.acs.org/doi/10.1021/acs.nanolett.6b04522}

\bibitem{Meyer2015}
Meyer S, Schlitz R, Gepr\"{a}gs S, Opel M, Huebl H, Gross R and Goennenwein
  S~T~B 2015 {\em Appl. Phys. Lett.\/} {\bf 106} 132402
  \urlprefix\url{http://dx.doi.org/10.1063/1.4916342}

\bibitem{limmer_angle-dependent_2006}
Limmer W, Glunk M, Daeubler J, Hummel T, Schoch W, Sauer R, Bihler C, Huebl H,
  Brandt M~S and Goennenwein S~T~B 2006 {\em Physical Review B\/} {\bf 74} ISSN
  1098-0121, 1550-235X
  \urlprefix\url{https://link.aps.org/doi/10.1103/PhysRevB.74.205205}

\bibitem{wang_comparative_2017}
Wang H, Du C, Hammel P~C and Yang F 2017 {\em Applied Physics Letters\/} {\bf
  110} 062402 ISSN 0003-6951, 1077-3118
  \urlprefix\url{http://aip.scitation.org/doi/10.1063/1.4975704}

\bibitem{Isasa2014}
Isasa M, Bedoya-Pinto A, V{\'{e}}lez S, Golmar F, S{\'{a}}nchez F, Hueso L~E,
  Fontcuberta J and Casanova F 2014 {\em Applied Physics Letters\/} {\bf 105}
  142402 \urlprefix\url{https://doi.org/10.1063/F1.4897544}

\bibitem{Wu2015}
Wu H, Zhang Q, Wan C, Ali S~S, Yuan Z, You L, Wang J, Choi Y and Han X 2015
  {\em {IEEE} Transactions on Magnetics\/} {\bf 51} 1--4
  \urlprefix\url{https://doi.org/10.1109/Ftmag.2015.2433060}

\bibitem{hui_spin_2016}
Hui Y~J, Cheng W~M, Zhang Z~B, Ji H~K, Cheng X~M, You L and Miao X~S 2016 {\em
  Applied Physics Express\/} {\bf 9} 073006 ISSN 1882-0778, 1882-0786
  \urlprefix\url{http://stacks.iop.org/1882-0786/9/i=7/a=073006?key=crossref.ea1bb1cebc836a2b9bcf6c487100161e}

\bibitem{shang_pure_2016}
Shang T, Zhan Q~F, Ma L, Yang H~L, Zuo Z~H, Xie Y~L, Li H~H, Liu L~P, Wang B~M,
  Wu Y~H, Zhang S and Li R~W 2016 {\em Scientific Reports\/} {\bf 5} ISSN
  2045-2322 \urlprefix\url{http://www.nature.com/articles/srep17734}

\bibitem{Meyer2014}
Meyer S, Althammer M, Gepr\"{a}gs S, Opel M, Gross R and Goennenwein S~T~B 2014
  {\em Appl. Phys. Lett.\/} {\bf 104} 242411
  \urlprefix\url{http://dx.doi.org/10.1063/1.4885086}

\bibitem{lotze_spin_2014}
Lotze J, Huebl H, Gross R and Goennenwein S~T~B 2014 {\em Physical Review B\/}
  {\bf 90} ISSN 1098-0121, 1550-235X
  \urlprefix\url{https://link.aps.org/doi/10.1103/PhysRevB.90.174419}

\bibitem{kamra_spin_2014}
Kamra A, Witek F~P, Meyer S, Huebl H, Geprägs S, Gross R, Bauer G~E~W and
  Goennenwein S~T~B 2014 {\em Physical Review B\/} {\bf 90} ISSN 1098-0121,
  1550-235X \urlprefix\url{https://link.aps.org/doi/10.1103/PhysRevB.90.214419}

\bibitem{Vlez2016}
V{\'{e}}lez S, Bedoya-Pinto A, Yan W, Hueso L~E and Casanova F 2016 {\em
  Physical Review B\/} {\bf 94}
  \urlprefix\url{https://doi.org/10.1103/physrevb.94.174405}

\bibitem{velez_hanle_2016}
Vélez S, Golovach V~N, Bedoya-Pinto A, Isasa M, Sagasta E, Abadia M, Rogero C,
  Hueso L~E, Bergeret F~S and Casanova F 2016 {\em Physical Review Letters\/}
  {\bf 116} ISSN 0031-9007, 1079-7114
  \urlprefix\url{https://link.aps.org/doi/10.1103/PhysRevLett.116.016603}

\bibitem{cho_large_2015}
Cho S, Baek S~h~C, Lee K~D, Jo Y and Park B~G 2015 {\em Scientific Reports\/}
  {\bf 5} 14668 ISSN 2045-2322
  \urlprefix\url{http://www.nature.com/articles/srep14668}

\bibitem{kim_spin_2016}
Kim J, Sheng P, Takahashi S, Mitani S and Hayashi M 2016 {\em Physical Review
  Letters\/} {\bf 116} ISSN 0031-9007, 1079-7114
  \urlprefix\url{https://link.aps.org/doi/10.1103/PhysRevLett.116.097201}

\bibitem{kobs_anisotropic_2011}
Kobs A, Heße S, Kreuzpaintner W, Winkler G, Lott D, Weinberger P, Schreyer A
  and Oepen H~P 2011 {\em Physical Review Letters\/} {\bf 106} ISSN 0031-9007,
  1079-7114
  \urlprefix\url{https://link.aps.org/doi/10.1103/PhysRevLett.106.217207}

\bibitem{avci_unidirectional_2015}
Avci C~O, Garello K, Ghosh A, Gabureac M, Alvarado S~F and Gambardella P 2015
  {\em Nature Physics\/} {\bf 11} 570--575 ISSN 1745-2473, 1745-2481
  \urlprefix\url{http://www.nature.com/articles/nphys3356}

\bibitem{Chen2016:tunneling}
Chen W, Sigrist M and Manske D 2016 {\em Physical Review B\/} {\bf 94} 104412
  \urlprefix\url{https://doi.org/10.1103/Fphysrevb.94.104412}

\bibitem{dong_spin_2018}
Dong B~W, Cramer J, Ganzhorn K, Yuan H~Y, Guo E~J, Goennenwein S~T~B and Kläui
  M 2018 {\em Journal of Physics: Condensed Matter\/} {\bf 30} 035802 ISSN
  0953-8984, 1361-648X
  \urlprefix\url{http://stacks.iop.org/0953-8984/30/i=3/a=035802?key=crossref.d27af4059e9e727599c01a68f7387c21}

\bibitem{Demokritov2001}
Demokritov S 2001 {\em Physics Reports\/} {\bf 348} 441--489
  \urlprefix\url{https://doi.org/10.1016/s0370-1573(00)00116-2}

\bibitem{Demokritov2008}
Demokritov S and Demidov V 2008 {\em {IEEE} Transactions on Magnetics\/} {\bf
  44} 6--12 \urlprefix\url{https://doi.org/10.1109/tmag.2007.910227}

\bibitem{Sebastian2015}
Sebastian T, Schultheiss K, Obry B, Hillebrands B and Schultheiss H 2015 {\em
  Frontiers in Physics\/} {\bf 3}
  \urlprefix\url{https://doi.org/10.3389/fphy.2015.00035}

\bibitem{Cheng2017}
Cheng Y, Chen K and Zhang S 2017 {\em Physical Review B\/} {\bf 96} 024449
  \urlprefix\url{https://doi.org/10.1103/physrevb.96.024449}

\bibitem{Li2016_Riverside}
Li J, Xu Y, Aldosary M, Tang C, Lin Z, Zhang S, Lake R and Shi J 2016 {\em
  Nature Communications\/} {\bf 7} 10858
  \urlprefix\url{https://doi.org/10.1038/ncomms10858}

\bibitem{WimmerMaster2016}
Wimmer T 2016 {\em Spin Transport in Magnetic Nanostructures\/} Master's thesis
  Technical University Munich
  \urlprefix\url{http://www.wmi.badw.de/publications/theses/Wimmer,Tobias_Masterarbeit_2016.pdf}

\bibitem{Cornelissen2016_temperaturedependence}
Cornelissen L~J, Shan J and van Wees B~J 2016 {\em Physical Review B\/} {\bf
  94} 180402 \urlprefix\url{https://doi.org/10.1103/physrevb.94.180402}

\bibitem{shan_influence_2016}
Shan J, Cornelissen L~J, Vlietstra N, Ben~Youssef J, Kuschel T, Duine R~A and
  van Wees B~J 2016 {\em Physical Review B\/} {\bf 94} ISSN 2469-9950,
  2469-9969 \urlprefix\url{https://link.aps.org/doi/10.1103/PhysRevB.94.174437}

\bibitem{Ganzhorn2017}
Ganzhorn K, Wimmer T, Cramer J, Schlitz R, Gepr\"{a}gs S, Jakob G, Gross R,
  Huebl H, Kl\"{a}ui M and Goennenwein S~T~B 2017 {\em {AIP} Advances\/} {\bf
  7} 085102 \urlprefix\url{https://doi.org/10.1063/1.4986848}

\bibitem{Ganzhorn2016Logic}
Ganzhorn K, Klingler S, Wimmer T, Gepr\"{a}gs S, Gross R, Huebl H and
  Goennenwein S~T~B 2016 {\em Applied Physics Letters\/} {\bf 109} 022405
  \urlprefix\url{https://doi.org/10.1063/1.4958893}

\bibitem{Cornelissen2016_fielddependence}
Cornelissen L~J and van Wees B~J 2016 {\em Physical Review B\/} {\bf 93} 020403
  \urlprefix\url{https://doi.org/10.1103/physrevb.93.020403}

\bibitem{Liu2017}
Liu J, Cornelissen L~J, Shan J, Kuschel T and van Wees B~J 2017 {\em Physical
  Review B\/} {\bf 95}
  \urlprefix\url{https://doi.org/10.1103/physrevb.95.140402}

\bibitem{Du2017}
Du C, van~der Sar T, Zhou T~X, Upadhyaya P, Casola F, Zhang H, Onbasli M~C,
  Ross C~A, Walsworth R~L, Tserkovnyak Y and Yacoby A 2017 {\em Science\/} {\bf
  357} 195--198 \urlprefix\url{https://doi.org/10.1126/science.aak9611}

\bibitem{thiery_nonlinear2017}
{Thiery} N, {Draveny} A, {Naletov} V~V, {Vila} L, {Attan{\'e}} J~P, {de
  Loubens} G, {Viret} M, {Beaulieu} N, {Ben Youssef} J, {Demidov} V~E,
  {Demokritov} S~O, {Slain} A~N, {Tiberkevich} V~S, {Anane} A, {Bortolotti} P,
  {Cros} V and {Klein} O 2017 {\em ArXiv e-prints\/} (\textit{Preprint}
  \eprint{1702.05226})

\bibitem{Wu2016_SputteredYIGVertical}
Wu H, Wan C~H, Zhang X, Yuan Z~H, Zhang Q~T, Qin J~Y, Wei H~X, Han X~F and
  Zhang S 2016 {\em Physical Review B\/} {\bf 93} 060403
  \urlprefix\url{https://doi.org/10.1103/physrevb.93.060403}

\bibitem{Shan2017_NFO}
Shan J, Bougiatioti P, Liang L, Reiss G, Kuschel T and van Wees B~J 2017 {\em
  Applied Physics Letters\/} {\bf 110} 132406
  \urlprefix\url{https://doi.org/10.1063/1.4979408}

\bibitem{hirobe_generation_2015}
Hirobe D, Shiomi Y, Shimada Y, Ohe J~i and Saitoh E 2015 {\em Journal of
  Applied Physics\/} {\bf 117} 053904 ISSN 0021-8979, 1089-7550
  \urlprefix\url{http://aip.scitation.org/doi/10.1063/1.4907040}

\bibitem{kamra_super-poissonian_2016}
Kamra A and Belzig W 2016 {\em Physical Review Letters\/} {\bf 116} ISSN
  0031-9007, 1079-7114
  \urlprefix\url{https://link.aps.org/doi/10.1103/PhysRevLett.116.146601}

\bibitem{safranski_spin_2017}
Safranski C, Barsukov I, Lee H~K, Schneider T, Jara A~A, Smith A, Chang H, Lenz
  K, Lindner J, Tserkovnyak Y, Wu M and Krivorotov I~N 2017 {\em Nature
  Communications\/} {\bf 8} ISSN 2041-1723
  \urlprefix\url{http://www.nature.com/articles/s41467-017-00184-5}

\bibitem{Takei2015_SpinSuperfluid}
Takei S and Tserkovnyak Y 2015 {\em Physical Review Letters\/} {\bf 115}
  \urlprefix\url{https://doi.org/10.1103/physrevlett.115.156604}

\bibitem{li_ballistic_2005}
Li S~Y, Taillefer L, Wang C~H and Chen X~H 2005 {\em Physical Review Letters\/}
  {\bf 95} ISSN 0031-9007, 1079-7114
  \urlprefix\url{https://link.aps.org/doi/10.1103/PhysRevLett.95.156603}

\end{thebibliography}
\end{document}